\documentclass[a4paper,10pt]{article}

\usepackage{multirow}	
\usepackage[latin1]{inputenc}
\usepackage[T1]{fontenc}
\usepackage[british]{babel}
\usepackage{graphicx}
\usepackage[dvips]{geometry}
\usepackage{float}
\usepackage{array}
\usepackage{amsmath}
\usepackage{amsfonts}
\usepackage{pgf}
\usepackage{tikz}
\usepackage{psfrag}

\usetikzlibrary{snakes}
\usetikzlibrary{arrows}
\usetikzlibrary{shapes}

\newcommand{\ve}[2]{
\begin{scope}[xshift=#1 cm,yshift=#2 cm,rotate=45,scale=1.414]
	\clip[draw] (-0.5,-0.5) rectangle (0.5,0.5);
	\draw[line width=3pt] (-0.5,-0.5) circle (0.5cm)  (0.5,0.5) circle (0.5cm);
\end{scope}
}

\newcommand{\vi}[2]{
\begin{scope}[xshift=#1 cm,yshift=#2 cm,rotate=45,scale=1.414]
	\clip[draw] (-0.5,-0.5) rectangle (0.5,0.5);
	\draw[line width=3pt] (-0.5,0.5) circle (0.5cm)  (0.5,-0.5) circle (0.5cm);
\end{scope}
}

\newcommand{\vd}[2]{
\begin{scope}[xshift=#1 cm,yshift=#2 cm,rotate=45,scale=1.414]
	\clip[draw] (-0.5,-0.5) rectangle (0.5,0.5);
	\draw[line width=3pt] (-0.5,-0.5) circle (0.5cm);
\end{scope}
}

\newcommand{\vu}[2]{
\begin{scope}[xshift=#1 cm,yshift=#2 cm,rotate=45,scale=1.414]
	\clip[draw] (-0.5,-0.5) rectangle (0.5,0.5);
	\draw[line width=3pt] (0.5,0.5) circle (0.5cm);
\end{scope}
}

\newcommand{\vr}[2]{
\begin{scope}[xshift=#1 cm,yshift=#2 cm,rotate=45,scale=1.414]
	\clip[draw] (-0.5,-0.5) rectangle (0.5,0.5);
	\draw[line width=3pt] (0.5,-0.5) circle (0.5cm);
\end{scope}
}

\newcommand{\vl}[2]{
\begin{scope}[xshift=#1 cm,yshift=#2 cm,rotate=45,scale=1.414]
	\clip[draw] (-0.5,-0.5) rectangle (0.5,0.5);
	\draw[line width=3pt] (-0.5,0.5) circle (0.5cm);
\end{scope}
}

\newcommand{\vul}[2]{
\begin{scope}[xshift=#1 cm,yshift=#2 cm,rotate=45,scale=1.414]
	\clip[draw] (-0.5,-0.5) rectangle (0.5,0.5);
	\draw[line width=3pt] (0,-1) -- (0,1);
\end{scope}
}

\newcommand{\vur}[2]{
\begin{scope}[xshift=#1 cm,yshift=#2 cm,rotate=45,scale=1.414]
	\clip[draw] (-0.5,-0.5) rectangle (0.5,0.5);
	\draw[line width=3pt] (-1,0) -- (1,0);
\end{scope}
}

\newcommand{\vv}[2]{
\begin{scope}[xshift=#1 cm,yshift=#2 cm,rotate=45,scale=1.414]
	\clip[draw] (-0.5,-0.5) rectangle (0.5,0.5);
\end{scope}
}

\newcommand{\ti}[2]{
\begin{scope}[xshift=#1 cm,yshift=#2 cm,scale=1.]
	\clip[draw] (-1,-1) -- (-1,1) -- (0,0) -- cycle;
	\draw[line width=3pt] (0,0) circle (0.707cm);
\end{scope}
}

\newcommand{\tv}[2]{
\begin{scope}[xshift=#1 cm,yshift=#2 cm,scale=1.]
	\clip[draw] (-1,-1) -- (-1,1) -- (0,0) -- cycle;
\end{scope}
}

\newcommand{\tb}[2]{
\begin{scope}[xshift=#1 cm,yshift=#2 cm,scale=1.]
	\clip[draw] (-1,-1) -- (-1,1) -- (0,0) -- cycle;
	\draw[line width=3pt] (0,0) circle (0.707cm);
	\filldraw[gray] (-0.707,0) circle (6pt);
\end{scope}
}

\newcommand{\tu}[2]{
\begin{scope}[xshift=#1 cm,yshift=#2 cm,scale=1.]
	\clip[draw] (-1,-1) -- (-1,1) -- (0,0) -- cycle;
	\draw[line width=3pt] (0,0) circle (0.707cm);
	\filldraw[gray] (-0.707,0) ++(-5pt,-5pt) rectangle ++(10pt,10pt);
\end{scope}
}

\newcommand{\honey}[2]{
	\begin{scope}[xshift=#1 cm, yshift=0.57735*(#2 cm), scale=0.6667]
	\draw[line width=1pt, gray] (0,0) -- ++(0.5,-0.866) -- ++(1,0) -- ++(0.5,0.886) -- ++(-0.5,0.886) -- ++(-1,0) -- cycle;
	\end{scope}
}

\newcommand{\blob}{
	\tikz \filldraw (0,0) circle (1.75pt);
}

\newcommand{\blobw}{
	\tikz \filldraw[white, draw=black, thick] (0,0) circle (1.65pt);
}

\newcommand{\unblob}{
	\tikz \filldraw (0,0) rectangle (3.2pt,3.2pt);
}

\newcommand{\unblobw}{
	\tikz \filldraw[white, draw=black, thick] (0,0) rectangle (3.1pt,3.1pt);
}

\newcommand{\q}{\tilde{q}}
\newcommand{\On}{\mathcal{O}(n)}
\newcommand{\Z}{\mathbb{Z}}
\newcommand{\V}{\mathcal{V}}
\newcommand{\D}{\mathcal{D}}

\newcommand{\ket}[1]{
	\left| #1 \right>
}

\begin{document}

\title{Conformal boundary conditions in the critical $\On$ model and dilute loop models}

\author{J\'er\^ome Dubail$^{1}$, Jesper Lykke Jacobsen$^{2,1}$ and Hubert Saleur$^{1,3}$ \\
[2.0mm]
  ${}^1$Institut de Physique Th\'eorique, CEA Saclay,
  91191 Gif Sur Yvette, France \\
  ${}^2$LPTENS, 24 rue Lhomond, 75231 Paris, France \\
  ${}^3$Department of Physics,
  University of Southern California, Los Angeles, CA 90089-0484}

\maketitle

\begin{abstract}

 We study the conformal boundary conditions of the dilute ${\cal O}(n)$
 model in two dimensions. A pair of mutually dual solutions to the boundary
 Yang-Baxter equations are found. They describe anisotropic special
 transitions, and can be interpreted in terms of symmetry breaking
 interactions in the ${\cal O}(n)$ model. We identify the corresponding
 boundary condition changing operators, Virasoro characters, and
 conformally invariant partition functions. We compute the entropies
 of the conformal boundary states, and organize the flows between the
 various boundary critical points in a consistent phase diagram. The
 operators responsible for the various flows are identified. Finally,
 we discuss the relation to open boundary conditions in the ${\cal O}(n)$
 model, and present new crossing probabilities for Ising domain walls.

\end{abstract}

\section{Introduction}

Loop models in statistical mechanics have been studied for a very long time, both for their physical and mathematical properties. They have enjoyed considerable interest recently as fundamental objects in the SLE approach to critical phenomena, and in the search for  systems with quasiparticles obeying non-abelian statistics \cite{Kitaev,Freedman,Fendley1}. 

Although a lot is known about these models, new, very basic features keep being discovered. An example of this concerns conformal boundary conditions in the two dimensional (classical) case. 

Conformal boundary conditions (CBCs) are an important tool in the understanding of conformal field theories (CFTs) and in many applications. They are systematically classified for minimal models, and usually complicated to implement physically, except for the simplest ones: ``free'' and ``fixed'', in the proper variable. Surprisingly, it turns out that for dense loop models (see the definition below), a continuous set of CBCs is obtained simply by deciding to give loops that touch the boundary a fugacity $n_1$ different from the fugacity $n$ for loops in the bulk. These CBCs---now dubbed ``JS boundary conditions''---have been studied in a series of papers using lattice and algebraic techniques \cite{JS1,JS2,DJSdense,deGier2BTL}. They have also been tackled in terms of matrix models \cite{Kostov,Bourgine}, and found applications in the study of logarithmic CFTs \cite{PearceRasmussenZuber, ReadSaleur}. 

There are basically two kinds of loop models: the loops are either dilute or dense (for recent reviews, see \cite{NienhuisHouches,JesperReview}). In the former case, the loops occupy only a critical fraction of the available lattice sites, while in the latter case, the loops cover more than this critical fraction. Bulk properties then are the same, independently of this fraction, and characterized by what is called the dense universality class (often, this universality class is most conveniently studied on the square lattice by forcing the loops to cover the whole lattice, giving rise to a fully packed loop (FPL) model. One has to be careful however that properties in the FPL case are not always universal). If $n$ is the fugacity of the loops, the dilute theory is believed to be in the same universality class as the ${\cal O}(n)$ spin model (once continued analytically to real values of $n$), and conformal for $n\in [-2,2]$. The dense theory on the other hand does not describe the low temperature phase of the generic  ${\cal O}(n)$ model, as self-intersections are absent, which now play a crucial role \cite{JacobsenReadSaleur}. The classification of conformal boundary conditions for ${\cal O}(n)$ models in their Goldstone phase is still an open problem.

The dilute model admits one more relevant parameter than the dense model: the fugacity of the basic monomers making up the loops. Only when this parameter $x$ is adjusted to a critical value $x_{c}$ does the model become conformally invariant in the bulk. This freedom translates into a more complicated boundary behaviour: not only can one adjust the weight of loops touching the boundary, but one can (and must) also adjust the weight of boundary monomers. Indeed, in the dilute model a generic loop touches the surface with probability zero, so the surface fugacity has to be critically enhanced to allow a finite fraction of boundary sites to be occupied by loops. This is not necessary in the dense model since the boundary is always covered with loops in the continuum limit. This has also a nice interpretation in the language of symmetry breaking boundary interactions in the ${\cal O}(n)$ model \cite{DiehlEisenriegler}, that we discuss shortly below, and is the key to physical applications to appear elsewhere \cite{DJSlett}. Once this feature is under control, the basic aspects of CBCs in the dilute case are formally similar to those in the dense case, after a proper redefinition of the parameters. Carrying this out in details gives us control of key combinatorial quantities and crossing probabilities that were not known up to now.

The paper is organized as follows. In section 2 we introduce the problem in more details and discuss the basic features of the phase diagram. In section 3 we find a new solution to the boundary Yang-Baxter equation, which generalizes the solution of \cite{BatchelorYung} in the special case to what we call the ``anisotropic special case''. This constitutes the new CBCs for the dilute loop model. In section 4 we discuss the basic features of these new CBCs, and  determine the associated critical exponents and annulus partition functions. In section 5, we discuss some features of the boundary phase diagram, in particular the existence of RG flows and their relations to the boundary entropy. In section 6, we discuss yet another type of CBC, where, instead of affecting the weights of loops touching the boundary, we introduce a boundary magnetic field, and thus ``open'' the loops on the boundary. In the dense case, it turned out that these open boundary conditions could be reformulated exactly as a particular case of the JS boundary conditions. The situation is a bit more complicated here, however similar. We finally apply our results to the determination of crossing probabilities of Ising clusters on an annulus. In section 7 we give a more detailed summary of our results, and we discuss some further generalizations and open directions.




\section{Special transitions in the $\On$ model}
\label{part:model}
We start by giving a small review of known results about the $\On$ model and its (bulk and surface) critical behaviour. We do not try to be exhaustive, and refer the interested reader to the vast  literature on the subject \cite{NienhuisHouches,JesperReview,Binder, NienhuisConjecture, NienhuisDL, CardyBook, BatchelorCardy} for further details. Nevertheless, we want to set up a self-contained physical framework in which  further discussions will appear most natural.

\subsection{The $\On$ model and its loop expansion on the honeycomb lattice}

The $\On$ model is usually presented as follows. It can be defined on an arbitrary lattice in $D$ dimensions, although we will soon restrict ourselves to $D=2$ and to the honeycomb lattice. It is a spin model, with $n$-components spins $S^\mu_i$ living on the sites $i$ of the lattice, with $\mu \in \left\{ 1,\dots,n \right\}$. The spins are subject to the constraint $\vec{S}_i.\vec{S}_i = \sum_{\mu =1}^n S^\mu_i S^\mu_i = n$, which means they actually live on a $(n-1)$-dimensional sphere. The partition function of the model is taken\footnote{Of course, a more natural $\On$ model would be $Z = \mathrm{Tr} \left\{ \prod_{<ij>} \exp \left( K \vec{S}_i. \vec{S}_j \right) \right\}$, and this is indeed the generic $\On$ model which has been studied for a very long time and in higher dimensions. When $D=2$ however, one usually prefers to consider the model $(\ref{eq:ZOn})$ as it is much simpler and yet it is believed to belong to the same universality class up to the critical point. In the Goldstone phase however, the two models are believed to have different critical behaviour generically (the Ising case $n=1$ is an noteworthy exception), as discussed briefly in the introduction (see also reference \cite{JacobsenReadSaleur}).} as
\begin{equation}
Z = \mathrm{Tr} \left\{ \prod_{<ij>} \left( 1+ x \vec{S}_i. \vec{S}_j \right) \right\}
\label{eq:ZOn}
\end{equation}
where the $\left<ij\right>$ are the lattice edges in the bulk. $x$ is a parameter interpreted as an inverse temperature. In particular, note that the $\mathcal{O}(n=1)$ model is the Ising model with $x= \tanh K$ if the partition function of the Ising model is written $\displaystyle Z_{\rm{Ising}} = \sum_{\rm{conf.} \; \left\{ S \right\} } \prod_{<ij>} e^{K S_i S_j}$.
 The trace $\rm{Tr} \left\{ . \right\}$ is proportional to $\displaystyle \prod_i \int d \vec{S}_i \delta ( \vec{S}_i. \vec{S}_i -n)$. It can be normalized such that it has the properties $\rm{Tr} \left\{ 1 \right\} = 1$, $\rm{Tr} \left\{ S^\mu_i S^\nu_j \right\} = \delta_{i j} \delta_{\mu \nu}$ and $\rm{Tr} \left\{ S^\mu_i  \right\} = \rm{Tr} \left\{ \left( S^\mu_i \right)^3 \right\} \; = \; \rm{Tr} \left\{ \left( S^\mu_i \right)^5 \right\} \; = \; \dots \; = \; 0$.
\paragraph{}
With these relations, the partition function $(\ref{eq:ZOn})$ allows a ``loop'' expansion. This loop formulation is well-known, and is obtained by a high-temperature (small $x$) expansion. When we take the trace, the terms with odd powers of $S^\mu_i$ cancel. The terms of the form $S_{i_1}^{\mu_1} S_{i_2}^{\mu_1} S_{i_2}^{\mu_2} S_{i_3}^{\mu_2} \dots S_{i_k}^{\mu_k} S_{i_1}^{\mu_k}$ subsist, where $\left< i_1 i_2 \right>$, $\left< i_2 i_3 \right>$, etc. and $\left< i_k i_1 \right>$ are links of the lattice. Each such term can be represented as a loop on the lattice, and gets a weight $n$ once the trace is taken. Other terms with an even power of $S_i^\mu$, such as $\left(S_i^\mu\right)^4$ can be nonzero. Such terms can be avoided if we restrict ourselves to trivalent lattices (with at most three edges connected to each site). This is the reason why the two-dimensional $\On$ model is usually defined on the hexagonal lattice. Of course, one could argue on universality grounds that in the end the critical behaviour of the model is lattice independent. In any case, for the honeycomb lattice the loop formulation is exact, and we end up with the celebrated loop partition function 
\begin{equation}
Z = \sum_{\rm{conf.}} x^{X} n^{N}
\label{eq:ZOnLoop}
\end{equation}
where the sum is taken over all the configurations of non-intersecting loops that one can draw on the honeycomb lattice. $X$ is the number of monomers for each configuration, while $N$ is the number of loops. Note that the loop model with the partition function $(\ref{eq:ZOnLoop})$ is more general than the spin model $(\ref{eq:ZOn})$, because in the former case $n$ is not restricted to be a positive integer. However, both interpretations of the $\On$ model, as a magnetism model or as a geometric loop model, are useful to get some intuition about its rich critical behaviour.

\paragraph{}
As it is a ferromagnetic spin model, we expect the $\On$ model to possess a disordered phase when $x$ is small, and an ordered phase when $x$ is big. In two dimensions, the transition between these two regimes is known to be of first order when $n > 2$. Therefore we restrict to $-2 <n \leq 2$, where the model exhibits a second-order phase transition. In this case, the following phase diagram holds
$$
\begin{tikzpicture}
  \draw [very thick] (0,0) -- (12,0);
  \draw [<<-,very thick] (2.5,0) -- (5,0);
  \draw [->>,very thick] (0,0) -- (7,0);
  \draw [<<-,very thick] (10,0) -- (12,0);
  \filldraw[black] (0,0) circle (1mm) +(0,0.4) node {No loops} +(0,-0.4) node {$x=0$};
  \filldraw[black] (5,0) circle (1mm) +(0,0.4) node {Dilute loops} +(0,-0.4) node {$x=x_c$};
  \filldraw[black] (8.5,0) circle (1mm) +(0,0.4) node {Dense loops} +(0,-0.4) node {$x=x_0$};
  \filldraw[black] (12,0) circle (1mm) +(0,0.4) node {Fully packed loops} +(0,-0.4) node {$x=\infty$};
  \draw[thick, snake=brace] (4.5,-1) -- (0.5,-1);
  \draw[thick, snake=brace] (11.5,-1) -- (5.5,-1);
  \draw (2.5,-1.4) node {Massive theory};
  \draw (2.5,-1.8) node {Disordered spins};
  \draw (8.5,-1.4) node {Dense loops phase (massless)};
  \draw (8.5,-1.8) node {Ordered spins};
\end{tikzpicture}
$$
In terms of loops, this can be understood as follows. For $x=0$ there is no loop at all. For small $x>0$, some small loops appear and as $x$ increases, the loops are more and more numerous and can become longer and longer. At some point ($x=x_c$) the mean length of the loops going through a randomly chosen point diverges. If one increases the fugacity $x$ a little more, the loops proliferate and they finally cover each edge of the lattice with finite probability (even when the lattice is infinite), so they fill the whole space in the continuum limit. Note that in the discrete setting however, even for $x>x_c$, the loops do not fill all the sites of the lattice. But when $x=\infty$, that is in the fully packed loop (FPL) phase, all the sites of the lattice are covered by a piece of loop. This is a rather strong constraint on the loop configurations, and in particular the universality class of the FPL phase is not lattice independent. We will not be concerned by this theory in what follows, however it should be noted that it corresponds (on the honeycomb lattice) to a non-trivial theory, different from the one of the dense phase. The dilute phase ($x=x_c$), the dense phase ($x_c<x<\infty$) and the FPL phase ($x=\infty$) are described by conformal field theories with central charges \cite{NienhuisDL, KdGNienhuis}
\begin{subequations}
\begin{eqnarray}
\displaystyle c_{\rm{dilute}} &=& 1 - 6\frac{(g-1)^2}{g} \qquad g \;=\; 1+\frac{\gamma}{\pi} \\
\displaystyle c_{\rm{dense}} &=& 1 - 6\frac{(g'-1)^2}{g'} \qquad g'\; =\; 1-\frac{\gamma}{\pi} \\
\displaystyle c_{\rm{FPL}} &=& 1 + c_{\rm{dense}}
\end{eqnarray}
\end{subequations}
where $n=2 \cos \gamma$ with $\gamma \in \left[ 0 , \pi \right)$. Note the consistency with Zamolodchikov's $c$-theorem \cite{cTheorem}, which states that the central charge always decreases along the RG flow\footnote{This theorem is true for unitary theories, which is not the case in general for the $\On$ model. Therefore the application of this theorem here is not justified and certainly not rigorous, but it still provides some insight in the behaviour of the model.}.

\paragraph{}
The exact localization of the points $x_0$ and $x_c$ for the honeycomb lattice is the point of Nienhuis' conjecture \cite{NienhuisConjecture}
\begin{equation}
x_c = \frac{1}{\sqrt{2+\sqrt{2-n}}} \qquad \qquad x_0 = \frac{1}{\sqrt{2-\sqrt{2-n}}}
\end{equation}
which we will discuss in greater detail in section $\ref{part:Sklyanin}$.

\paragraph{}
At this point one might be worried by the special case $n=1$, which looks a bit different from the traditional results about the Ising model. However, it is not difficult to see that everything fits together here if we recall that $x$ is related to the usual Ising coupling by $x=\tanh K$. Then $x=x_0=1$ corresponds to $K=+\infty$, and $x > x_0$ gives a complex coupling $K$. Hence, the zero temperature point of the Ising model is $x=x_0$, and the part $x>x_0$ is irrelevant when we deal with the model with $K$ real. Also, the Ising model is known to be dual to an Ising model on the dual lattice, so the theories at $K=0$ and $K=+\infty$ should be dual to each other. Again there is no contradiction here, since $c_{\rm{dense}}=0$, as expected if it has to be related to a massive theory.

\subsection{Surface critical behaviour : ordinary, extraordinary and special transitions}
\label{sec:special}
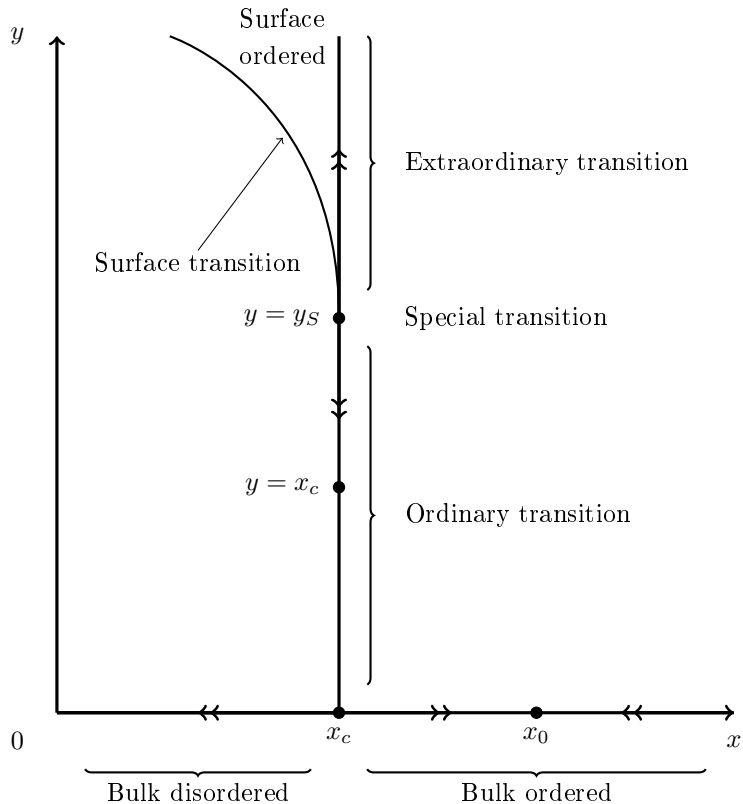
\begin{figure}[h]
\centering
\begin{tikzpicture}[scale=0.75]
  \draw [very thick, ->] (0,0) -- (12,0);
  \draw [<<-,very thick] (2.5,0) -- (5,0);
  \draw [->>,very thick] (0,0) -- (7,0);
  \draw [<<-,very thick] (10,0) -- (12,0);
  \draw [very thick, ->] (0,0) -- (0,12);
  \draw [thick] (5,7) .. controls (5,11) and (2,12) .. (2,12);
  \draw (-0.7,12) node {$y$};
  \draw (12,-0.5) node {$x$};
  \draw (-0.7,-0.5) node {$0$};
  \draw [very thick] (5,0) -- (5,12);
  \draw [<<-,very thick] (5,5.2) -- (5,7);
  \draw [->>,very thick] (5,7) -- (5,10);
  \filldraw[black] (5,4) circle (1mm) +(-1,0) node {$y=x_c$};
  \filldraw[black] (5,7) circle (1mm) +(-1,0) node {$y=y_S$};
  \filldraw[black] (5,0) circle (1mm) +(0,0.4) +(0,-0.4) node {$x_c$};
  \filldraw[black] (8.5,0) circle (1mm) +(0,0.4) +(0,-0.4) node {$x_0$};
  \draw[thick, snake=brace] (5.5,6.5) -- (5.5,0.5);
  \draw[thick, snake=brace] (5.5,12) -- (5.5,7.5);
  \draw[thick, snake=brace] (4.5,-1) -- (0.5,-1);
  \draw[thick, snake=brace] (11.5,-1) -- (5.5,-1);
  \draw (6,7) node[right] {Special transition};
  \draw (6,9.75) node[right] {Extraordinary transition};
  \draw (6,3.5) node[right] {Ordinary transition};
  \draw (2.5,8) node {Surface transition};
  \draw (4,12) node[above] {Surface} node[below] {ordered};
  \draw [->] (2.5,8.2) -- (4,10.2);
  \draw (2.5,-1.4) node {Bulk disordered};
  \draw (8.5,-1.4) node {Bulk ordered};
\end{tikzpicture}
\caption{Phase diagram of the $\On$ model with bulk spin coupling $x$ and boundary spin coupling $y$. In this paper we will focus on the line $x=x_c$, which corresponds to the dilute phase for the loops.}
\label{fig:phasebulkboundary}
\end{figure}

So far, we have focused on the bulk critical behaviour of the $\On$ model. Now let us turn to its behaviour near a boundary. We consider the partition function
\begin{equation}
Z = \mathrm{Tr} \left\{ \prod_{\rm{bulk} \; <ij>} \left( 1+ x \vec{S}_i. \vec{S}_j \right) \prod_{\rm{boundary} \; <ij>} \left( 1+ y \vec{S}_i. \vec{S}_j \right) \right\}
\label{eq:ZOnboundary}
\end{equation}
which allows the possibility to enhance the coupling $y$ between two boundary spins above its bulk value $x$. To begin with, let us drop this possibility and assume that the spin coupling is the same for the spins living on the boundary and those in the bulk: $y=x$. When one goes from $x<x_c$ to $x>x_c$, the bulk spins order. However, the boundary spins have less close neighbours and therefore the spontaneous magnetization vanishes near the boundary, on a typical length of the order of the bulk correlation length. This transition to ordered spins in the bulk, with a spontaneous magnetization in the bulk which drops to zero near the boundary, is called the \textit{ordinary transition}. Now, consider $x<x_c$: the bulk spins are disordered. If the boundary coupling $y$ is sufficiently enhanced above $x$, then the boundary spins order, despite the fact that the bulk is disordered. This transition from disordered to ordered spins in a one-dimensional layer along the boundary of width of order of the bulk correlation length, while the bulk remains disordered, is called the \textit{surface transition}. It is expected to belong to the universality class of the one-dimensional $\On$ model\footnote{Of course there can be such a boundary transition at finite coupling $K$ only if the lower critical dimension $D_{LC}$ of the $\On$ model is such that $D_{LC} \leq 1$. In particular, this would imply $n<1$. However, as in the case of the Ising model above, we will actually never work with the real coupling $K$ or its boundary equivalent $K_b$, but rather with $x=\tanh K$ and $y = \tanh K_b$. Thus nothing should prevent us from considering situations where $y > 1$, even if it does not correspond to a real coupling $K_c$. In the loop model, the parameters $x$ and $y$ are the natural parameters, and the surface transition exists in this model at finite $y$. Thus in this paper we will always avoid this discussion about the lower critical dimension, and we will not restrict ourselves to $n < 1$ (as the authors of \cite{BatchelorCardy} did), because even if the surface and extraordinary transition do not appear in the $\On$ spin model at finite $K_b$, they exist as geometric transitions in the loop formulation of the model. We keep discussing these transitions in the spin language because it still gives some intuition about the critical behaviour.}. If we start from a configuration with the surface spins ordered and the bulk disordered $(x<x_c)$, and go to $x>x_c$ so that the bulk spins order, the transition is called \textit{extraordinary transition}. If we follow the surface transition line $y(x)$ when $x \rightarrow x_c$, the boundary and the bulk correlation lengths both diverge, and this point is known as the \textit{special transition}. At this point, the coupling $y$ is critically enhanced, in the sense that bulk and surface order simultaneously. In other words, $y$ precisely compensates for the lack of nearest neighbours at the boundary. This bulk/boundary critical behaviour is summarized in figure $\ref{fig:phasebulkboundary}$.

\begin{figure}[htbp]
\centering
a.\begin{tikzpicture}[scale=0.75]
\foreach \x in {0,2,4}
\foreach \y in {-2,0,2,4,6,8}
{
	\honey{\x}{\y}
}
\foreach \x in {1,3,5}
\foreach \y in {-1,1,3,5,7}
{
	\honey{\x}{\y}
}
\draw (-0.8,1.7) node[rotate=90] {Boundary};
\draw[line width=2pt, red, scale=0.6667] (0,0) ++(1.5,0.866) -- ++(0.5,0.866) -- ++(1,0) -- ++(0.5,0.866) -- ++(-0.5,0.866) -- ++(0.5,0.866) -- ++(1,0) -- ++(0.5,0.866) -- ++(1,0) -- ++(0.5,-0.866) -- ++(1,0) -- ++(0.5,-0.866) -- ++(-0.5,-0.866) -- ++(0.5,-0.866) -- ++(-0.5,-0.866) -- ++(0.5,-0.866) -- ++(-0.5,-0.866) -- ++(-1,0) -- ++(-0.5,0.866) -- ++(-1,0) -- ++(-0.5,-0.866) -- ++(-1,0) -- ++(-0.5,0.866) -- ++(-1,0) -- cycle;
\draw[line width=2pt, red, scale=0.6667] (1.5,6.062) -- ++(0.5,0.866) -- ++(1,0) -- ++(0.5,0.866) -- ++(1,0) -- ++(0.5,-0.866) -- ++(-0.5,-0.866) -- ++(-1,0) -- ++(-0.5,-0.866) -- ++(-1,0) -- cycle;
\draw[line width=2pt, red, scale=0.6667] (4.5,2.598) -- ++(0.5,0.866) -- ++(1,0) -- ++(0.5,-0.866) -- ++(-0.5,-0.866) -- ++(-1,0) -- cycle;
\end{tikzpicture} \qquad \qquad b.
\begin{tikzpicture}[scale=0.75]
\foreach \x in {0,2,4}
\foreach \y in {-2,0,2,4,6,8}
{
	\honey{\x}{\y}
}
\foreach \x in {1,3,5}
\foreach \y in {-1,1,3,5,7}
{
	\honey{\x}{\y}
}
\draw (-0.8,1.7) node[rotate=90] {Boundary};
\draw[line width=2pt, red, scale=0.6667] (0.5,-2.598) -- ++(-0.5,0.866) -- ++(0.5,0.866) -- ++(-0.5,0.866) -- ++(0.5,0.866) -- ++(-0.5,0.866) -- ++(0.5,0.866) -- ++(-0.5,0.866) -- ++(0.5,0.866) -- ++(-0.5,0.866) -- ++(0.5,0.866) -- ++(-0.5,0.866) -- ++(0.5,0.866);
\draw[line width=2pt, red, scale=0.6667] (0,0) ++(1.5,0.866) -- ++(0.5,0.866) -- ++(1,0) -- ++(0.5,0.866) -- ++(-0.5,0.866) -- ++(0.5,0.866) -- ++(1,0) -- ++(0.5,0.866) -- ++(1,0) -- ++(0.5,-0.866) -- ++(1,0) -- ++(0.5,-0.866) -- ++(-0.5,-0.866) -- ++(0.5,-0.866) -- ++(-0.5,-0.866) -- ++(0.5,-0.866) -- ++(-0.5,-0.866) -- ++(-1,0) -- ++(-0.5,0.866) -- ++(-1,0) -- ++(-0.5,-0.866) -- ++(-1,0) -- ++(-0.5,0.866) -- ++(-1,0) -- cycle;
\draw[line width=2pt, red, scale=0.6667] (1.5,6.062) -- ++(0.5,0.866) -- ++(1,0) -- ++(0.5,0.866) -- ++(1,0) -- ++(0.5,-0.866) -- ++(-0.5,-0.866) -- ++(-1,0) -- ++(-0.5,-0.866) -- ++(-1,0) -- cycle;
\draw[line width=2pt, red, scale=0.6667] (4.5,2.598) -- ++(0.5,0.866) -- ++(1,0) -- ++(0.5,-0.866) -- ++(-0.5,-0.866) -- ++(-1,0) -- cycle;
\end{tikzpicture} \qquad \qquad
\caption{The ordinary (a) and extraordinary (b) boundary conditions for the dilute loop model. The ordinary b.c corresponds to a free b.c for the loops. With extraordinary b.c, one loop is adsorbed on the surface, preventing the remaining loops to touch it. The latter then behave exactly as if there was a free b.c.}
\label{fig:ordextra}
\end{figure}
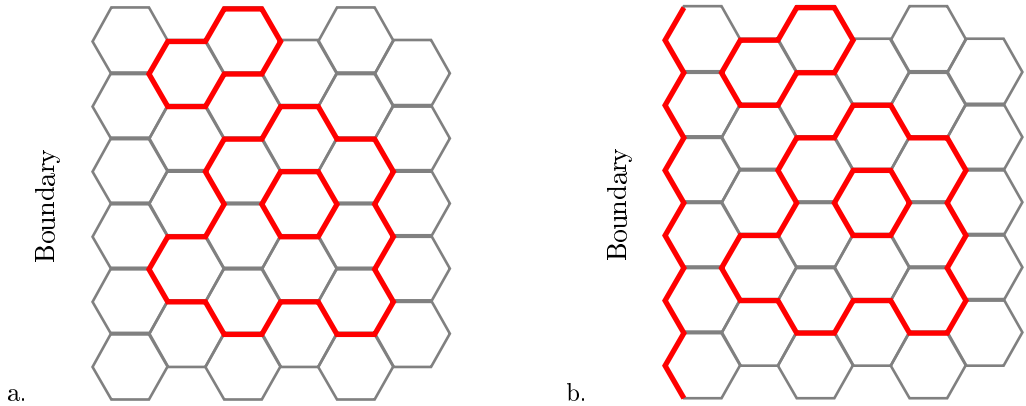

\paragraph{}
The loop interpretation of the special transition goes as follows. The partition function $(\ref{eq:ZOnboundary})$ becomes
\begin{equation}
Z = \sum_{\rm{conf.}} x^{X} y^{Y} n^{N}
\label{eq:ZOnLoopboundary0}
\end{equation}
where $N$ is the number of loops, $X$ is the number of bulk monomers, and $Y$ is the number of monomers at the boundary (for each configuration). When we focus on the line $x=x_c$, as we will do in the rest of this article, we have a model of dilute loops with fugacity $n$ which get different energies per unit length when they come to touch the boundary. Since we now work with the following one-dimensional phase diagram, there is no point in talking about ordinary or extraordinary \textit{transitions}, and we will adopt the terminology of ordinary and extraordinary \textit{boundary conditions} (b.c) instead.
$$
\begin{tikzpicture}[>=stealth]
  \draw [very thick] (0,0) -- (12,0);
  \draw [<<-,very thick] (5,0) -- (12,0);
  \draw [->>,very thick] (0,0) -- (9.5,0);
  \filldraw[black] (0,0) circle (1mm) +(0,-0.4) node {$y=0$};
  \filldraw[black] (3,0) circle (1mm) +(0,-0.4) node {$y=x_c$};
  \filldraw[black] (7,0) circle (1mm) +(0,-0.4) node {$y=y_S$};
  \filldraw[black] (12,0) circle (1mm) +(0,-0.4) node {$y=\infty$};
  \draw[thick, snake=brace] (6.5,-1) -- (0,-1);
  \draw[thick, snake=brace] (12,-1) -- (7.5,-1);
  \draw (7,0.4) node {Special b.c}; 
  \draw (3.3,-1.4) node {Ordinary b.c};
  \draw (9.7,-1.4) node {Extraordinary b.c}; 
\end{tikzpicture}
$$
When $y$ is small, the loops avoid the boundary, and behave as if they lived on a lattice with one less row on which we put free boundary conditions. Hence the theories with $y=0$ or $y=x=x_c$ are obviously the same\footnote{That is the reason why we do not indicate a RG flow between the points $y=0$ and $y=x_c$.}, and correspond to free boundary conditions for the loops. On the contrary, if $y$ is big enough, then there is one polymer which is adsorbed along the boundary, and it prevents the other loops from touching the surface. These loops are then not affected by the surface, and they behave as in the case of ordinary b.c. Hence the RG fixed points describing the ordinary and extraordinary b.c correspond to the same theory, except for the fact that there is one adsorbed polymer on the surface in the extraordinary case. This is illustrated in figure $\ref{fig:ordextra}$.

\paragraph{}
The transition between the free regime and the adsorbed one is the special transition. The special point corresponds to a non-trivial scaling limit, where the loops can touch the boundary without being completely glued on it. To our knowledge, the exact localization of the special point on the honeycomb lattice has first been conjectured by Batchelor \& Yung \cite{BatchelorYung}
\begin{equation}
y_S = \left( 2-n \right)^{-1/4}
\end{equation}
and again, we will come back to this result in section $\ref{part:Sklyanin}$.

\begin{figure}[t]
\centering
a. \includegraphics[width=0.9\textwidth]{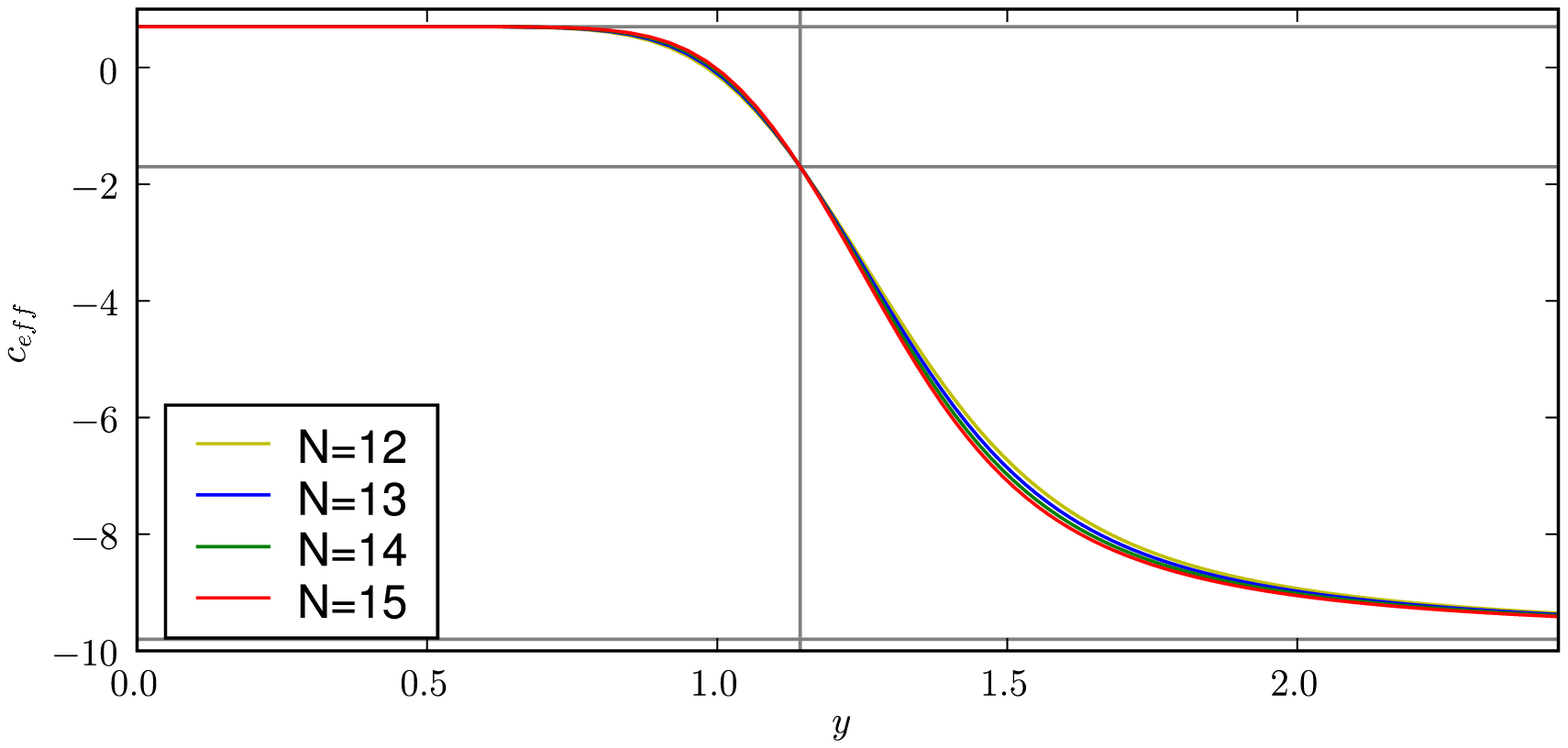} \\
b. \includegraphics[width=0.9\textwidth]{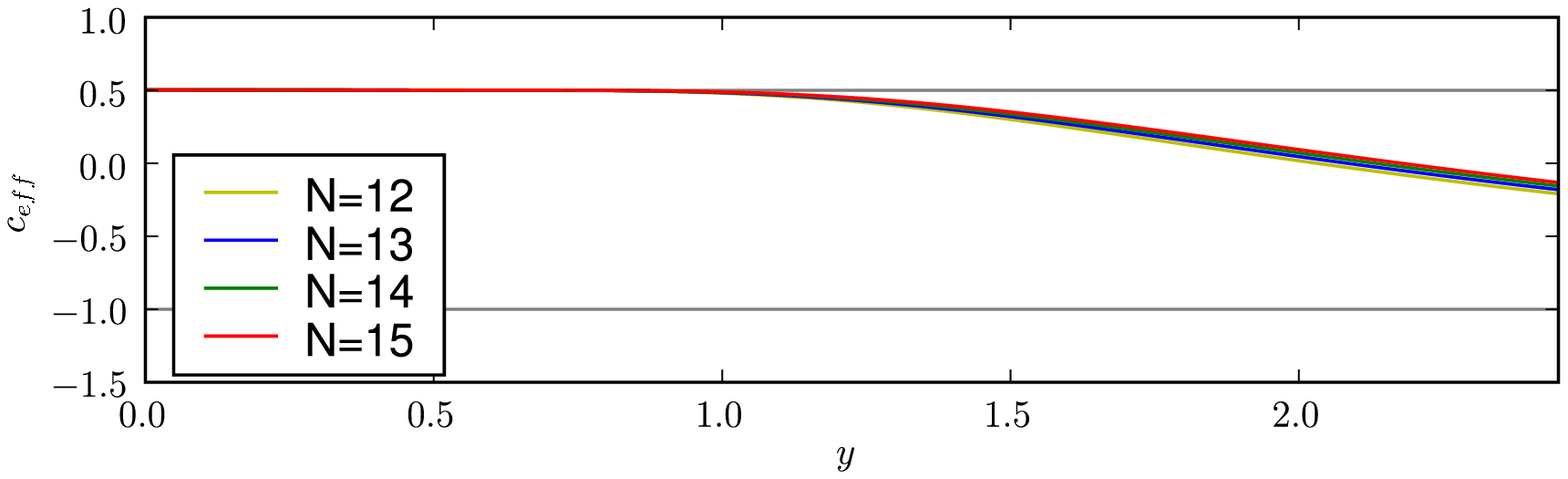}
\caption{Influence of the surface monomer fugacity $y$ in the dilute (a) and dense phase (b) of the $\On$ loop model. Here we have $n=\sqrt{2}$, $x=x_c \approx 0.60$ for (a), $x=x_0 \approx 0.90$ for (b). In (a) the CFT has central charge $7/10$. We see that there are three regimes: $y<y_S$ where $c_{\rm{eff}}=c=0.7$, $y=y_S$ with $c_{\rm{eff}}=c-24 h_{1,2}=-1.7$ and $y>y_S$ with $c_{\rm{eff}}=c- 24 h_{2,1}= -9.8$ (one-arm exponent in the dilute phase). In (b) the central charge is $1/2$. Although the numerical precision is not good, our conclusion is that $c_{\rm{eff}}=c=0.5$ for every finite $y$ (indeed, when the size $N$ increases, $c_{\rm{eff}}(N)$ goes to $c=0.5$). For $y=\infty$ it is $c_{\rm{eff}}=c-24 h_{1,2}=-1$ (one-arm exponent in the dense phase).}
\label{fig:SpDenseDilute}
\end{figure}

\paragraph{}
We would like to conclude this section with a remark about the dense phase of the $\On$ loop model. For $x > x_c$ one could also consider a model where $y$ is enhanced, and wonder whether there is a special transition in the dense loop model or not. The answer is no. The reason for this is that in dense loop models, the boundary is already covered with loops, so one cannot change the critical behaviour by encouraging the loops to touch the boundary. We expect that as soon as $y<\infty$, the coupling renormalizes towards $y=x$. However, in the present model on the honeycomb lattice, the situation $y=\infty$ still corresponds to a loop adsorbed on the boundary, exactly as in the dilute phase ($x=x_c$). This means that, when we diagonalize the transfer matrix of the model, one half-loop is adsorbed on the boundary while the other half-loop moves freely in the system. Then the effective central charge we compute should be $c_{\rm{eff}} = c_{\rm{dense}} - 24 h_{1,2}$ rather than $c_{\rm{dense}}$ ($h_{1,2}$ being the one-arm exponent in the dense phase). A numerical check of these observations is presented in figure $\ref{fig:SpDenseDilute}$. It is obtained by computing the free energy per site $f_L$ for successive $L=8,9,\dots, 15$, and relating the finite-size corrections of $f_L$ to the effective central charge by the well-known relation
\begin{equation}
f_L = f_{\rm{bulk}} + \frac{f_{\rm{boundary}}}{L} - \frac{\pi c_{\rm{eff}}}{24 L^2} + \mathcal{O} \left( \frac{1}{L^3} \right)
\end{equation}
up to order $\mathcal{O} \left( \frac{1}{L^4} \right)$. More generally, the behavior discussed in \cite{JS1,JS2} was technically obtained only for loops covering the whole square lattice, that is a maximally dense situation. It turns out that, if one relaxes this constraint to the more general dense case, where loops cover more than the critical fraction of available lattice sites (that is, $x_c<x<\infty$), the same behavior is obtained: the same formulas for conformal weights apply, independently of the fugacity of surface monomers, and of bulk monomers, that is, independently of $x$ and $y$.

\subsection{Anisotropic special transition}
\label{sec:anisospecial}
Following Diehl \& Eisenriegler \cite{DiehlEisenriegler}, we introduce an anisotropic generalization of $(\ref{eq:ZOnboundary})$. The idea is to break the $\On$ symmetry of the interaction at the boundary down to $\mathcal{O}(n_1) \times \mathcal{O}(n-n_1)$, which allows the coupling to be different for the $n_1$ first components of the spin and for the $n-n_1$ other ones. This can be written
\begin{eqnarray}
\label{eq:ZOnanisotropic}
Z&=&\mathrm{Tr} \left\{ \prod_{\mathrm{bulk}\;  <ij>} \left( 1 + x_c \sum_{\mu \in \left\{ 1,\dots,n \right\} }S^\mu_i S_j^\mu \right) \right.\\
\nonumber &&\qquad \times \left.  \prod_{\mathrm{boundary} \; <ij> } \left( 1 + y_{n_1} \sum_{\alpha \in \left\{ 1,\dots,n_1 \right\}} S_i^\alpha S_j^\alpha  + y_{n-n_1} \sum_{\beta \in \left\{ n_1+1,\dots,n \right\}} S_i^\beta S_j^\beta  \right)\right\}
\end{eqnarray}
which of course makes sense only when $n$ and $n_1$ are both integers. The loop framework, however, provides an analytic continuation to non-integer numbers of components $n_1$, exactly as it did for the bulk parameter $n$.

\paragraph{}
The boundary critical behaviour of this model was studied perturbatively by Diehl \& Eisenriegler \cite{DiehlEisenriegler}. They used scaling arguments, mean-field theory and $\epsilon$-expansion below the upper critical dimension $D=4$ to deduce the phase diagram and the crossover exponents appearing in their model. Let us sketch some of there results as follows. The case when $y_{n_1}=y_{n-n_1}$ is called \textit{isotropic} and was discussed in section $\ref{sec:special}$: when the bulk crosses the line $x=x_c$ and goes from its disordered to its ordered phase, the boundary behaviour depends on the boundary coupling $y$ and three different transitions are possible : ordinary, extraordinary or special. In the anisotropic case, things are quite similar. Let us start with $y_{n_1}$ and $y_{n-n_1}$ both small. Clearly, this cannot be very different from the isotropic case when $y$ is small : because of the lack of close neighbours, the (anisotropic) coupling between the boundary spins is weaker than in the bulk and the spontaneous magnetization cancels at the boundary. This means that in the space of parameters $(y_{n_1},y_{n-n_1})$, the RG fixed point of the ordinary transition is attractive.

\begin{figure}[h]
\centering
\psfrag{Anisotropy}[c]{Anisotropy}
\psfrag{isotropic line}[c]{Isotropic coupling}
\psfrag{Sp}[l]{$Sp$}
\psfrag{ord}[l]{$Ord$}
\psfrag{w_u}[c]{$y_{n-n_1}$}
\psfrag{w_b}[c]{$y_{n_1}$}
\psfrag{AS1}[c]{$AS_{n_1}$}
\psfrag{AS2}[l]{$AS_{n-n_1}$}
\psfrag{Ordinary}[l]{Ordinary transition}
\psfrag{Extraordinary}[l]{Extraordinary transition}
\includegraphics[width=0.9\textwidth]{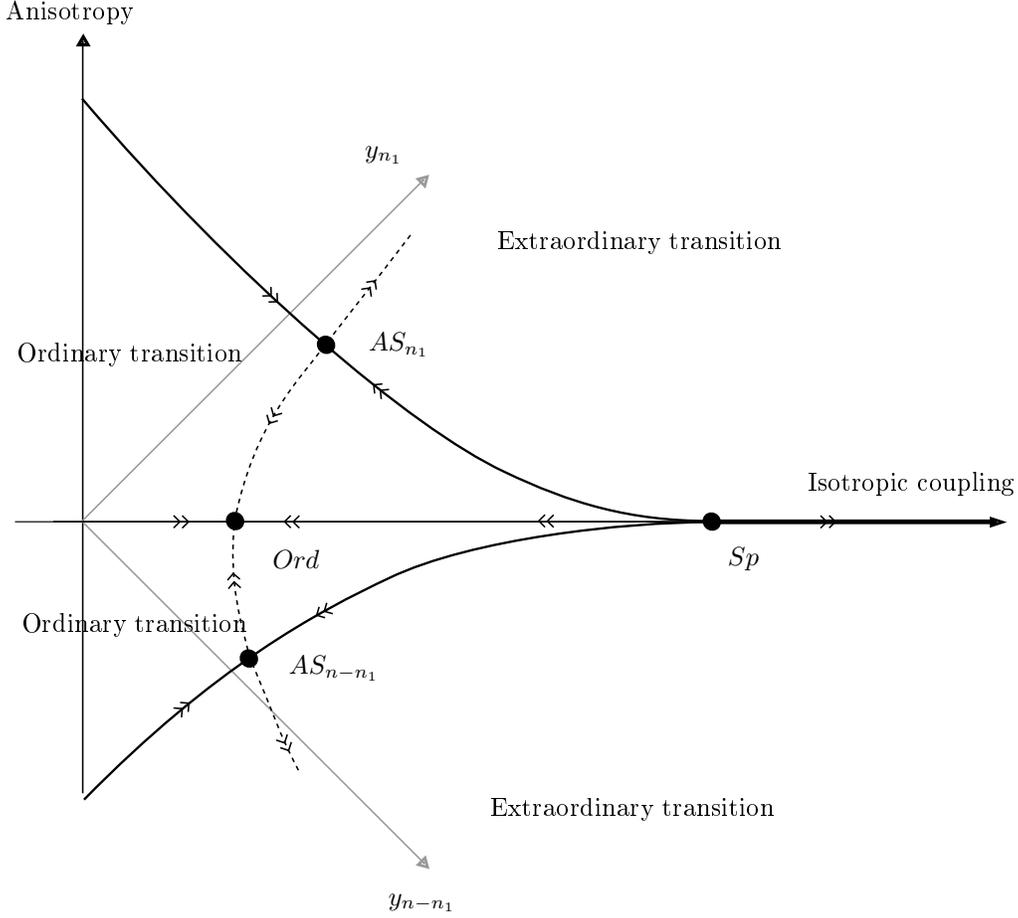}
\caption{Generic phase diagram for $0 < n_1< n$ in the rotated $(y_{n_1}, y_{n-n_1})$ plane.}
\label{fig:phase}
\end{figure}

\paragraph{}
In the opposite case, when $y_{n_1}$ and $y_{n-n_1}$ are both large, the boundary spins are ordered even when the bulk is disordered, and there is a spontaneous magnetization at the boundary. This has already been discussed in the isotropic case, and corresponds to the extraordinary transition when $x \rightarrow x_c$. When $y_{n_1} > y_{n-n_1}$ and $y_{n_1}, \; y_{n-n_1} \rightarrow \infty$, the boundary spins all point in the same direction $\vec{S}$, where $S^\beta=0$ for $\beta \in \left\{n_1+1,\dots,n \right\}$. The dual case $y_{n-n_1} > y_{n_1}$ would lead to a ground state with all the boundary spins having their components $\left\{ 1,\dots , n_1 \right\}$ all equal to zero. Crossing the isotropic half-line $y_{n_1}=y_{n-n_1} > y_S$ corresponds to switching from one ground state to another, and the transition is of the first order.

\paragraph{}

What happens when we go from the regime of the ordinary transition to the (anisotropic) extraordinary one? Let us fix $y_{n-n_1}$ to some value (smaller than $y_S$). When $y_{n_1}$ is sufficiently small the theory flows towards the attractive point corresponding to the ordinary transition, and when $y_{n_1}$ is sufficiently large it goes to the (anisotropic) extraordinary transition with the last $n-n_1$ components of the spins all equal to zero. The transition between the two regimes happens when the $n_1$ first components of the boundary spins order. This transition is called the \textit{anisotropic special transition}. For fixed values of $n$ and $n_1$, we can consider also the dual case $y_{n_1}< y_{n-n_1}$ which leads to a different anisotropic transition. Then, in the space of parameters $(y_{n_1},y_{n-n_1})$ there are three regions : two of them correspond to the (anisotropic) extraordinary transitions, and one to the ordinary transition. The three of them are separated by : the isotropic line between the two extraordinary transitions (and this is a first-order transition), and the two anisotropic special transitions between the ordinary and extraordinary transitions. These three lines all meet at the special transition point $Sp$ (figure $\ref{fig:phase}$).
\paragraph{}
RG flows can be organized as follows. As argued above, there is a fixed point for the ordinary transition, which we call $Ord$, and this point is stable. There are two extraordinary points at infinity, which are stable. On each of the two special transition lines, there is one RG fixed point $AS_{n_1}$ or $AS_{n-n_1}$. We will argue in the following parts of this article that these points are stable for a perturbation along the line, but unstable in the other direction. Note that, all along the line the theory is attracted by this point and hence is in its universality class. The special transition point $Sp$ is unstable in both directions. This is all summarized in figure $\ref{fig:phase}$.

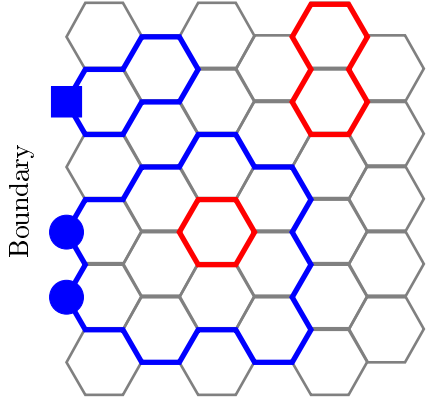
\begin{figure}[h]
\center
\begin{tikzpicture}[scale=0.75]
\foreach \x in {0,2,4}
\foreach \y in {-2,0,2,4,6,8}
{
	\honey{\x}{\y}
}
\foreach \x in {1,3,5}
\foreach \y in {-1,1,3,5,7}
{
	\honey{\x}{\y}
}
\draw (-0.8,1.7) node[rotate=90] {Boundary};
\draw[line width=2pt, blue, scale=0.6667] (0,0) -- ++(0.5,0.866) -- ++(-0.5,0.866) -- ++(0.5,0.866) -- ++(1,0) -- ++(0.5,0.866) -- ++(1,0) -- ++(0.5,0.866) -- ++(1,0) -- ++(0.5,-0.866) -- ++(1,0) -- ++(0.5,-0.866) -- ++(-0.5,-0.866) -- ++(0.5,-0.866) -- ++(-0.5,-0.866) -- ++(0.5,-0.866) -- ++(-0.5,-0.866) -- ++(-1,0) -- ++(-0.5,0.866) -- ++(-1,0) -- ++(-0.5,-0.866) -- ++(-1,0) -- ++(-0.5,0.866) -- ++(-1,0) -- cycle;
\draw[line width=2pt, blue, scale=0.6667] (0,5.196) -- ++(0.5,0.866) -- ++(1,0) -- ++(0.5,0.866) -- ++(1,0) -- ++(0.5,-0.866) -- ++(-0.5,-0.866) -- ++(-1,0) -- ++(-0.5,-0.866) -- ++(-1,0) -- cycle;
\draw[line width=2pt, red, scale=0.6667] (4.5,2.598) ++(1.5,2.598) -- ++(0.5,0.866) -- ++(-0.5,0.866) -- ++(0.5,0.866) -- ++(1,0) -- ++(0.5,-0.866) -- ++(-0.5,-0.866) -- ++(0.5,-0.866) -- ++(-0.5,-0.866) -- ++(-1,0) -- cycle;
\draw[line width=2pt, red, scale=0.6667] (4.5,2.598) ++(-1.5,-0.866) -- ++(0.5,0.866) -- ++(1,0) -- ++(0.5,-0.866) -- ++(-0.5,-0.866) -- ++(-1,0) -- cycle;
\filldraw[blue, scale=0.6667] (0,0) circle (0.45);
\filldraw[blue, scale=0.6667] (0,1.732) circle (0.45);
\filldraw[blue, scale=0.6667] (0,5.196) ++(-0.4,-0.4) rectangle ++(0.8,0.8);
\end{tikzpicture}
\caption{Two different kinds of loops can touch the boundary. They are marked with a circle (blob) or with a square. Blobbed loop (with circles) have a fugacity $n_1$, and unblobbed ones (with a square) a fugacity $n-n_1$. In the bulk, the loops still have a fugacity $n$.}
\label{fig:blob1b}
\end{figure}

\paragraph{}
To discuss the anisotropic transitions in the the loop language, we have to introduce a simple object which we use to distinguish some loops when they touch the boundary. This object is the \textit{blob} (see figure \ref{fig:blob1b}). It is used to mark the loops coming from the trace of the terms $S_{i_1}^{\mu_1} S_{i_2}^{\mu_1} S_{i_2}^{\mu_2} S_{i_3}^{\mu_2} \dots S_{i_k}^{\mu_k} S_{i_1}^{\mu_k}$ ($\left< i_1 i_2 \right>$, $\left< i_2 i_3 \right>$, etc. and $\left< i_k i_1 \right>$ are links of the lattice) restricted to the components $\alpha \in \left\{1,\dots,n_1\right\}$ in $(\ref{eq:ZOnanisotropic})$. These \textit{blobbed} loops have a fugacity $n_1$ instead of $n$. Of course, there is an equivalent object for the $n-n_1$ remaining components : these loops will be called the \textit{unblobbed} ones, and have a fugacity $n-n_1$. Each time a blobbed loop touches the boundary it takes an additional blob. This does not change its fugacity, but the weight of the monomers touching the boundary is different if they are blobbed or unblobbed. To simplify a bit the formulae to come in what follows, let us introduce the weights
\begin{subequations}
\label{eq:wblob}
\begin{eqnarray}
w_{\blob} &=& \left( \frac{y_{n_1}}{x} \right)^2 \\
w_{\unblob} &=& \left( \frac{y_{n-n_1}}{x} \right)^2
\end{eqnarray}
\end{subequations}
which can be viewed as the fugacity of the circles and squares (figure $\ref{fig:blob1b}$) when all the monomers get a weight $x$, without distinction for the boundary ones. The partition function $(\ref{eq:ZOnanisotropic})$ becomes
\begin{equation}
Z = \sum_{\rm{conf.}} x^{X} \; w_{\blob}^{W_{\blob}} \; w_{\unblob}^{W_{\unblob}} \; n^{N} \; n_1^{N_{\blob}} \; \left(n-n_1\right)^{N_{\unblob}}.
\label{eq:ZOnLoopboundary}
\end{equation}
In this formula, $N$ is the number of bulk loops, $N_{\blob}$ is the number of blobbed loops and $N_{\unblob}$ the number of unblobbed loops. $X$ is the total number of monomers (boundary and bulk ones), $W_{\blob}$ is the number of blobs and $W_{\unblob}$ is the number of ``unblobs'' (squares). When we work within this loop picture, the phase diagram shown in figure $\ref{fig:phase}$ must be interpreted as follows. Let us again fix $x=x_c$, so we are dealing with a critical dilute loop model in the bulk, and we adopt the terminology of ordinary/extraordinary b.c instead of transitions. For $n$ and $n_1$ fixed, there are three different regimes: the ordinary b.c when the loops avoid the boundary, and two (anisotropic) extraordinary b.c when there is one polymer (blobbed when $w_{\blob}>w_{\unblob}$, unblobbed when $w_{\unblob} > w_{\blob}$) adsorbed all along the boundary (see figure $\ref{fig:ordextra}$). The transition between the two extraordinary b.c is of the first order. The transition between the ordinary and one of the extraordinary b.c is the anisotropic special transition. Again, it corresponds to a non-trivial scaling limit, where there are two kinds of loops (blobbed and unblobbed) which touch the boundary without being adsorbed on it. We expect the geometric characteristics to be different for the blobbed and unblobbed loops. 
The universality class of this transition is completely characterized by the RG fixed point which sits on the transition line, and is stable under perturbations along the line. This point was named $AS_{n_1}$ (or $AS_{n-n_1}$) in the context of the spin model, let us rename it $AS_{\blob}$ (or $AS_{\unblob}$) for convenience when we deal with the loops.  The two anisotropic transition lines and the first order transition meet at the (isotropic) special point $Sp$. We already discussed that point in the context of the dilute loop model in section $\ref{sec:special}$. In section $\ref{part:Sklyanin}$, we will arrive to the following conjecture for the exact position of the points $AS_{\blob}$ and $AS_{\unblob}$ for the model on the honeycomb lattice
\begin{subequations}
\label{eq:conjectureAS}
\begin{eqnarray}
\displaystyle (AS_{\blob}) &&  \left\{ \begin{array}{rcl}
\displaystyle w_{\blob} &=& 1 + \frac{1}{2} \sqrt{2-n} + \frac{n_1-n/2+\sqrt{n_1 (n_1-n) +1}}{\sqrt{2-n} }\\ 
\displaystyle w_{\unblob} &=& 1 + \frac{1}{2} \sqrt{2-n} - \frac{n_1-n/2+\sqrt{n_1 (n_1-n) +1}}{\sqrt{2-n} } 
\end{array} \right. \\
\displaystyle (AS_{\unblob}) && \left\{ \begin{array}{rcl}
\displaystyle w_{\blob} &=& 1 + \frac{1}{2} \sqrt{2-n} + \frac{n_1-n/2-\sqrt{n_1 (n_1-n) +1}}{\sqrt{2-n} }\\ 
\displaystyle w_{\unblob} &=& 1 + \frac{1}{2} \sqrt{2-n} - \frac{n_1-n/2-\sqrt{n_1 (n_1-n) +1}}{\sqrt{2-n} } 
\end{array} \right. 
\end{eqnarray}
\end{subequations}
In the next sections, we will be interested in the anisotropic special transition in a more quantitative way.




\section{Integrable and critical points in $2D$ : from bulk to boundary}
\label{part:Sklyanin}
Statistical models in two dimensions can be exactly solved when they are related to  solutions of the Yang-Baxter equation. This equation implies the existence of a continuous family of commuting periodic transfer matrices (or Hamiltonians) which can be used to compute many quantities - such as the low-energy spectrum  - using Bethe-ansatz techniques. In the case when there are boundaries, the equivalent of the Yang-Baxter equation is the Sklyanin reflection equation \cite{Sklyanin}, sometimes called boundary Yang-Baxter equation. With a $R$-matrix solution of the Yang-Baxter equation in the bulk and a $K$-matrix solution of Sklyanin's equation at the boundary, one can build a continuous family of transfer matrix with boundaries \cite{Sklyanin}.

\paragraph{}
Although there is no general result relating integrability and criticality of a model, they can sometimes coincide. Integrable points often play some particular role in the phase diagram of a model, and in general the information they provide is a key point to understand the critical behaviour. In the particular case of loop models, some signs of a deeper relation between integrability and criticality have appeared very recently in the literature \cite{CardyIkhlef, CardyRiva}. The relation between integrability in the sense of Yang-Baxter and the lattice holomorphicity of certain discrete observables, which would lead to critical models in the continuum limit, is a fascinating subject. We hope that future work will develop this approach and provide a straightforward and rigorous way to relate lattice parameters to the characteristics of the objects showing up in the continuum limit \cite{Smirnov}. But for now, let us use the more traditional way of studying a lattice model: we look for (bulk and boundary) integrable solutions, and conjecture that they correspond to critical points of the loop model.

\subsection{The loop $\On$ model on the square lattice}
As explained in part $\ref{part:model}$, the $\On$ spin model can be reformulated as a dilute loop model on the honeycomb lattice. To study the integrability of the model, it is useful to relax the constraint that it lives on the honeycomb lattice, and to define a dilute loop model on the square lattice \cite{NienhuisOnSquare}. The model contains $9$ different plaquettes in the bulk. Two neighboring faces share a common edge, which can be crossed or not by a single loop. A loop cannot be stopped on an edge (\textit{ie} such a configuration is given a weight zero). The total Boltzmann weight of a configuration is given by the product of the weights of the plaquettes and a factor $n^{N}$, with $N$ the number of loops.   

$$
\begin{array}{cccccc}
\begin{tikzpicture}
\vv{0}{0}
\draw[yshift=-2.6cm] node{$\omega_1$};
\end{tikzpicture} &
\begin{tikzpicture}
\vl{0}{-1.2}
\vr{0}{1.2}
\draw[yshift=-2.6cm] node{$\omega_2$};
\end{tikzpicture} &
\begin{tikzpicture}
\vd{0}{-1.2}
\vu{0}{1.2}
\draw[yshift=-2.6cm] node{$\omega_3$};
\end{tikzpicture} &
\begin{tikzpicture}
\vur{0}{-1.2}
\vul{0}{1.2}
\draw[yshift=-2.6cm] node{$\omega_4$};
\end{tikzpicture} &
\begin{tikzpicture}
\vi{0}{0}
\draw[yshift=-2.6cm] node{$\omega_5$};
\end{tikzpicture} &
\begin{tikzpicture}
\ve{0}{0}
\draw[yshift=-2.6cm] node{$\omega_6$};
\end{tikzpicture}
\end{array}
$$
Vertices related by a horizontal or vertical reflection are given the same weight, so there are only $6$ independent weights $\omega_1,\dots ,\omega_6$. An $R$-matrix solution to the Yang-Baxter equation is given by the sum of the above $9$ diagrams with weights \cite{NienhuisOnSquare}
\begin{equation}
\label{eq:YBweights}
\begin{array}{rcl}
\omega_1 &=& \sin 2\Phi \sin 3\Phi + \sin u \sin (3\Phi-u)\\
\omega_2 &=& \sin 2\Phi \sin (3\Phi-u) \\
\omega_3 &=& \sin 2\Phi \sin u \\
\omega_4 &=& \sin u \sin (3\Phi-u) \\
\omega_5 &=& \sin (2\Phi-u) \sin (3\Phi-u) \\
\omega_6 &=& - \sin u \sin (\Phi-u) \\
\end{array}
\end{equation}
where $u$ is the spectral parameter and $\Phi$ is related to the weight $n$ of a closed loop by the parameterization\footnote{We use this parameterization only in this part, because it is more convenient and somewhat usual for the dilute loop model on the square lattice. In the rest of the paper we use $n=2 \cos \gamma$, which is more convenient when we deal with various critical exponents and conformal field theory.}
\begin{equation}
n = -2 \cos 4 \Phi.
\end{equation}
It is convenient to draw the $R$-matrix as
$$
\begin{tikzpicture}
\draw (-1.9,1.1) node {$R(u) \;=$};
\begin{scope}[scale=1.5]
\draw[rotate=45] (0,0) rectangle ++(1,1) ++(-0.5,-0.5) node{$u$}; 
\draw[rotate=45] (3pt,0) arc (0:90:3pt);
\end{scope}
\end{tikzpicture}
$$
where it is understood that the marked angle stands for the orientation of the plaquettes. Note that, in particular, the weights $(\ref{eq:YBweights})$ imply that
$$
\begin{tikzpicture}
\draw (0,1.1) node {$=$};
\draw (-3,1.1) node {$=$};
\draw (3,1.1) node {$=$};
\begin{scope}[xshift=-4.5cm, scale=1.5]
\draw[rotate=45] (0,0) rectangle ++(1,1) ++(-0.5,-0.5) node{$3 \Phi-u$}; 
\draw[rotate=45] (3pt,0) arc (0:90:3pt);
\end{scope}
\begin{scope}[xshift=-1.5cm, scale=1.5]
\draw[rotate=45] (0,0) rectangle ++(1,1) ++(-0.5,-0.5) node{$3 \Phi-u$}; 
\draw[rotate=45] (1,1) ++(-3pt,0) arc (180:270:3pt);
\end{scope}
\begin{scope}[xshift=1.5cm, scale=1.5]
\draw[rotate=45] (0,0) rectangle ++(1,1) ++(-0.5,-0.5) node{$u$}; 
\draw[rotate=45] (0,1) ++(0,-3pt) arc (270:360:3pt);
\end{scope}
\begin{scope}[xshift=4.5cm, scale=1.5]
\draw[rotate=45] (0,0) rectangle ++(1,1) ++(-0.5,-0.5) node{$u$}; 
\draw[rotate=45] (1,0) ++(0,3pt) arc (90:180:3pt);
\end{scope}
\end{tikzpicture}
$$
and that when the spectral parameter $u$ goes to zero, the $R$-matrix is proportional to the identity
$$
\begin{tikzpicture}
\draw (1,1.1) node {$= \; \sin 2 \Phi \sin 3 \Phi$};
\begin{scope}[xshift=-1.5cm, scale=1.5]
\draw[rotate=45] (0,0) rectangle ++(1,1) ++(-0.5,-0.5) node{$0$}; 
\draw[rotate=45] (3pt,0) arc (0:90:3pt);
\end{scope}
\begin{scope}[xshift=3.4cm, scale=1.5]
\draw[rotate=45] (0,0) rectangle ++(1,1) ++(-0.5,-0.5) node{$Id$}; 
\draw[rotate=45] (3pt,0) arc (0:90:3pt);
\end{scope}
\end{tikzpicture}
$$
Other relations are satisfied by this $R$-matrix, but we will not need them in what follows. We refer the interested reader to \cite{BatchelorYung, NienhuisOnSquare} for more information.

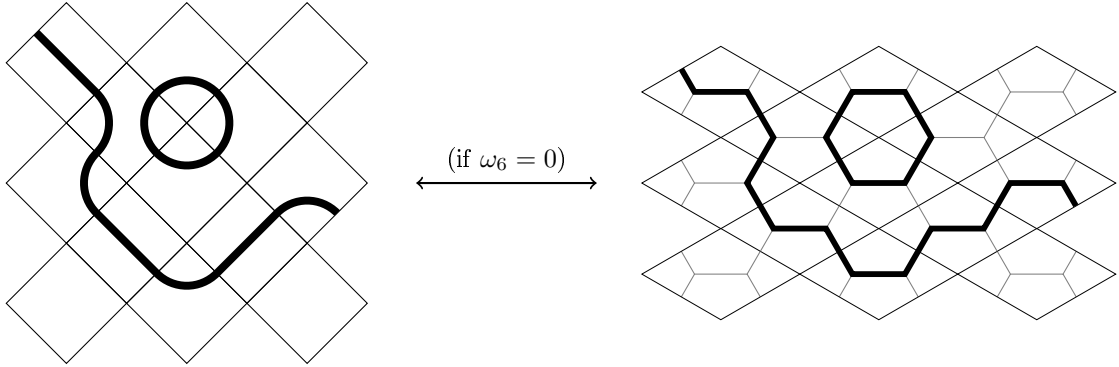
\begin{figure}[htbp]
\begin{tikzpicture}
\begin{scope}[scale=0.8]
\vv{0}{-2}
\vv{4}{-2}
\vu{2}{-2}
\vul{1}{-1}
\vur{3}{-1}
\vr{0}{0}
\vi{1}{1}
\vu{2}{0}
\vl{3}{1}
\vd{4}{0}
\vul{0}{2}
\vd{2}{2}
\vv{4}{2}
\end{scope}
\draw[thick, <->] (4.65,0) -- ++(1.2,0) node[anchor=south] {(if $\omega_6=0$)} -- ++(1.2,0);
\begin{scope}[xshift=6.6cm,scale=2.1]
\foreach \x in {1,...,3}
\foreach \y in {-1,...,1}
{
	\draw[thin, gray] (\x-0.25,0.577*\y+0.1443) -- ++(0.0833,-0.1443) -- ++(0.333,0) -- ++(0.0833,0.1443) ++(0,-0.2886) -- ++(-0.0833,0.1443) ++(-0.333,0) -- ++(-0.0833,-0.1443);
	\draw[very thin] (\x-0.5,0.577*\y) -- ++(0.5,0.289) -- ++(0.5,-0.289) -- ++(-0.5,-0.289) -- cycle;
}

\foreach \x in {1.5,2.5}
\foreach \y in {-0.5,0.5}
{
	\draw[thin, gray] (\x-0.25,0.577*\y+0.1443) -- ++(0.0833,-0.1443) -- ++(0.333,0) -- ++(0.0833,0.1443) ++(0,-0.2886) -- ++(-0.0833,0.1443) ++(-0.333,0) -- ++(-0.0833,-0.1443);
}
\draw[line width=2.2pt] (0.75,0.7217) -- ++(0.0833,-0.1443) -- ++(0.333,0) -- ++(0.1666,-0.2886) -- ++(-0.1666,-0.2886) -- ++(0.1666,-0.2886) -- ++(0.333,0) -- ++(0.1666,-0.2886) -- ++(0.333,0) -- ++(0.1666,0.2886) -- ++(0.333,0) -- ++(0.1666,0.2886) -- ++(0.333,0) -- ++(0.0833,-0.1443);
\draw[line width=2.2pt] (0.75,0.7217) ++(0.0833,-0.1443) ++(0.333,0) ++(0.1666,-0.2886) ++(0.333,0) -- ++(0.1666,-0.2886) -- ++(0.333,0) -- ++(0.1666,0.2886) -- ++(-0.1666,0.2886) -- ++(-0.333,0) -- cycle;
\end{scope}
\end{tikzpicture}
\caption{When the weight $\omega_6=0$, the dilute loop model on the square lattice is equivalent to the $O(n)$ model on the honeycomb lattice.}
\label{fig:squarehoney}
\end{figure}

\paragraph{}
We want to turn back to the honeycomb lattice, the present digression about the square lattice being only a trick to catch an integrable model (a solution to the Yang-Baxter equation). In the above model, the square lattice can be viewed as the honeycomb one for a specific choice of the spectral parameter. This can be done only if one of the two weights $\omega_5$ or $\omega_6$ is zero. In the latter case, figure $\ref{fig:squarehoney}$ shows how to go from the square to the honeycomb lattice.

There are two possible choices leading to $\omega_6=0$: $u=0$ or $u=\Phi$ (and other ones which are equivalent up to some reparameterization). The solution $u=0$ is trivial as $R(0) \propto Id$. It is not difficult to see that the solution $u=\Phi$ is related to the $O(n)$ model on the honeycomb lattice with two physical parameters, namely the weight of a closed loop $n=-2 \cos 4\Phi$ and the fugacity of each monomer $x=1/(2 \cos \Phi)$. Hence, these two parameters are related by
\begin{equation}
x=\frac{1}{\sqrt{2 \pm \sqrt{2-n}}}
\end{equation}
where the $\pm$ sign appears when we consider the different values of $\Phi$ which give the same $n$. Now we are ready to state Nienhuis' conjecture \cite{NienhuisConjecture}. The two critical points of the $O(n)$ model on the honeycomb lattice, corresponding to the dilute and dense phases, are the two integrable points. In particular, the critical fugacity $x_c$ at the dilute point is
\begin{equation}
x_c=\frac{1}{\sqrt{2 + \sqrt{2-n}}}
\end{equation}
as claimed in the first part of this article. Actually this was not the argument first used by Nienhuis in \cite{NienhuisConjecture}, but he formulated it in \cite{NienhuisOnSquare}.

\subsection{Blob operators}

We will work with three boundary plaquettes
$$
\begin{array}{ccccccc}
\begin{tikzpicture}
\tv{0}{0}
\draw[xshift=-0.5cm,yshift=-1.4cm] node{$\beta_1$};
\end{tikzpicture} & & &
\begin{tikzpicture}
\ti{0}{0}
\draw[xshift=-0.5cm,yshift=-1.4cm] node{$\beta_2$};
\end{tikzpicture} & & &
\begin{tikzpicture}
\tb{0}{0}
\draw[xshift=-0.5cm,yshift=-1.4cm] node{$\beta_3$};
\end{tikzpicture}
\end{array}
$$
The piece of loop on the third plaquette carries a blob. It does not matter whether a loop is marked with one or several blobs : once a loop is marked, it is blobbed once and for all, and additional blobs do not affect this status. In that sense, the blob operator acting on a loop is a projector. The physical effect of the blob on a loop is that it changes its fugacity. A blobbed closed loop is given a weight $n_1$ instead of $n$, where\footnote{This parameterization is used only in this part dealing with integrability. In the rest of the paper we use $n_1=\frac{\sin (r_1+1) \gamma}{\sin r_1 \gamma}$ and $n=2 \cos \gamma$ with $\gamma \in \left[ 0, \pi \right)$, $r_1 \in \left(0, \pi/\gamma \right)$.}
\begin{equation}
n_1 = - \frac{\sin 4(\kappa-1)\Phi}{\sin 4 \kappa \Phi}.
\end{equation}

\begin{figure}[htbp]
\centering
\begin{tikzpicture}
\begin{scope}[scale=0.7]
\vu{0}{-2}
\vr{4}{-2}
\vu{2}{-2}
\ve{5}{-1}
\tb{0}{-1}
\ve{1}{-1}
\vur{3}{-1}
\ve{0}{0}
\vi{1}{1}
\ti{0}{1}
\vl{2}{0}
\vul{3}{1}
\vur{5}{1}
\vi{4}{0}
\ve{0}{2}
\vi{2}{2}
\vv{4}{2}
\tb{0}{3}
\vd{1}{3}
\vl{3}{3}
\vv{5}{3}
\vur{0}{4}
\vul{2}{4}
\vu{4}{4}
\tv{0}{5}
\vd{1}{5}
\vul{3}{5}
\vd{5}{5}
\end{scope}
\end{tikzpicture}
\caption{Blobs in the dilute loop model on the square lattice. A blobbed loop is a loop which carries at least one blob. A blobbed loop has a fugacity $n_1$, whereas a loop without blob gets a fugacity $n$.}
\label{fig:bloblattice}
\end{figure}
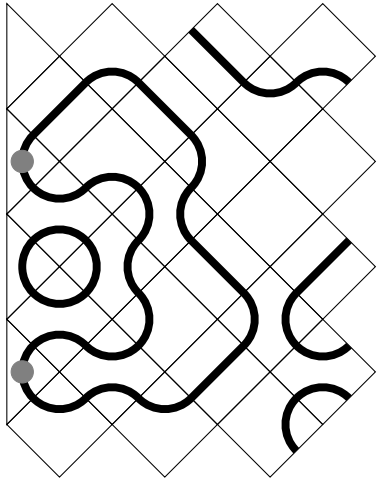

Note that a blobbed loop can touch the boundary without receiving an additional blob each time it touches it. It just needs to get at least one blob. If we sum over the possible configurations of boundary plaquettes, the total weight of a blobbed loop which touches $k$ times the boundary is $n_1 \left[ \left( \beta_2 + \beta_3 \right)^k - \beta_2^k \right] $. A loop which is never blobbed would get a weight $n \beta_2^k$. One can redistribute those two weights and make things more symmetric if we consider a linear combination of the two kinds of loops (blobbed and not blobbed) which carries a weight $(n-n_1) \beta_2^k$. We already know what such an object is : it is the unblobbed loop, which we marked with a square in part $\ref{part:model}$. There is a slightly more algebraic way to reformulate this. Since the blob acts as a projector on the loops, it is natural to introduce the orthogonal projector (or \textit{unblob operator})
$$
\begin{tikzpicture}
\tu{0}{0}
\draw (0.5,0) node{$=$};
\ti{2}{0}
\draw (2.5,0) node{$-$};
\tb{4}{0}
\end{tikzpicture}
$$ 
and if the three previous boundary plaquettes come with the weights $\beta_1$, $\beta_2$ and $\beta_3$, it is a simple change of basis to switch to the blobbed/unblobbed plaquettes. The unblob operator then gets a weight $\beta_2$, but the weight of the blob operator is $\beta_2 + \beta_3$ instead of $\beta_2$. We will use this in section $\ref{sec:sklyaninblob}$, when we derive the relation $(\ref{eq:conjectureAS})$ for the integrable weights $w_{\blob}$ and $w_{\unblob}$ on the honeycomb lattice.

\paragraph{}
There is a nice mathematical object hidden behind this blob operator. The rich underlying algebraic structure of loop models is well-known to be the celebrated Temperley-Lieb algebra \cite{TL}. The extension to loop models with blobs at the boundary is the so-called \textit{blob algebra} or \textit{one-boundary Temperley-Lieb algebra}. We do not want to elaborate too much about this object here, and we refer the reader to references \cite{MartinSaleur, deGierNichols, deGier2BTL, DJSprep} for detailed discussions of this subject.

\subsection{Solutions to Sklyanin's reflection equation from blob operators}
\label{sec:sklyaninblob}
Recall that Sklyanin's equation for the $K$-matrix can be written
$$
\begin{tikzpicture}
\begin{scope}[scale=1.5,xshift=-1.2cm]
\draw[rotate=45] (0,0) rectangle ++(1,1) ++(-0.5,-0.5) node{$R(u+v)$};
\draw[rotate=45] (0,0) ++(3pt,0) arc (0:90:3pt);
\draw[rotate=45] (1,1) rectangle ++(1,1) ++(-0.5,-0.5) node{$R(u-v)$};
\draw[rotate=45] (1,1) ++(3pt,0) arc (0:90:3pt);
\draw[rotate=45] (0,0) -- ++(0,1) -- ++(-1,-1) -- cycle ++(0.71,0.29) node{$K(v)$};
\draw[rotate=45] (1,1) -- ++(0,1) -- ++(-1,-1) -- cycle ++(0.71,0.29) node{$K(u)$};
\draw[rotate=45, line width=3pt, gray, style=dashed] (1,0.5) arc (-90:0:0.5);
\end{scope}
\draw (0,1.7) node{$=$};
\begin{scope}[scale=1.5,xshift=1.5cm,yshift=0.717cm]
\draw[rotate=45] (0,0) rectangle ++(1,1) ++(-0.5,-0.5) node{$R(u+v)$};
\draw[rotate=45] (0,0) ++(3pt,0) arc (0:90:3pt);
\draw[rotate=45] (-1,-1) rectangle ++(1,1) ++(-0.5,-0.5) node{$R(u-v)$};
\draw[rotate=45] (-1,-1) ++(3pt,0) arc (0:90:3pt);
\draw[rotate=45] (0,0) -- ++(0,1) -- ++(-1,-1) -- cycle ++(0.71,0.29) node{$K(u)$};
\draw[rotate=45] (1,1) -- ++(0,1) -- ++(-1,-1) -- cycle ++(0.71,0.29) node{$K(v)$};
\draw[rotate=45, line width=3pt, gray, style=dashed] (0,-0.5) arc (-90:0:0.5);
\end{scope}
\end{tikzpicture}
$$
where the dashed lines stand for the identity (or equivalently a contraction of the two edges joined by the dashed line). Two $K$-matrices solutions to Sklyanin's equation have been found by Batchelor \& Yung \cite{BatchelorYung} for the above dilute loop model without the blob operator ($\beta_3=0$). These two solutions are
\begin{equation}
\label{eq:BY1}
\left\{\begin{array}{rcl}
\beta_1 &=& \sin \left( \frac{3}{2} \Phi + u \right) \\
\beta_2 &=& \pm \sin \left( \frac{3}{2} \Phi - u \right) \\
\beta_3 &=& 0
\end{array}\right.
\end{equation}
and
\begin{equation}
\label{eq:BY2}
\left\{\begin{array}{rcl}
\beta_1 &=& \cos \left( \frac{3}{2} \Phi + u \right) \\
\beta_2 &=& \pm \cos \left( \frac{3}{2} \Phi - u \right) \\
\beta_3 &=& 0
\end{array}\right.
\end{equation}
The $K$-matrices found by Batchelor \& Yung also satisfy the so-called \textit{boundary crossing relation}, which plays some role in what follows. This relation is also satisfied by the two $K$-matrices $(\ref{eq:SklyaninBlob1})$ and $(\ref{eq:SklyaninBlob2})$ to appear below.
\begin{equation}
\label{eq:boundarycrossing}
\begin{tikzpicture}
\begin{scope}[scale=1.5,xshift=-1.8cm]
\draw[rotate=45] (0,-1) rectangle ++(1,1) ++(-0.5,-0.5) node{$2u$};
\draw[rotate=45] (0,-1) ++(3pt,0) arc (0:90:3pt);
\draw[rotate=45] (0,0) -- ++(0,1) -- ++(-1,-1) -- cycle ++(0.75,0.45) node{$3 \Phi$} ++(-0.2,-0.2) node{$-u$};
\draw[rotate=45, line width=3pt, gray, style=dashed] (0,0.5) arc (90:0:0.5);
\draw[rotate=45, line width=3pt, gray, style=dashed] (0,-0.5) arc (-90:-180:0.5);
\end{scope}
\draw (0,0) node{$\propto$};
\begin{scope}[scale=1.5,xshift=1.2cm]
\draw[rotate=45] (0,0) -- ++(0,1) -- ++(-1,-1) -- cycle ++(0.71,0.29) node{$u$};
\end{scope}
\end{tikzpicture}
\end{equation}

We still have to explain how we can interpret these plaquettes in the honeycomb limit. Using $R$-matrices and $K$-matrices, one can build the family of Sklyanin's transfer matrices for arbitrary $\omega_1, \omega_2, \dots, \omega_N$ (we fix $N=6$ for simplicity)
\begin{equation}
\label{eq:SklyaninTr}
\begin{tikzpicture}
\begin{scope}[scale=1.5]
\draw[rotate=45] (0.707,0.707) -- ++(0,1) -- ++(-1,-1) -- cycle ++(0.75,0.45) node{$3 \Phi$} ++(-0.2,-0.2) node{$-u$};
\draw[xshift=6cm,rotate=45] (0.707,0.707) -- ++(1,0) -- ++(-1,-1) -- cycle ++(0.3,0.7) node{$u$};
\draw (0,0) grid +(6,2);
\draw (0.5,0.5) node{$u+\omega_1$};
\draw (0.5,1.5) node{$u-\omega_1$};
\draw (1.5,0.5) node{$u+\omega_2$};
\draw (1.5,1.5) node{$u-\omega_2$};
\draw (2.5,0.5) node{$u+\omega_3$};
\draw (2.5,1.5) node{$u-\omega_3$};
\draw (3.5,0.5) node{$u+\omega_4$};
\draw (3.5,1.5) node{$u-\omega_4$};
\draw (4.5,0.5) node{$u+\omega_5$};
\draw (4.5,1.5) node{$u-\omega_5$};
\draw (5.5,0.5) node{$u+\omega_6$};
\draw (5.5,1.5) node{$u-\omega_6$};
\foreach \x in {0,...,5}
	{
	\draw[xshift=\x cm] (3pt,0) arc (0:90:3pt);
	\draw[yshift=1cm,xshift=\x cm] (1,0) ++(-3pt,0) arc (0:-90:-3pt);
	}
\draw[gray, line width=3pt, style=dashed] (0,1.5) arc (90:135:0.5cm);
\draw[gray, line width=3pt, style=dashed] (0,.5) arc (-90:-135:0.5cm);
\draw[gray, line width=3pt, style=dashed] (6,1.5) arc (90:45:0.5cm);
\draw[gray, line width=3pt, style=dashed] (6,.5) arc (-90:-45:0.5cm);
\end{scope}
\end{tikzpicture}
\end{equation}
which commute with each other for any value of the spectral parameter $u$. Now we choose $\omega_1=\omega_3=\omega_5=u$ and $\omega_2=\omega_4=\omega_6=-u$, then half of the $R$-matrix appearing in Sklyanin's transfer matrix $(\ref{eq:SklyaninTr})$ are proportional to the identity.
$$
\begin{tikzpicture}
\begin{scope}[scale=1.5]
\draw[rotate=45] (0.707,0.707) -- ++(0,1) -- ++(-1,-1) -- cycle ++(0.75,0.45) node{$3 \Phi$} ++(-0.2,-0.2) node{$-u$};
\draw[xshift=6cm,rotate=45] (0.707,0.707) -- ++(1,0) -- ++(-1,-1) -- cycle ++(0.3,0.7) node{$u$};
\draw (0,0) -- ++(0,1) -- ++(1,0) -- ++(0,1) -- ++(1,0) -- ++(0,-1) -- ++(1,0) -- ++(0,1) -- ++(1,0) -- ++(0,-1) -- ++(1,0) -- ++(0,1) -- ++(1,0) -- ++(0,-1) -- ++(-1,0) -- ++(0,-1) -- ++(-1,0) -- ++(0,1) -- ++(-1,0) -- ++(0,-1) -- ++(-1,0) -- ++(0,1) -- ++(-1,0) -- ++(0,-1) -- ++(-1,0);
\draw (0.5,0.5) node{$2u$};
\draw (1.5,1.5) node{$2u$};
\draw (2.5,0.5) node{$2u$};
\draw (3.5,1.5) node{$2u$};
\draw (4.5,0.5) node{$2u$};
\draw (5.5,1.5) node{$2u$};
\draw (0,0) ++(3pt,0) arc (0:90:3pt);
\draw (2,1) ++(-3pt,0) arc (0:-90:-3pt);
\draw (2,0) ++(3pt,0) arc (0:90:3pt);
\draw (4,1) ++(-3pt,0) arc (0:-90:-3pt);
\draw (4,0) ++(3pt,0) arc (0:90:3pt);
\draw (6,1) ++(-3pt,0) arc (0:-90:-3pt);
\draw[gray, line width=3pt, style=dashed] (0.5,1) arc (0:135:0.5cm);
\draw[gray, line width=3pt, style=dashed] (0,.5) arc (-90:-135:0.5cm);
\draw[gray, line width=3pt, style=dashed] (6,1.5) arc (90:45:0.5cm);
\draw[gray, line width=3pt, style=dashed] (4,1.5) arc (90:0:0.5cm);
\draw[gray, line width=3pt, style=dashed] (6,0.5) arc (-90:-45:0.5cm);
\draw[gray, line width=3pt, style=dashed] (1,1.5) arc (-90:-180:0.5cm);
\draw[gray, line width=3pt, style=dashed] (3,1.5) arc (-90:-180:0.5cm);
\draw[gray, line width=3pt, style=dashed] (2,1.5) arc (90:0:0.5cm);
\draw[gray, line width=3pt, style=dashed] (5,1.5) arc (-90:-180:0.5cm);
\draw[gray, line width=3pt, style=dashed] (1,0.5) arc (-90:0:0.5cm);
\draw[gray, line width=3pt, style=dashed] (1.5,0) arc (180:90:0.5cm);
\draw[gray, line width=3pt, style=dashed] (3,0.5) arc (-90:0:0.5cm);
\draw[gray, line width=3pt, style=dashed] (3.5,0) arc (180:90:0.5cm);
\draw[gray, line width=3pt, style=dashed] (5,0.5) arc (-90:0:0.5cm);
\draw[gray, line width=3pt, style=dashed] (5.5,0) arc (180:90:0.5cm);
\end{scope}
\end{tikzpicture}
$$
The left $K$-matrix can be contracted with the leftmost remaining $R$-matrix using $(\ref{eq:boundarycrossing})$. We end up with the so-called \textit{diagonal-to-diagonal} transfer matrix (we change $u$ into $u/2$ for convenience)
$$
\begin{tikzpicture}
\draw (-3,1.5) node{$T(u) \; =$};
\begin{scope}[scale=1.5]
\draw[rotate=45] (0,0) rectangle ++(1,1) ++(-0.5,-0.5) node{$R(u)$};
\draw[rotate=45] (0,0) ++(3pt,0) arc (0:90:3pt);
\draw[rotate=45] (0,1) rectangle ++(1,1) ++(-0.5,-0.5) node{$R(u)$};
\draw[rotate=45] (0,1) ++(3pt,0) arc (0:90:3pt);
\draw[rotate=45] (1,0) rectangle ++(1,1) ++(-0.5,-0.5) node{$R(u)$};
\draw[rotate=45] (1,0) ++(3pt,0) arc (0:90:3pt);
\draw[rotate=45] (1,-1) rectangle ++(1,1) ++(-0.5,-0.5) node{$R(u)$};
\draw[rotate=45] (1,-1) ++(3pt,0) arc (0:90:3pt);
\draw[rotate=45] (2,-1) rectangle ++(1,1) ++(-0.5,-0.5) node{$R(u)$};
\draw[rotate=45] (2,-1) ++(3pt,0) arc (0:90:3pt);
\draw[rotate=45] (0,1) -- (0,2) -- (-1,1) -- cycle (0,1) ++(-0.29,0.29) node{$K(\frac{u}{2})$};
\draw[rotate=45] (2,-1) -- (3,-1) -- (2,-2) -- cycle (2,-1) ++(0.29,-0.29) node{$K(\frac{u}{2})$};
\end{scope}
\end{tikzpicture}
$$
and now it is easy to go to the honeycomb lattice. The important point for us now is that the boundary plaquettes take weights $\beta_1 (u/2)$ and $\beta_2(u/2)$. When we take the honeycomb limit $u=\Phi$, the second plaquette can be viewed as two boundary half-monomers on the honeycomb lattice. The remaining two half-monomers also live on the boundary of the honeycomb lattice, but they come from the square lattice model with the bulk plaquettes. To take this into account, it is not difficult to check that the weight $y$ of a boundary monomer has to be related to $\omega_1$ and $\omega_2$ by $y^2/x_c=\beta_2(\Phi/2) / \beta_1(\Phi/2)$. For the two solutions of Batchelor \& Yung, this leads to $y^2/x_c=\pm 1/(2 \cos \Phi)$ or $y^2/x_c= \cos \Phi / \cos 2 \Phi$, which simplifies to
\begin{equation}
y = x_c
\end{equation}
or
\begin{equation}
y = y_S = (2-n)^{-1/4}.
\end{equation}
The first point corresponds to the ordinary phase (see part $\ref{part:model}$). The second one is conjectured to be the exact localization of the special transition on the honeycomb lattice.

\paragraph{}
The addition of the blob operator in this dilute loop model allows us to build more solutions to Sklyanin's reflection equation. Plugging the three boundary plaquettes in the equations, we find two solutions with $\beta_3 \neq 0$
\begin{equation}
\label{eq:SklyaninBlob1}
\left\{\begin{array}{rcl}
\beta_1 &=& \sin \left((2 \kappa +\frac{1}{2}) \Phi-u\right) \sin \left((2 \kappa - \frac{1}{2})\Phi +u\right) \\
\beta_2 &=& \sin \left((2 \kappa +\frac{1}{2}) \Phi-u\right) \sin \left((2 \kappa - \frac{1}{2})\Phi -u\right) \\
\beta_3 &=& \sin(2 u) \sin (4 \kappa \Phi)
\end{array}\right.
\end{equation}
and
\begin{equation}
\label{eq:SklyaninBlob2}
\left\{\begin{array}{rcl}
\beta_1 &=& \cos \left((2 \kappa +\frac{1}{2}) \Phi-u\right) \cos \left((2 \kappa - \frac{1}{2})\Phi +u\right) \\
\beta_2 &=& \cos \left((2 \kappa +\frac{1}{2}) \Phi-u\right) \cos \left((2 \kappa - \frac{1}{2})\Phi -u\right) \\
\beta_3 &=& -\sin(2 u) \sin (4 \kappa \Phi)
\end{array}\right.
\end{equation}
Taking the honeycomb limit $u=\Phi$, the weight of a blobbed monomer on the boundary is related to the plaquette weights by $y_{n_1}^2 / x = \left(\beta_2(\Phi/2) + \beta_3(\Phi/2) \right) / \beta_1 (\Phi/2)$. For unblobbed monomers one has $y_{n-n_1}^2 / x = \beta_2 (\Phi / 2) / \beta_1 (\Phi/2)$. Using the notations of the first part (see equation $(\ref{eq:wblob})$), one has 
\begin{subequations}
\begin{eqnarray}
\left\{ \begin{array}{rcl}
w_{\blob} &=& \frac{2 \cos \Phi \sin (2 \kappa+1) \Phi}{\sin 2 \kappa \Phi} \\
w_{\unblob} &=& \frac{2 \cos \Phi \sin (2 \kappa-1) \Phi}{ \sin 2 \kappa \Phi}
\end{array} \right. \\
\left\{ \begin{array}{rcl}
w_{\blob} &=& \frac{2 \cos \Phi \cos (2 \kappa+1) \Phi}{ \cos 2 \kappa \Phi} \\
w_{\unblob} &=& \frac{2 \cos \Phi \cos (2 \kappa-1) \Phi}{ \cos 2 \kappa \Phi}
\end{array} \right.
\end{eqnarray}
\end{subequations}
We conjecture that these two points are the points $AS_{\blob}$ and $AS_{\unblob}$, so they catch the whole universal behaviour of the anisotropic special transition. This finally leads to the relations $(\ref{eq:conjectureAS})$.




\section{Boundary conformal field theory of the anisotropic special transition}
\label{partBCFT}
From now on, we turn to the conformal field theory description of the model with the foregoing boundary conditions. The conformal field theory of the dilute $\On$ model in the bulk has been known for a long time \cite{NienhuisDL}. In particular, Coulomb gas methods have played a crucial role in this development. The Coulomb gas approach is a  non-rigorous (but powerful) way of constructing arguments which lead to exact results, in particular the critical exponents of correlation functions. It is a convenient way to deal with loop models, because loops can be interpreted as level lines of a height model, which is argued to flow towards a Gaussian free field with some background charge in the scaling limit. The Coulomb gas approach to loop models is thus heuristically straightforward. The Coulomb gas with boundaries, however, is still not well understood \cite{CardyOnAnnulus,CardyPercolationAnnulus,DJSdense}. In \cite{DJSdense} we mixed Coulomb gas with algebraic arguments related to boundary extensions of the Temperley-Lieb algebra to derive some results about a dense loop model (\textit{ie} the dense phase of the $\On$ model) with blobbed boundary conditions. The purpose of this section is to extend this approach to the dilute loop model.

\subsection{Coulomb gas framework}
\label{sec:Coulomb}
\subsubsection{The $\On$ model on an infinite cylinder}
Let us start by the usual Coulomb gas approach to the $\On$ model on an infinite cylinder. This will fix the coupling constant $g$ of the Gaussian free field as a function of the loop weight $n$. Consider a loop configuration on the cylinder, and choose an orientation for each loop. Now turn this non-locally interacting model (loops are non-local objects) into a local one. Following an oriented loop all along, give a weight $e^{i \gamma \rm d \alpha /(2 \pi)}$ to each infinitesimal piece of loop when it turns $\rm d \alpha$, where $\alpha$ is the winding angle of the curve. Depending on its orientation, a contractible loop gets a weight $e^{\pm i \gamma}$. The trace over the orientations gives back an unoriented loop with weight $n = e^{i \gamma} + e^{- i \gamma}$.
Oriented loops are then viewed as level lines of a height field $h$. The height varies by $\Delta h = \pm \pi$ when one crosses a level line. Then it is generally argued that this model renormalizes towards a Gaussian free field $h$ with action
\begin{equation}
\label{eq:actionGFF}
\mathcal{S}= \frac{g}{4 \pi} \int (\partial h)^2 d^2 x.
\end{equation}
However, this does not take properly into account the loops which wrap around the annulus, because the integrated winding angle $\int d \alpha$ is equal to $0$ in that case, not $\pm 2 \pi$. This problem can be solved by adding two charges $e^{\pm (\gamma/\pi) h}$ at the ends of the cylinder. This modifies the scaling dimension of the vertex operator $e^{i \alpha h}$ to
\begin{equation}
\Delta_\alpha = \frac{g}{4} \left\{ (\alpha + \gamma/\pi)^2 - (\gamma/ \pi)^2 \right\}.
\end{equation}
The coupling constant $g$ is then usually fixed by the following argument. We started from a discrete model in which the height $h$ was a integer multiple of $\Delta h = \pi$, so the operator $\cos 2 h$ should be marginal. This requires $\Delta_2 = 2$ or $\Delta_{-2}=2$, so $g = 1 \pm \frac{\gamma}{\pi}$. So far, we have not specified if we were working with dense or dilute loops. This is what this undetermined sign takes into account. The solution $g<1$ corresponds to the dense phase of the loop model, and $g>1$ to the dilute phase. Thus, in what follows we will always use the relation between the coupling constant $g$ and the loop weight $n$
\begin{equation}
\label{eq:coupling}
g = 1 + \frac{\gamma}{\pi} \qquad n=2 \cos \gamma \quad \gamma \in [0,\pi).
\end{equation}

\subsubsection{Boundaries in the height model}

The arguments here are copied from \cite{DJSdense}. To set up a well-behaved Coulomb gas framework, we put the model on an infinite strip. The strip has two boundaries, and the key point is to put different boundary conditions on the left and right sides. Each time a loop touches the left boundary it can get a black blob (or not). A black blobbed loop has a fugacity
\begin{equation}
\label{eq:n1}
n_1 = \frac{\sin (r_1+1)\gamma}{\sin r_1 \gamma} \qquad r_1 \in (0,\frac{\pi}{\gamma})
\end{equation}
instead of $n=2 \cos \gamma$. On the right side, loops can get a white blob (or not). A white loop has a fugacity
\begin{equation}
\label{eq:n2}
n_2 = \frac{\sin (r_2+1)\gamma}{\sin r_2 \gamma} \qquad r_2 \in (0,\frac{\pi}{\gamma}).
\end{equation}

\begin{figure}[h]
\center
\begin{tikzpicture}
\begin{scope}[xshift=-4cm]
	\filldraw[fill=red!15!white, draw=white] (-2,-4) rectangle (2,4);
	\draw[very thick, gray] (-2,-4) -- (-2,4); 
	\draw[very thick, gray] (2,-4) -- (2,4); 
	\draw[smooth, red] plot coordinates{(0,0) (0.3,0.2) (0.5,0.8) (0.7,0.6) (0.5,0.5) (0.8,0.2) (0.3,-0.4) (-0.2,-0.3) (-0.6,-1) (-0.9,-0.5) (-1.3,0.2) (-1.6,0) (-1.4,0.5) (-1.2,1.2) (-0.8,0.3) (0,0)};
	\draw[smooth, red] plot coordinates{(1,2) (-0.5,1.8) (-1,2.1) (-1.2,2.7) (-1,3) (-1.5,3.6) (-1,3.8) (-0.5,3.6) (0.5,3) (1.2,3.5) (1.5,2.5) (1,2)};
	\draw[smooth, red] plot coordinates{(0,2.5) (-0.5,3) (0.3,2.8) (0.6,2.2) (0,2.5)};
	\draw[smooth, red] plot coordinates{(1.2,-2.4) (1.8,-2.3) (1.5,-1.6) (1.3,-1.8) (1.2,-2.4)};
	\draw[smooth, red] plot coordinates{(-2,-1.8) (-1.7,-1.6) (-1.3,-2.5) (-1,-2.2) (-0.8,-2.5) (-1.5,-2.8) (-2,-3) (-1.8,-2.6) (-1.5,-2.5) (-1.8,-2.1) (-2,-1.8)};
	\draw[smooth, red] plot coordinates{(-2,-3.5) (-1.7,-3.8) (-1.3,-3.6) (-1.5,-3) (-2,-3.5)};
	\draw[smooth, red] plot coordinates{(-2,0) (-1.5,1.5) (-2,3) (-1.8,3.2) (-1.3,2) (-1, 1.5) (1,1.9) (2,1.8) (1.6,1.1) (2,0.6) (1.7,0.7) (1.3,0.5) (2,0.1) (1.5,0) (1.7,-0.4) (2,-0.5) (1,-0.6) (0.5,-0.4) (0,-1) (-1,-1.8) (-1.8, -1) (-2,0)};
	\draw[smooth, red] plot coordinates{(2,-1) (1.7,-1.3) (1,-1.7) (0.8,-2.4) (1.4,-3) (2,-3.4) (1.8,-3.6) (1.2,-3.2) (0.5,-3.4) (0.5,-2.8) (0,-2.3) (1.3,-1.2) (1.8,-0.8) (2,-1)};
	\draw[fill, black] (-2,0) circle (3pt);
	\draw[fill, black] (-2,-3) circle (3pt);
	\filldraw[white, draw=black, thick] (2,-3.5) circle (3pt);
	\filldraw[white, draw=black, thick] (2,0.6) circle (3pt);
	\filldraw[white, draw=black, thick] (2,-0.5) circle (3pt);
	\filldraw[white, draw=black, thick] (2,-1) circle (3pt);
\end{scope}
\draw[very thick, ->] (-1,0) -- (1,0);
\begin{scope}[xshift=4cm]
	\filldraw[fill=red!15!white, draw=white] (-2,-4) rectangle (2,4);
	\draw[very thick, gray] (-2,-4) -- (-2,4); 
	\draw[very thick, gray] (2,-4) -- (2,4); 
	\draw[smooth, red] plot coordinates{(0,0) (0.3,0.2) (0.5,0.8) (0.7,0.6) (0.5,0.5) (0.8,0.2) (0.3,-0.4) (-0.2,-0.3) (-0.6,-1) (-0.9,-0.5) (-1.3,0.2) (-1.6,0) (-1.4,0.5) (-1.2,1.2) (-0.8,0.3) (0,0)};
	\draw[smooth, red] plot coordinates{(1,2) (-0.5,1.8) (-1,2.1) (-1.2,2.7) (-1,3) (-1.5,3.6) (-1,3.8) (-0.5,3.6) (0.5,3) (1.2,3.5) (1.5,2.5) (1,2)};
	\draw[smooth, red] plot coordinates{(0,2.5) (-0.5,3) (0.3,2.8) (0.6,2.2) (0,2.5)};
	\draw[smooth, red] plot coordinates{(1.2,-2.4) (1.8,-2.3) (1.5,-1.6) (1.3,-1.8) (1.2,-2.4)};
	\draw[smooth, red] plot coordinates{(-2,-1.8) (-1.7,-1.6) (-1.3,-2.5) (-1,-2.2) (-0.8,-2.5) (-1.5,-2.8) (-2,-3) (-1.8,-2.6) (-1.5,-2.5) (-1.8,-2.1) (-2,-1.8)};
	\draw[smooth, red] plot coordinates{(-2,-3.5) (-1.7,-3.8) (-1.3,-3.6) (-1.5,-3) (-2,-3.5)};
	\draw[smooth, red] plot coordinates{(-2,0) (-1.5,1.5) (-2,3) (-1.8,3.2) (-1.3,2) (-1, 1.5) (1,1.9) (2,1.8) (1.6,1.1) (2,0.6) (1.7,0.7) (1.3,0.5) (2,0.1) (1.5,0) (1.7,-0.4) (2,-0.5) (1,-0.6) (0.5,-0.4) (0,-1) (-1,-1.8) (-1.8, -1) (-2,0)};
	\draw[smooth, red] plot coordinates{(2,-1) (1.7,-1.3) (1,-1.7) (0.8,-2.4) (1.4,-3) (2,-3.4) (1.8,-3.6) (1.2,-3.2) (0.5,-3.4) (0.5,-2.8) (0,-2.3) (1.3,-1.2) (1.8,-0.8) (2,-1)};
	\filldraw[gray] (-2,0) circle (1.5pt);
	\filldraw[gray] (-2,-3) circle (1.5pt);
	\filldraw[gray] (2,-3.5) circle (1.5pt);
	\filldraw[gray] (2,0.6) circle (1.5pt);
	\filldraw[gray] (2,-0.5) circle (1.5pt);
	\filldraw[gray] (2,-1) circle (1.5pt);
	\draw (-0.6,-0.2) node {$0$};
	\draw (0,0.8) node {$1$};
	\draw (-0.5,-3) node {$0$};
	\draw (-1.8,1.5) node {$2$};
	\draw (-1.7,-1.9) node {$1$};
	\draw (-1.8,-2.4) node {$0$};
	\draw (-1.7,-3.5) node {$-1$};
	\draw (1.6,-2.7) node {$-2$};
	\draw (0.8,-3) node {$-1$};
	\draw (1.5,-2.15) node {$-3$};
	\draw (0.5,3.5) node {$2$};
	\draw (-0.5,2.3) node {$3$};
	\draw (0.15,2.65) node {$4$};
	\draw (1.8,1.1) node {$2$};
	\draw (1.8,0.45) node {$0$};
	\draw (1.8,-0.25) node {$0$};
	\draw[thick, xshift=-0.3cm, yshift=0.1cm, rotate=-100] (-0.1,-0.15) -- (0,0) -- (0.1,-0.15);
	\draw[thick, xshift=-1.2cm, yshift=-2.7cm, rotate=-70] (-0.1,-0.15) -- (0,0) -- (0.1,-0.15);
	\draw[thick, xshift=0cm, yshift=-1cm, rotate=-45] (-0.1,-0.15) -- (0,0) -- (0.1,-0.15);
	\draw[thick, xshift=0.8cm, yshift=-1.6cm, rotate=-50] (-0.1,-0.15) -- (0,0) -- (0.1,-0.15);
	\draw[thick, xshift=0.8cm, yshift=-2.2cm, rotate=0] (-0.1,-0.15) -- (0,0) -- (0.1,-0.15);
	\draw[thick, xshift=1.35cm, yshift=-1.7cm, rotate=-25] (-0.1,-0.15) -- (0,0) -- (0.1,-0.15);
	\draw[thick, xshift=-1.65cm, yshift=-3.1cm, rotate=-40] (-0.1,-0.15) -- (0,0) -- (0.1,-0.15);
	\draw[thick, xshift=1.45cm, yshift=0.4cm, rotate=65] (-0.1,-0.15) -- (0,0) -- (0.1,-0.15);
	\draw[thick, xshift=-0.4cm, yshift=1.6cm, rotate=-75] (-0.1,-0.15) -- (0,0) -- (0.1,-0.15);
	\draw[thick, xshift=-0.2cm, yshift=2.67cm, rotate=-130] (-0.1,-0.15) -- (0,0) -- (0.1,-0.15);
	\draw[thick, xshift=-0.2cm, yshift=3.4cm, rotate=55] (-0.1,-0.15) -- (0,0) -- (0.1,-0.15);
\end{scope}
\end{tikzpicture}
\caption{Mapping of the loop model with the blobs onto an oriented loop model, and a corresponding height configuration $h/\pi$.}
\label{fig:striporiented}
\end{figure}
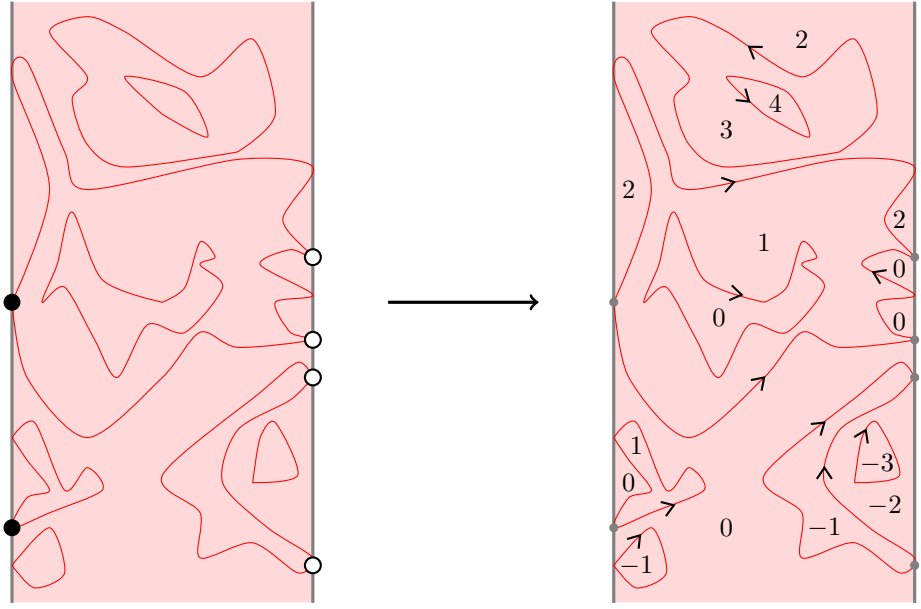
Such a configuration is shown in figure $\ref{fig:striporiented}$. The first step is now to map this model on a height model, by choosing orientations for the loops. The question is : what do we do with the blobs? Remember that blobs are projectors so if they act on the orientation of the loops, they must be projectors on some linear combination of the two possible orientations for one loop. In particular, a blob cannot act on the orientation by simply adding a phase shift. Writing that the blob has to be a projector, and using $(\ref{eq:n1})-(\ref{eq:n2})$ we find that it can be represented as follows
\begin{subequations}
\label{eq:blobsorient}
\begin{eqnarray}
\begin{tikzpicture}
\draw[line width=1pt] (0,-0.4) -- (0,0.4) (-1,-0.03) node{$2 i \sin r_1 \gamma$};
\draw[fill] (0,0) circle (3pt) ++(1,0) node {$\; = \; - e^{-i r_1 \gamma}$};
\draw[thick, ->] (2,-0.3) -- (2,0.3);
\draw[thick, ->] (2,-0.4) -- (2,-0.2);
\draw[very thick] (2,-0.4) -- (2,0.4);
\draw[fill] (2,0) circle (1.5pt) ++(1,0) node {$ + \; \; i e^{-i \varphi_1}$};
\draw[thick, ->] (3.8,0.4) -- (3.8,0.2);
\draw[thick, ->] (3.8,-0.4) -- (3.8,-0.2);
\draw[very thick] (3.8,-0.4) -- (3.8,0.4);
\draw[fill] (3.8,0) circle (1.5pt) ++(1,0) node {$ + \; \;  e^{i r_1 \gamma}$};
\draw[thick, ->] (5.6,0.4) -- (5.6,0.2);
\draw[thick, ->] (5.6,-0) -- (5.6,-0.3);
\draw[very thick] (5.6,-0.4) -- (5.6,0.4);
\draw[fill] (5.6,0) circle (1.5pt) ++(1,0) node {$ + \; \; i e^{i \varphi_1}$};
\draw[thick, <->] (7.4,-0.3) -- (7.4,0.3);
\draw[very thick] (7.4,-0.4) -- (7.4,0.4);
\draw[fill] (7.4,0) circle (1.5pt);
\end{tikzpicture} \\
\begin{tikzpicture}
\draw[line width=1pt] (0,-0.4) -- (0,0.4) (-1,-0.03) node{$2 i \sin r_2 \gamma$};
\filldraw[white, draw=black, thick] (0,0) circle (3pt);
\draw (1,0) node {$\; = \; \;  e^{i r_2 \gamma}$};
\draw[thick, ->] (2,-0.3) -- (2,0.3);
\draw[thick, ->] (2,-0.4) -- (2,-0.2);
\draw[very thick] (2,-0.4) -- (2,0.4);
\draw[fill] (2,0) circle (1.5pt) ++(1,0) node {$ + \; \; i e^{-i \varphi_2}$};
\draw[thick, ->] (3.8,0.4) -- (3.8,0.2);
\draw[thick, ->] (3.8,-0.4) -- (3.8,-0.2);
\draw[very thick] (3.8,-0.4) -- (3.8,0.4);
\draw[fill] (3.8,0) circle (1.5pt) ++(1,0) node {$ - \; \;  e^{-i r_2 \gamma}$};
\draw[thick, ->] (5.6,0.4) -- (5.6,0.2);
\draw[thick, ->] (5.6,-0) -- (5.6,-0.3);
\draw[very thick] (5.6,-0.4) -- (5.6,0.4);
\draw[fill] (5.6,0) circle (1.5pt) ++(1,0) node {$ + \; \; i e^{i \varphi_2}$};
\draw[thick, <->] (7.4,-0.3) -- (7.4,0.3);
\draw[very thick] (7.4,-0.4) -- (7.4,0.4);
\draw[fill] (7.4,0) circle (1.5pt);
\end{tikzpicture}
\end{eqnarray}
\end{subequations}
It is an easy exercise to check that, tracing over the two possible orientations, a blobbed loop takes a total weight $n_1$ (or $n_2$). This works with the convention that a loop has a weight $e^{-i \gamma}$ when it is clockwise oriented. The parameters $\varphi_1$ and $\varphi_2$ are still free. Actually our problem has a global gauge invariance which can be fixed by setting, for example $\varphi_1 + \varphi_2=0$. The difference $\varphi_1 - \varphi_2$ is, however, very important. It appears when one computes the weight $n_{12}$ of a loop which carries the two blobs (black and white). Four terms contribute to $n_{12}$, because each half-loop can have two orientations. This is possible only because the two blobs $(\ref{eq:blobsorient})$ are used here. One gets
\begin{equation}
	\psfrag{eq}[l]{$\displaystyle = \; \frac{-1}{4 \sin r_1 \gamma \sin r_2 \gamma}$}
	\psfrag{alpha2}[l]{$e^{-i(r_1+r_2+1) \gamma}$}
	\psfrag{alpha3}[l]{$-  e^{-i (\varphi_1-\varphi_2)}$}
	\psfrag{alpha4}[l]{$+  e^{i(r_1+r_2+1) \gamma}$}
	\psfrag{alpha5}[l]{$- \; e^{i (\varphi_1- \varphi_2)}$}
	\psfrag{par1}[l]{$\left\{ \begin{array}{c} \\ \\ \\ \\ \\ \\ \end{array} \right.$}
	\psfrag{par2}[l]{$\left. \begin{array}{c} \\ \\ \\ \\ \\ \\ \end{array} \right\}$}
	\includegraphics[width=0.85\textwidth]{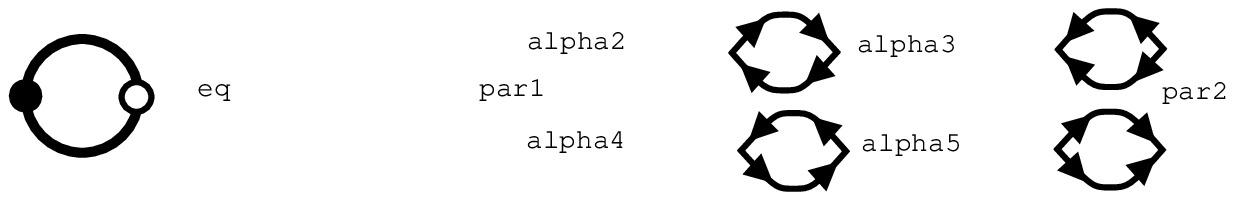}
\end{equation}
which gives
\begin{equation}
\label{eq:n12}
n_{12} = \frac{\sin \left( \frac{r_1+r_2+1+r_{12}}{2} \gamma \right) \sin \left( \frac{r_1+r_2+1-r_{12}}{2} \gamma \right)}{\sin r_1 \gamma \sin r_2 \gamma} 
\end{equation}
where we defined $r_{12} \gamma = \varphi_1- \varphi_2$ for later convenience.

\subsubsection{Flow towards the Gaussian free field}
\label{subsubsec:flowgaussian}
\begin{figure}[htbp]
\center
a. \includegraphics[width=0.4\textwidth]{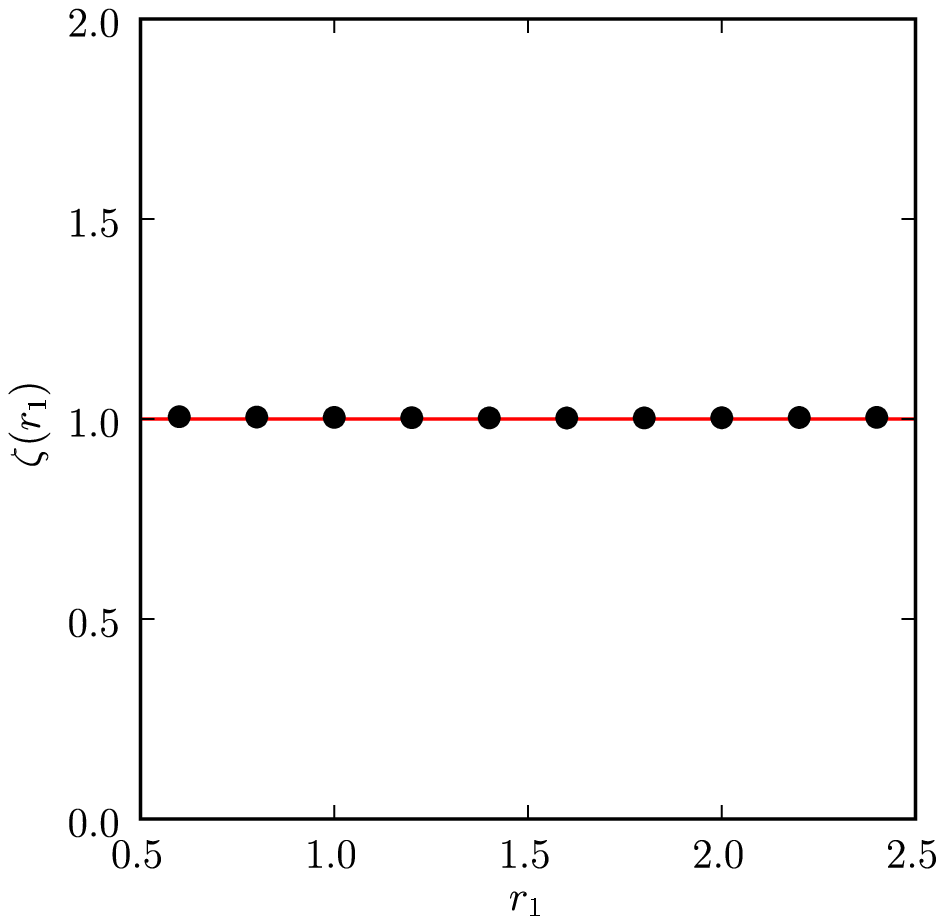}
\qquad  b.\includegraphics[width=0.4\textwidth]{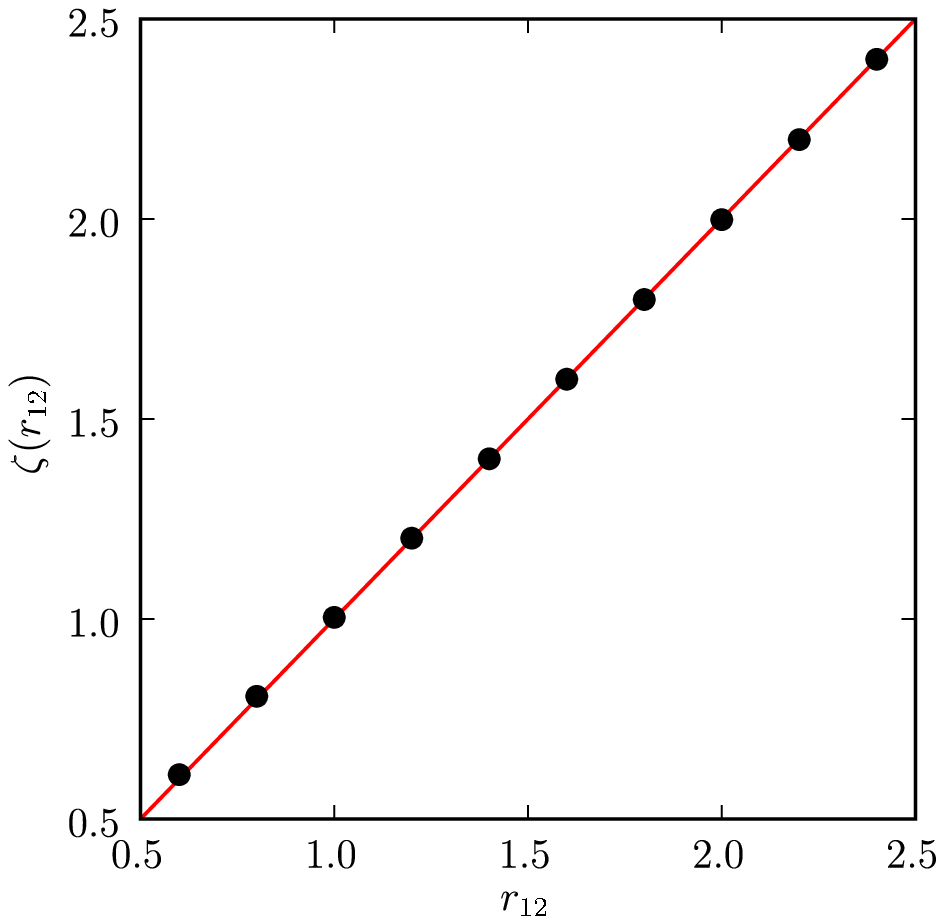}
\caption{The leading exponent $h_0$ in the sector without strings is written $h_0=\left(\zeta^2 -1\right) \frac{(g-1)^2}{4g}$. We plot $\zeta$ vs. $r_1$ (a) and $r_{12}$ (b). Numerically, it is clear here that $\Phi$ does not depend on $r_1$ at all, and that $\zeta=r_{12}$, as predicted from the Coulomb gas arguments.}
\label{fig:checkr1r2}
\end{figure}

We have translated the blobs into boundary vertices acting on the orientation of the loops (figure $\ref{fig:striporiented}$). The height model obtained in this way involves complicated local weights on the boundaries given by the relations $(\ref{eq:blobsorient})$. Fortunately only a few of these weights will be relevant when we go to the continuum limit. Our argument for that goes as follows. At the position of a blob, the orientation of a loop can be conserved or not (see figure $\ref{fig:striporiented}$). When it is conserved, relations $(\ref{eq:blobsorient})$ tell us such configurations come with weights proportional to $e^{\pm i r_1 \gamma}$ or $e^{\pm i r_2 \gamma}$. How do we take these into account in the continuum limit? Our guess is that we do not need to take care of them because they do not contribute to the continuum limit. Following \cite{DJSdense}, we argue that since this corresponds to the diagonal action of the blobs on the orientations, it can be viewed equivalently as a field living on the boundary. Such a boundary condition (with a coupling to a field living on the boundary) is expected to renormalize towards a fixed boundary condition under RG, independently of the microscopic details about the interaction or the strength of the field. A numerical check of the fact that $r_1$ and $r_2$ do not appear in the continuum limit will be discussed below (figure $\ref{fig:checkr1r2}$). On the contrary, the parameter $r_{12} = (\varphi_1 - \varphi_2)/\gamma$ is important. This parameter appears with pairs of vertices (one on each boundary) which do not conserve the orientation of the loops (figure $\ref{fig:striporiented}$). A weight $e^{\pm i r_{12} \gamma}$ must be given to each pair of half-loops going from one boundary to another. Then we expect that we end up with a Gaussian free field on the strip with action $(\ref{eq:actionGFF})$, with Neumann boundary conditions on both sides, where we have to take properly into account these factors $e^{\pm i r_{12} \gamma}$. This is the crucial point, which allows us to identify the operator changing from one blobbed boundary condition to another.

\begin{figure}[h]
\center
	\begin{tikzpicture}[scale=1.2]
		\begin{scope}[xshift=4.5cm, yshift=-0.7cm]
			\clip (-2.3,-0.6) rectangle (2.3,2.3);
			\filldraw[fill=red!15!white, draw=red!80!white, thick] (-10,0) rectangle (10,10);
			\filldraw[white] (0,0) circle (0.1cm) ++(0,-0.4) node[black] {$0$};
			\begin{scope}[scale=0.8]
				\draw[thick,gray] (1.5,0) arc (0:180:1.5cm);
				\foreach \x in {-60,-30,...,60} {
					\begin{scope}[rotate=\x]
						\draw[gray,thick,->] (0,1.3) -- (0,1.7);
					\end{scope}
				}
				\draw (1.5,1.5) node {$L_0$};
			\end{scope}
		\end{scope}
		\draw[->, thick] (-1,0) -- (1,0);
		\draw (0,0.2) node[above] {$z \mapsto \exp \left( -\frac{i \pi}{L} z \right)$};
		\begin{scope}[xshift=-3cm]
			\clip (-1,-2) rectangle (1,3);
			\filldraw[fill=red!15!white, draw=red!80!white, thick] (-0.8,-10) rectangle (0.8,10);
			\draw[<->,thick] (-0.7,-1) -- (0.7,-1); \draw (0,-1.1) node[below] {$L$};
			\draw[gray,thick] (-0.8,0.5) -- (0.8,0.5);
			\foreach \x in {-0.6,-0.3,...,0.6} \draw[->,gray,thick] (\x,0.3) -- (\x,0.7);
			\draw (0,0.7) node[above] {$H$}; 
		\end{scope}
	\end{tikzpicture}
\caption{The infinite strip is conformally equivalent to the half-plane. When the boundary condition is different on both sides of the strip, a B.C.C. operator is sitting at the point $0$.}
\label{fig:HL0}
\end{figure}
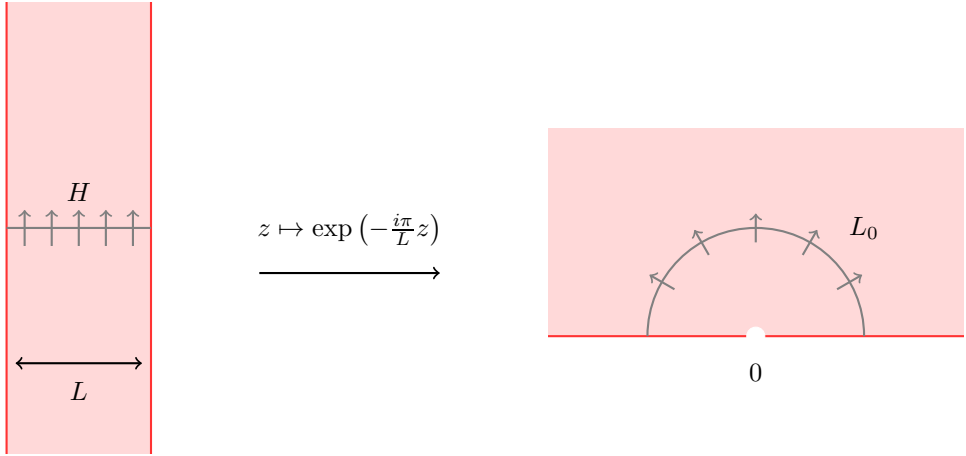

\paragraph{}
Let us be slightly more quantitative by looking at our loop model with blobs on an annulus instead of the strip. We want to compute the partition function of this model using the above formalism. On the annulus, there can be $p$ pairs of half-loops going from the left boundary to the right boundary. This introduces a defect $\Delta h = 2 \pi p$ when one turns around the annulus. In other words, the field $h(x,y)$ can be written as 
\begin{equation}
h(x,y)=\tilde{h}(x,y) + 2 \pi p y/T
\end{equation}
where $\tilde{h}(x,y+T)=\tilde{h}(x,y)$ and $\partial_x \tilde{h}(x=0,y) = \partial_x \tilde{h}(x=L,y)=0$. The integration over the fluctuations of $\tilde{h}$ gives the usual factor $Z_0=q^{-1/24} / P(q)$, with $q=e^{- \pi T/L}$ and $P(q)=\prod_{k \geq 1} \left(1-q^k\right)$. Note that this is just the partition function of the Gaussian free field with Neumann boundary conditions on both sides. The other contribution comes from the height defects $2 \pi p y/T$, which must be counted with a phase $e^{ i p r_{12} \gamma}$. The partition function $K_0$ is then\footnote{This is actually not really a partition function for the loop model, but rather a Virasoro character. Hence we call it $K_0$ and not $Z$. We will come back to this in section $\ref{sec:annulusZ}$.}
\begin{equation}
K_0 \propto Z_0 \sum_{p \in \Z} e^{i p r_{12}\gamma} e^{-(g/4\pi) p^2 (2\pi/T)^2 (L T)} = Z_0 \sum_{p \in  \Z} e^{i p r_{12} \gamma} e^{-(\pi g / \tau) p^2} 
\end{equation}
where $\tau = T/L$. We use the Poisson formula $\sum_p \rightarrow \sum_n \int dp e^{-i 2 \pi n p}$ and perform the integration over $p$. This gives
\begin{equation}
\label{eq:interm}
K_0 \propto Z_0 (\tau/g)^{1/2} \sum_{n \in \Z} e^{-(\pi \tau /4g) (r_{12} \gamma/\pi -2n)^2}.
\end{equation}
From general arguments we know that such a partition function on a very long annulus ($T \gg L$) should behave as $K_0 \sim q^{h_0-c/24}$ where $h_0$ is the lowest exponent in the spectrum of the Hamiltonian $L_0$. We use this to normalize properly our result $(\ref{eq:interm})$. Using the expression of the central charge
\begin{equation}
\label{eq:c}
c = 1 - 6 \frac{(g-1)^2}{g}
\end{equation}
and Kac' parameterization for the exponents
\begin{equation}
\label{eq:kac}
h_{r,s} = \frac{(g r - s)^2 -(g-1)^2}{4 g}
\end{equation}
the relation $(\ref{eq:interm})$ becomes
\begin{equation}
\label{eq:K0}
K_0(q) = \frac{q^{-c/24}}{P(q)} \sum_{n \in \Z} q^{h_{r_{12},r_{12}-2 n}}
\end{equation}
which should be interpreted in boundary conformal field theory (BCFT, see \cite{CardyBCFT, CardyBCFTVerlinde}) as follows. First, we interpret our Gaussian free field as a $1+1$-quantum field theory quantized on a segment of length $L$. In this language, the Hamiltonian $H$ is the generator of translations in the $T$-direction on the annulus or on the strip. The strip can be mapped by the conformal transformation $z \mapsto \exp \left( -\frac{i \pi}{L} z \right)$ onto the half-plane (figure $\ref{fig:HL0}$), where $H$ is related to the usual Virasoro generator $L_0$ (the dilatation operator) by the well-known formula $H = \frac{\pi}{L} \left( L_0 - \frac{c}{24} \right)$. The spectrum of $H$ and the spectrum of $L_0$ are then related. But $L_0$ is a Virasoro generator: it acts on some (highest-weight) representations of the Virasoro algebra. In conformal field theory, the highest-weight representations are the primary operators. Where are these operators in the half-plane? They must be situated at the origin, in order for the boundary conditions to change between the positive and negative real axes. Such operators are \textit{boundary-condition-changing-operators} (B.C.C) and were introduced by Cardy in \cite{CardyBCFTVerlinde}.
Computing the partition function $(\ref{eq:K0})$ on an annulus of size $L \times T$ is equivalent to taking the trace of the evolution operator $e^{- T H}$. Recalling that we defined $q = e^{- \pi T/L}$, such a trace should have the form $tr \left\{ e^{- T H} \right\} = (1/P(q)) \sum_{h_\alpha} q^{h_\alpha - c/24}$ where $\left\{ h_\alpha \right\}$ are all the primary operators appearing in the spectrum of $L_0$. The factor $1/P(q)$ comes from the trace over the descendants of each primary operator. This is indeed the form of formula $(\ref{eq:K0})$.

\paragraph{}
We have thus identified the spectrum of primary operators which are necessary to change the boundary condition from the black blobs on the left to the white blobs on the right. These primary operators depend on the loop weights $n_1$, $n_2$ and $n_{12}$ only through the parameter $r_{12}$ and the relations $(\ref{eq:n1}),(\ref{eq:n2}),(\ref{eq:n12})$, according to the above argument. We have checked these results numerically by transfer-matrix diagonalization of the dilute loop model on the honeycomb lattice. The free energy per unit site $f_L$ is computed for successive widths $L=8,9,\dots,12$. We then use the well-known relation
\begin{equation}
\label{eq:numerics}
f_L = f_{\rm{bulk}} + \frac{f_{\rm{boundary}}}{L} + \frac{\pi\left(h_0 - c/24 \right)}{L^2} + \mathcal{O} \left(\frac{1}{L^3} \right)
\end{equation}
up to order $\mathcal{O}\left(\frac{1}{L^4}\right)$ to extract the lowest exponent $h_0$ appearing in the spectrum of $L_0$. The result predicted above $(\ref{eq:K0})$ is that $h_0 = h_{r_{12},r_{12}}$ for $r_{12} \in [ 0 , \pi / \gamma )$. Results are shown in figure $\ref{fig:checkr1r2}$ for $n=\sqrt{2}$ (so $\gamma=\frac{1}{4}$ and $c=\frac{7}{10}$) and for anisotropic special boundary conditions $AS_{\unblob}$ and $AS_{\unblobw}$ on the left and right sides. They are in very good agreement with the Coulomb gas prediction.

\subsection{Boundary operators : B.C.C versus string-creating operators}

\subsubsection{From ordinary to anisotropic special boundary conditions}
\label{subsec:BCC}
In section $\ref{sec:Coulomb}$ we computed the spectrum of the B.C.C operator changing from one type of blobs to another (formula $(\ref{eq:K0})$), and we saw that the weight $n_{12}$ given to loops touching both boundaries was a crucial ingredient. We would like to use this information to understand a simpler B.C.C operator : the operator going from the ordinary boundary condition to the anisotropic special one. To do this, we must give to the (white) blobbed loop the weight $n_2=n$, and to the loops carrying the two blobs the weight $n_{12}=n_1$. In that case there are no closed white unblobbed loops, because they would get a weight $n_2-n=0$. If a loop is blobbed it does not get an additional weight each time it gets a new blob on the boundary because $w_{\blobw}=1$ (see relation $(\ref{eq:conjectureAS})$ for the point $AS_{\unblobw}$). Thus we identify the ordinary b.c with the point $AS_{\unblobw}$ when $n_2=n$. Now recall the parameterizations $(\ref{eq:n2}),(\ref{eq:n12})$ to see that $r_2=1$ and $r_{12}=r_1$. Then the leading exponent appearing in $(\ref{eq:K0})$ is $h_{r_1,r_1}$. Our guess is that the primary operator $\Phi_{r_1,r_1}$ is the B.C.C operator from ordinary to anisotropic special. Note that, since in general $r_1$ does not need to be an integer, $\Phi_{r_1,r_1}$ is the highest-weight state of a generic Verma module, \textit{ie} without null states.

\paragraph{}
However there is something we did not specify yet: what anisotropic special b.c we are talking about. Is it $AS_{\blob}$ or $AS_{\unblob}$? To answer this, note that all the different quantities in our model are supposed to vary continuously when we vary $n_1$. Taking the limit $n_1 \rightarrow n$ we see that the B.C.C operator is the identity so the boundary condition is actually the same on both sides of the B.C.C operator. $AS_{\blob}$ is not the ordinary b.c. in that case (actually it is the isotropic special one), whereas $AS_{\unblob}$ fits. Hence our conclusion is that for arbitrary $n_1$ the B.C.C operator $\left( AS_{\unblob} / Ord \right)$ is $\Phi_{r_1,r_1}$.

\paragraph{}
The operator changing from ordinary to the other anisotropic special b.c can be identified by symmetry. Indeed, note that the model $(\ref{eq:ZOnanisotropic})$ is invariant under the change $n_1 \mapsto n-n_1$ and $\blob \rightarrow \unblob$ in all the formulas. For example, the points $AS_{\blob}$ and $AS_{\unblob}$ are exchanged. Thus the B.C.C operator we are looking for is the same as the previous one if we replace $n_1$ by $n-n_1$. This is the same as taking $\gamma/\pi - r_1$ instead of $r_1$. Using Kac' parameterization $(\ref{eq:kac})$ we find that $h_{\gamma/\pi - r_1, \gamma/\pi -r_1} = h_{r_1, r_1+1}$ so we conclude that the B.C.C operator $\left( AS_{\blob} / Ord \right)$ is $\Phi_{r_1,r_1+1}$.

\subsubsection{Strings}
\label{subsec:strings}
In a loop model, a very natural boundary operator is the operator creating a piece of loop (a \textit{string}) at a point on the boundary.  A string cannot stop in the the bulk or at a boundary point, therefore it must propagate to infinity. The string-creating operator is known to be $\Psi_{2,1}$ with scaling dimension $h_{2,1}$ \cite{CardyOnAnnulus,BauerSaleur,BauerBernard}. It has a null state\footnote{We use the convention that a primary operator $\Phi_{r,s}$ has no null state (and hence corresponds to a generic Verma module) whereas $\Psi_{r,s}$ with $r,s \in \mathbb{N}$ has a null state at level $rs$ (hence a submodule of the generic Verma module is quotiented out).} at level $2$
\begin{equation}
\label{eq:level2}
\left( L_{-2} - \frac{1}{g} L_{-1}^2 \right) \ket{\Psi_{2,1}}=0
\end{equation}
the link between this operator, the relation $(\ref{eq:level2})$ and the Schramm-Loewner evolution $SLE_\kappa$, $\kappa=\frac{4}{g}$ is presented in \cite{BauerBernard}. One can generate more than one string at a boundary point. However, when two strings are generated, they can be contracted into a closed loop, without propagating to infinity. On the contrary, one can ask that they both reach infinity, but this corresponds to a different representation of the Virasoro algebra. This can be formulated in terms of fusion as
\begin{equation}
\label{eq:fusionstring}
\Psi_{2,1} \otimes_f \Psi_{2,1} = \Psi_{1,1} \oplus \Psi_{3,1}
\end{equation} 
where the operator $\Psi_{1,1}$ is just the usual vacuum of the theory (this is the case when the two strings are contracted) and $\Psi_{3,1}$ creates two strings, conditioned not to annihilate with each other. Note that $\Psi_{3,1}$ has a null state at level $3$. Let us give a more pictorial view of this:

$$
\begin{tikzpicture}
		\begin{scope}[xshift=-2cm, yshift=-0.7cm,scale=0.7]
			\clip (-2,-0.6) rectangle (2,2);
			\filldraw[fill=red!15!white, draw=red!80!white, thick] (-10,0) rectangle (10,10);
			\draw[smooth, red] plot coordinates{(0,0) (-0.4,0.2) (-0.2,0.4) (-0.6,0.7) (-1,0.3) (-1.2,0.3) (-1.5,0.2) (-1.4,0.4) (-1.7,0.3) (-1.6,0.5) (-1.8,0.6) (-1.7,0.9) (-1.5,1.2) (-1.2,1.1) (-1.1,1.3) (-1.3,1.6) (-1.5,1.6) (-1.7,1.8) (-1.4,2)};
			\filldraw[gray] (0,0) circle (0.1cm) ++(0,-0.4) node[black] {$\Psi_{2,1}$};
		\end{scope}
		\draw (0.1,0 )node{$\otimes_f$};
		\begin{scope}[xshift=2cm, yshift=-0.7cm,scale=0.7]
			\clip (-2,-0.6) rectangle (2,2);
			\filldraw[fill=red!15!white, draw=red!80!white, thick] (-10,0) rectangle (10,10);
			\draw[smooth, red] plot coordinates{(0,0) (0,0.2) (0.1,0.3) (0.3,0.3) (0.4,0.1) (0.5,0.2) (0.8,0.8) (1.2,0.4) (1.5,0.9) (1.6,1.2) (1.8,1) (1.7,1.4) (1.5,1.6) (1.2,1.3) (1.1,1.4) (1.3,1.6) (1,1.5) (1,1.8) (2.2,4)};
			\filldraw[gray] (0,0) circle (0.1cm) ++(0,-0.4) node[black] {$\Psi_{2,1}$};
		\end{scope}
		\draw (4,0) node{$=$};
		\begin{scope}[xshift=6cm, yshift=-0.7cm,scale=0.7]
			\clip (-2,-0.6) rectangle (2,2);
			\filldraw[fill=red!15!white, draw=red!80!white, thick] (-10,0) rectangle (10,10);
			\draw[smooth, red] plot coordinates{(0,0) (0,0.2) (0.1,0.3) (0.3,0.3) (0.4,0.1) (0.5,0.2) (0.8,0.8) (0,1) (-0.6,0.7) (-0.2,0.4)  (-0.4,0.2) (0,0)};
			\filldraw[gray] (0,0) circle (0.1cm) ++(0,-0.4) node[black] {$\Psi_{1,1}$};
		\end{scope}
		\draw (8,0) node{$\oplus$};
		\begin{scope}[xshift=10cm, yshift=-0.7cm,scale=0.7]
			\clip (-2,-0.6) rectangle (2,2);
			\filldraw[fill=red!15!white, draw=red!80!white, thick] (-10,0) rectangle (10,10);
			\draw[smooth, red] plot coordinates{(0,0) (-0.4,0.2) (-0.2,0.4) (-0.6,0.7) (-1,0.3) (-1.2,0.3) (-1.5,0.2) (-1.4,0.4) (-1.7,0.3) (-1.6,0.5) (-1.8,0.6) (-1.7,0.9) (-1.5,1.2) (-1.2,1.1) (-1.1,1.3) (-1.3,1.6) (-1.5,1.6) (-1.7,1.8) (-1.4,2)};
			\draw[smooth, red] plot coordinates{(0,0) (0,0.2) (0.1,0.3) (0.3,0.3) (0.4,0.1) (0.5,0.2) (0.8,0.8) (1.2,0.4) (1.5,0.9) (1.6,1.2) (1.8,1) (1.7,1.4) (1.5,1.6) (1.2,1.3) (1.1,1.4) (1.3,1.6) (1,1.5) (1,1.8) (2.2,4)};
			\filldraw[gray] (0,0) circle (0.1cm) ++(0,-0.4) node[black] {$\Psi_{3,1}$};
		\end{scope}
\end{tikzpicture}
$$
The operator creating $L$ strings can be found by induction, iterating the fusion with $\Psi_{2,1}$. It is of course $\Psi_{1+L,1}$ (with a null state at level $1+L$).

\paragraph{}
What become the strings when we put the anisotropic special b.c $AS_{\blob}$ or $AS_{\unblob}$ on the left part of the boundary? The result has to be interpreted in terms of the fusion of the string-creating operator $\Psi_{2,1}$ and the B.C.C operator $\Phi_{r_1,r_1+1}$ (or $\Phi_{r_1,r_1}$ in the case of $AS_{\unblob}$). Since $\Psi_{2,1}$ has a null-state at level two, the fusion can be computed by writing down the differential equation satisfied by the three-point function $\left<\Phi_{r_1,r_1+1}(z_1) \Phi_{\delta}(z_2) \Psi_{2,1}(z_3) \right>$ to identify the possible scaling dimensions of $\Phi_{\delta}$. The result is well-known
\begin{equation}
\label{eq:fusionblob}
\Phi_{r_1,r_1+1} \otimes_f \Psi_{2,1} = \Phi_{r_1+1,r_1+1} \oplus \Phi_{r_1-1,r_1+1}
\end{equation}
The two terms on the right correspond to the two possible states of the string generated by $\Psi_{2,1}$, which can now be either blobbed or unblobbed. In the case of the b.c $AS_{\blob}$, the weight given to a blobbed monomer is bigger than the one of an unblobbed monomer. Then the scaling dimension of the operator creating a blobbed string should be smaller than the one of the operator creating an unblobbed string in that case. In general, $h_{r_1+1, r_1+1} < h_{r_1-1,r_1+1}$ so we get the following pictorial view
$$
\begin{tikzpicture}
		\begin{scope}[xshift=-2cm, yshift=-0.7cm,scale=0.7]
			\clip (-2,-0.6) rectangle (2,2);
			\filldraw[fill=red!15!white, draw=red!80!white, thick] (-10,0) rectangle (10,10);
			\draw[line width=2.5pt] (-2.5,0) -- (0,0);
			\filldraw[gray] (0,0) circle (0.1cm) ++(0,-0.4) node[black] {$\Phi_{r_1,r_1+1}$};
			\filldraw[gray] (-1.5,0.4) node[black] {$AS_{\blob}$};
			\filldraw[gray] (1.5,0.4) node[black] {$Ord$};
		\end{scope}
		\draw (0.1,0 )node{$\otimes_f$};
		\begin{scope}[xshift=2cm, yshift=-0.7cm,scale=0.7]
			\clip (-2,-0.6) rectangle (2,2);
			\filldraw[fill=red!15!white, draw=red!80!white, thick] (-10,0) rectangle (10,10);
			\draw[smooth, red] plot coordinates{(0,0) (-0.4,0.2) (-0.2,0.4) (-0.6,0.7) (-1,0.3) (-1.2,0.3) (-1.5,0.2) (-1.4,0.4) (-1.7,0.3) (-1.6,0.5) (-1.8,0.6) (-1.7,0.9) (-1.5,1.2) (-1.2,1.1) (-1.1,1.3) (-1.3,1.6) (-1.5,1.6) (-1.7,1.8) (-1.4,1.9) (-1,1.7) (-0.5,1.3) (-0.2,1.5) (0.2,2)};
			\filldraw[gray] (0,0) circle (0.1cm) ++(0,-0.4) node[black] {$\Psi_{2,1}$};
		\end{scope}
		\draw (4,0) node{$=$};
		\begin{scope}[xshift=6cm, yshift=-0.7cm,scale=0.7]
			\clip (-2,-0.6) rectangle (2,2);
			\filldraw[fill=red!15!white, draw=red!80!white, thick] (-10,0) rectangle (10,10);
			\draw[smooth, red] plot coordinates{(0,0) (-0.4,0.2) (-0.2,0.4) (-0.6,0.7) (-1,0.) (-1.2,0.3) (-1.5,0.2) (-1.4,0.4) (-1.7,0.3) (-1.6,0.5) (-1.8,0.6) (-1.7,0.9) (-1.5,1.2) (-1.2,1.1) (-1.1,1.3) (-1.3,1.6) (-1.5,1.6) (-1.7,1.8) (-1.4,1.9) (-1,1.7) (-0.5,1.3) (-0.2,1.5) (0.2,2)};
			\draw[line width=2.5pt] (-2.5,0) -- (0,0);
			\filldraw[gray] (0,0) circle (0.1cm) ++(0,-0.4) node[black] {$\Phi_{r_1+1,r_1+1}$};
			\filldraw (-1.7,0.9) circle (0.1); 
		\end{scope}
		\draw (8,0) node{$\oplus$};
		\begin{scope}[xshift=10cm, yshift=-0.7cm,scale=0.7]
			\clip (-2,-0.6) rectangle (2,2);
			\filldraw[fill=red!15!white, draw=red!80!white, thick] (-10,0) rectangle (10,10);
			\draw[smooth, red] plot coordinates{(0,0) (-0.4,0.2) (-0.2,0.4) (-0.6,0.7) (-1,0.) (-1.2,0.3) (-1.5,0.2) (-1.4,0.4) (-1.7,0.3) (-1.6,0.5) (-1.8,0.6) (-1.7,0.9) (-1.5,1.2) (-1.2,1.1) (-1.1,1.3) (-1.3,1.6) (-1.5,1.6) (-1.7,1.8) (-1.4,1.9) (-1,1.7) (-0.5,1.3) (-0.2,1.5) (0.2,2)};
			\draw[line width=2.5pt] (-2.5,0) -- (0,0);
			\filldraw[gray] (0,0) circle (0.1cm) ++(0,-0.4) node[black] {$\Phi_{r_1-1,r_1+1}$};
			\filldraw (-1.8,0.8) rectangle ++(0.2,0.2); 
		\end{scope}
\end{tikzpicture}
$$
One can add more strings by iterating the fusion procedure. However, note that only the leftmost string can be blobbed or unblobbed, because the other ones never touch the boundary. The operator creating $L$ strings with $AS_{\blob}$ b.c on the left and with the leftmost string blobbed is then $\Phi_{r_1+L,r_1+1}$. When the leftmost string is unblobbed, it is $\Phi_{r_1-L,r_1+1}$. In the case of $AS_{\unblob}$ b.c on the left, the operator is $\Phi_{r_1+L,r_1}$ when the leftmost string is blobbed, and $\Phi_{r_1-L,r_1}$ when it is unblobbed. Note that again, the $n_1 \leftrightarrow n-n_1$ duality shows up, because all these operators are exchanged when one changes all the $\blob$ into $\unblob$ and $r_1$ into $\pi/\gamma - r_1$.

\subsubsection{Virasoro characters}
So far, we have identified the B.C.C operators going from ordinary to one of the anisotropic special boundary conditions, as well as their fusion with the string-creator operator $\Psi_{2,1}$. Taking the trace of $q^{L_0-c/24}$ over the Verma modules of these operators, we obtain the corresponding Virasoro characters. $L$ is the number of strings and we indicate the blob status of the leftmost string when necessary. $K_0$ is the character without string ($L=0$). For completeness we indicate the characters for anisotropic special b.c on the left and right side. These are conjectured from a combination of results of \cite{DJSdense} and from algebraic arguments coming from the two-boundary Temperley-Lieb algebra. We do not want to develop this here, and relegate it to \cite{DJSprep}.

\begin{equation}
\label{eq:Kordord}
Ord / Ord: \qquad \begin{array}{rcl}
K_L(q) &=& \frac{q^{-c/24}}{P(q)} \left( q^{h_{1+L,1}} - q^{h_{1+L,-1}}\right)
\end{array}
\end{equation}

\begin{equation}
\label{eq:Kas1ord}
AS_{\blob} / Ord: \qquad \left\{\begin{array}{rcl}
K_0(q) &= & \frac{q^{h_{r_1,r_1+1}-c/24}}{P(q)} \\
K_L^{\blob}(q) &= & \frac{q^{h_{r_1+L,r_1+1}-c/24}}{P(q)} \\
K_L^{\unblob}(q) &= & \frac{q^{h_{r_1-L,r_1+1}-c/24}}{P(q)}
\end{array} \right.
\end{equation}

\begin{equation}
\label{eq:Kas2ord}
AS_{\unblob} / Ord: \qquad \left\{\begin{array}{rcl}
K_0(q) &= & \frac{q^{h_{r_1,r_1}-c/24}}{P(q)} \\
K_L^{\blob}(q) &= & \frac{q^{h_{r_1+L,r_1}-c/24}}{P(q)} \\
K_L^{\unblob}(q) &= & \frac{q^{h_{r_1-L,r_1}-c/24}}{P(q)}
\end{array} \right.
\end{equation}

\begin{equation}
\label{eq:Kasas}
AS_{\blob} / AS_{\blobw}: \qquad \left\{\begin{array}{rcl}
K_0(q) &= & \frac{q^{-c/24}}{P(q)} \sum_{n \in \Z} q^{h_{r_{12},r_{12}-2n}} \\
K_L^{\blob \; \blobw}(q) &=& \frac{q^{-c/24}}{P(q)} \sum_{n \geq 0} q^{h_{r_1+r_2-1+L,r_1+r_2+1-2n}} \\
K_L^{\blob \; \unblobw}(q) &=& \frac{q^{-c/24}}{P(q)} \sum_{n \geq 0} q^{h_{r_1-r_2-1+L,r_1-r_2-1-2n}} \\
K_L^{\unblob \; \blobw}(q) &=& \frac{q^{-c/24}}{P(q)} \sum_{n \geq 0} q^{h_{-r_1+r_2-1+L,-r_1+r_2-1-2n}} \\
K_L^{\unblob \; \unblobw}(q) &=& \frac{q^{-c/24}}{P(q)} \sum_{n \geq 0} q^{h_{-r_1-r_2-1+L,-r_1-r_2-3-2n}} \\
\end{array} \right.
\end{equation}

The cases of $AS_{\blob}/AS_{\unblobw}$, $AS_{\unblob}/AS_{\blobw}$, $AS_{\unblob}/AS_{\unblobw}$ can be deduced from the $AS_{\blob}/AS_{\blobw}$ case, by the duality transformations $n_1 \rightarrow n-n_1$ and/or $n_2 \rightarrow n-n_2$, as explained previously.

\subsection{Some remarks about fractal dimensions}
\label{sec:fractal_dimensions}

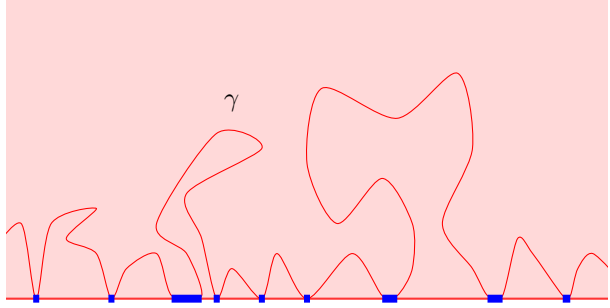
\begin{figure}[h]
\centering
\begin{tikzpicture}
	\begin{scope}[scale=2]
			\clip (-2,-0.6) rectangle (2,2);
			\filldraw[fill=red!15!white, draw=red!80!white, thick] (-10,0) rectangle (10,10);
			\draw[smooth, red] plot coordinates{(2.1,0.1) (1.9,0.3) (1.8,0.2) (1.75,0.02) (1.7,0.02) (1.5,0.3) (1.4,0.4) (1.3,0.02) (1.2,0.02) (1.1,0.2) (0.9,0.5) (1.1,1) (1,1.5) (0.6,1.2) (0.1,1.4) (0,0.9) (0.2,0.5) (0.5,0.8) (0.7,0.5) (0.7,0.2) (0.6,0.02) (0.5,0.02) (0.3,0.3) (0,0) (-0.2,0.3) (-0.3,0) (-0.5,0.2) (-0.6,0) (-0.7,0.4) (-0.8,0.7) (-0.3,1) (-0.6,1.1) (-1,0.5) (-0.8,0.3) (-0.7,0.02) (-0.9,0.02) (-1,0.3) (-1.2,0.2) (-1.3,0) (-1.4,0.3) (-1.6,0.4) (-1.4,0.6) (-1.7,0.5) (-1.8,0) (-1.9,0.5) (-2.1,0.3)};
			\draw (-0.5,1.3) node {$\gamma$};
			\draw[blue,line width=3pt] (1.75,0) -- (1.7,0) (1.3,0) -- (1.2,0) (0.6,0) -- (0.5,0) (-0.02,0) -- (0.02,0) (-0.32,0) -- (-0.28,0) (-0.62,0) -- (-0.58,0) (-0.7,0) -- (-0.9,0) (-1.32,0) -- (-1.28,0) (-1.82,0) -- (-1.78,0);  
		\end{scope}
\end{tikzpicture}
\caption{The contact set (blue) of a loop is the intersection of its curve $\gamma$ and the boundary of the upper half-plane (the real axis). Its fractal dimension is related to the scaling dimension of the $2$-arm exponent.}
\label{fig:contactset}
\end{figure}

The link between scaling dimensions of operators and fractal dimensions of some geometric objects is well-known. For example, in the dilute loop model with which we are dealing here, the fractal dimension $d_f$ of the loops in the bulk can be computed very easily. It is the RG eigenvalue of the $2$-arm exponent $h_{1,0}$, which gives the celebrated formula $d_f=2-2 h_{1,0} = 1+\frac{1}{2g}$. The boundary operators we have identified above might be used to derive such formulae. The boundary condition cannot change the local geometry (and especially the fractal dimension) of an object at a point in the bulk, so here the fractal dimension of the loops is not the quantity we are interested in. We should rather look at the set of points on the boundary which are visited by a chosen loop. Let us call this the \textit{contact set} of a given loop. As above, the fractal dimension $d^{(\rm{b.c.})}_{f}$ of such a set must be given by a (boundary) $2$-arm exponent, which depends on the boundary condition (the same on the left and right sides of the point where the operator is inserted). But we have just identified these operators for all the boundary conditions we are interested in. For example, for ordinary boundary conditions, one has
\begin{equation}
d^{(Ord)}_{f}=1-h_{3,1} \leq 0
\end{equation}
for $0 \leq n \leq 2$. This means that, in the scaling limit, the contact set of a given loop is almost surely empty: loops never touch the boundary. On the contrary, at the (isotropic) special point, this set has a non-trivial scaling-limit with fractal dimension
\begin{equation}
d^{(Sp)}_f=1-h_{3,3}=1 - 2 \frac{(g-1)^2}{g}
\end{equation}
When the b.c is $AS_{\blob}$ there are two kinds of loops: blobbed and unblobbed ones. The contact set of both types are different. For a blobbed loop the fractal dimension is
\begin{equation}
d^{(AS_{\blob})}_f \left( \blob \; \rm{loop} \right)=1-h_{2r_1+1,2r_1+1}=1 - r_1(r_1+1) \frac{(g-1)^2}{g}
\end{equation}
while for an unblobbed loop
\begin{equation}
d^{(AS_{\blob})}_f \left( \unblob \; \rm{loop} \right)=1-h_{-2r_1+1,-2r_1-3}\leq 0
\end{equation}
for $0 \leq n_1 \leq n \leq 2$ so in the scaling limit, an unblobbed loop never touches the boundary. We conjecture that these fractal dimensions might be of some relevance for $SLE_{\kappa,\rho}$ when $8/3 \leq \kappa \leq 4$. So far (to our knowledge), rigorous results have been obtained only for the contact set of $SLE_{\kappa}$ when $4 <\kappa <8$ \cite{AlbertsSheffield}.

\subsection{Annulus partition functions and boundary entropies}
We use the above Virasoro characters to derive expressions for the annulus partition functions of the $\On$ model with ordinary and anisotropic special boundary conditions. The geometry of the annulus plays a role in BCFT similar to the one of the torus in CFT. A partition function can be seen either as a periodic evolution of a one-dimensional quantum theory with two boundaries (also called the \textit{open-string channel}) or as the evolution of a periodic theory between two boundary states (this is the \textit{closed string channel}) \cite{CardyBCFTVerlinde}. So far we have dealt only with the open-string channel, working with B.C.C operators in the half-plane, then deducing the corresponding Virasoro characters. In this section we conclude this work by gathering all the pieces of the puzzle to obtain the annulus partition functions. Then we go to the closed-string channel and compute the boundary entropies of the boundary states $\ket{Ord}$, $\ket{AS_{\blob}}$, etc.

\subsubsection{A loop partition function is a Markov trace (not a trace)}

A loop configuration on the annulus can be drawn as a planar rectangular diagram with half-loops attached on the top and the bottom of the diagram. When a closed loops appear in the diagram, it can be removed if we give a weight $n$ to the diagram.
With this rule we get a set of diagrams connecting $N$ sites on the bottom to $N$ sites at the top of the diagram. Two diagrams are equivalent iff they represent the same connectivities between the sites. For example 
\begin{center}
\begin{tikzpicture}
	\draw (-3.4,1.2) node{$M \;=$};
	\begin{scope}[scale=1.2, xshift=-2cm]
		\draw[smooth, thick] plot coordinates{(0,0) (0,0.3) (0.3,0.4) (0.4,0.7) (0.3,1) (0.5,1.5) (0,2)};
		\draw[smooth, thick] (0,1.3) circle (0.15);
		\draw[smooth, thick] (0.1,0.6) circle (0.07);
		\draw[smooth, thick] plot coordinates{(0.6,0) (0.6,0.5) (1.5,1.5) (1.5,2)};
		\draw[smooth, thick] plot coordinates{(1.2,0) (1.2,0.5) (1.3,0.8) (1.7,0.7) (1.5,0.4) (1.5,0)};
		\draw[smooth, thick] plot coordinates{(0.6,2) (0.6,1.7) (0.8,1.5) (1.2,1.5) (1.3,1.7) (1.2,2)};
		\draw[smooth, thick] (0.8,1.2) circle (0.13);
		\draw[gray] (-0.3,0) rectangle (1.8,2);
	\end{scope}
	\draw (0.6,1.2) node{$= \quad n^3$};
	\begin{scope}[scale=1.2, xshift=1.5cm]
		\draw[smooth, thick] plot coordinates{(0,0)(0,2)};
		\draw[smooth, thick] plot coordinates{(0.6,0) (1.5,2)};
		\draw[smooth, thick] plot coordinates{(1.2,0) (1.3,0.3) (1.4,0.3) (1.5,0)};
		\draw[smooth, thick] plot coordinates{(0.6,2) (0.8,1.7) (1,1.7) (1.2,2)};
		\draw[gray] (-0.3,0) rectangle (1.8,2);
	\end{scope}
\end{tikzpicture}
\end{center}

If we define the product of two diagrams as a concatenation, putting one diagram above the other and counting the closed loops with a weight $n$, then this defines the Temperley-Lieb algebra $TL_N(n)$. The generic representation theory of this algebra is very simple\footnote{Generic means when $n$ is not of the form $2 \cos \left( \frac{\pi}{m} \right)$ with $m$ an integer. In the non-generic case most of the string representations become indecomposable. The irreducible representations obtained by taking quotients of the latter ones are related to the RSOS models \cite{PasquierADE}. This comes out in the Markov trace because some coefficients are zero in that case.}. A generic irreducible representation is given by the set of half-diagrams connecting $N$ points. The diagrams of the $TL$ algebra act on these by concatenation. For example the set
$$
\V_{0} = \left\{ \begin{array}{lll}
\begin{tikzpicture}
	\draw[thick] (0,0.4) arc (-180:0:0.15 and 0.4);
	\draw[thick] (0.6,0.4) arc (-180:0:0.15 and 0.4);
\end{tikzpicture} &, &
\begin{tikzpicture}
	\draw[thick] (0,0.4) arc (-180:0:0.45 and 0.4);
	\draw[thick] (0.3,0.4) arc (-180:0:0.15 and 0.25);
\end{tikzpicture}
\end{array}
\right\}
$$
is a representation of the algebra $TL_4(n)$. Other irreducible representations are obtained when one introduces strings. A pair of strings cannot be contracted: a $TL$ diagram which makes a connection between two strings takes a weight zero. With this rule, one obtains two additional representations for $TL_4(n)$
$$
\V_{2} = \left\{ \begin{array}{lllll}
\begin{tikzpicture}
	\draw[thick] (0,0.4) arc (-180:0:0.15 and 0.4);
	\draw[thick] (0.6,0) -- (0.6,0.4);
	\draw[thick] (0.9,0) -- (0.9,0.4);
\end{tikzpicture} &, &
\begin{tikzpicture}
	\draw[thick] (0.3,0.4) arc (-180:0:0.15 and 0.4);
	\draw[thick] (0,0) -- (0,0.4);
	\draw[thick] (0.9,0) -- (0.9,0.4);
\end{tikzpicture} &,&
\begin{tikzpicture}
	\draw[thick] (0.6,0.4) arc (-180:0:0.15 and 0.4);
	\draw[thick] (0,0) -- (0,0.4);
	\draw[thick] (0.3,0) -- (0.3,0.4);
\end{tikzpicture}
\end{array}
\right\} 
\qquad \rm{and} \qquad
\V_{4} = \left\{ \begin{array}{l}
\begin{tikzpicture}
	\draw[thick] (0,0) -- (0,0.4);
	\draw[thick] (0.3,0) -- (0.3,0.4);
	\draw[thick] (0.6,0) -- (0.6,0.4);
	\draw[thick] (0.9,0) -- (0.9,0.4);
\end{tikzpicture} \end{array} \right\}
$$
Now we are ready to the introduce the Markov trace. The Markov trace of a diagram is the number obtained when the top and the bottom of the diagram are identified and each closed loop is given a weight $n$. For example, the Markov trace of the above diagram $M$ is $\rm{Tr}\; M = n^5$. The Markov trace extends to the combinations of $TL$ diagrams by linearity. Note that it is also invariant under cyclic permutation of the diagrams: for two diagrams $A$ and $B$ which are multiplied with the concatenation rule, $\rm{Tr}\; AB= \rm{Tr} \; BA$. Note that the Markov trace is exactly the object we need to compute a partition function of a loop model on an annulus. The configurations of the loop model are given by the $TL$ diagrams with suitable weights, and the geometry of the annulus is obtained when one identify the top and the bottom of each diagram, with a weight given by the Markov trace.
\paragraph{}
The Markov trace of a diagram is very different from the usual trace over a representation of the $TL$ algebra. However, there is a well-known relation between these objects
\begin{equation}
\label{eq:Markov}
\rm{Tr} \; A = \sum_{L \geq 0} \D_{L} tr_{\V_L} A
\end{equation}
where the $\D_{L}$ are Chebyshev polynomials of the second kind $\D_0 = 1$, $\D_1=n$, $\D_2=n^2-1$, $\D_3=n(n^2-2)$, etc. In full generality $\D_L = \frac{\sin (L+1) \gamma}{\sin \gamma}$, with $n =2 \cos \gamma$. Of course, we use the convention that a representation $\V_L$ of $TL_N(n)$ is empty if $L$ and $N$ do not have the same parity, or if $L>N$. Again consider the example of the diagram drawn above: it is easy to check that $tr_{\V_0} M = n^3$, $tr_{\V_2}=n^3$ and $tr_{\V_4}=0$ and all the other traces are zero by definition, so using the formula $(\ref{eq:Markov})$ one finds $\rm{Tr} \; M = n^5$ as expected.
\paragraph{}
It is clear now why one has to be interested in the relation $(\ref{eq:Markov})$ in view of the preceding sections. Indeed, we have identified the natural candidates for the different string representations of the $TL(n)$ algebra in the scaling limit. These are the Verma modules of the string-creating operators $\Psi_{1+L,1}$. Thus in the scaling limit, the trace over the different $TL(n)$ representations of the evolution operator (on the lattice this is some power of the transfer matrix), when properly renormalized, converges to the trace over a Verma module of the BCFT evolution operator $q^{L_0-c/24}$. These traces are the Virasoro characters $(\ref{eq:Kordord})$. Thus we get the partition function of the $\On$ model on an annulus with ordinary b.c on both sides for free
\begin{equation}
\label{eq:Zordord}
Z_{Ord/Ord} = \sum_{L \geq 0} \D_L K_L(q).
\end{equation}
Interestingly enough, the fact that the coefficients $\D_L$ are Chebyshev polynomials of the second kind has a nice interpretation in terms of fusion. Indeed, we have
\begin{equation}
\label{eq:chebyshev}
\D_L \D_1 = \D_{L+1} + \D_{L-1}
\end{equation}
and this looks like the fusion rule (see $(\ref{eq:fusionstring}$) for the case $L=1$) of the operators $\Psi_{1+L,1}$ and $\Psi_{2,1}$. Another way to say this is that the coefficients $\D_L$, also named \textit{quantum dimensions}, satisfy the fusion algebra of $\Psi_{1+L,1}$. This gives a simple and powerful way of computing the coefficients $\D_L$.

\paragraph{}
So far, we have not talked about the blobs in the present section. Actually the generalization of formula $(\ref{eq:Markov})$ to the case with blobs is straightforward. The whole framework of the Temperley-Lieb algebra can be generalized to a diagrammatic algebra where some of the loops can carry blobs
\begin{center}
\begin{tikzpicture}
	\begin{scope}[scale=1.2, xshift=-2cm]
		\draw[smooth, thick] plot coordinates{(0,0) (0,0.3) (0.3,0.4) (0.4,0.7) (0.3,1) (0.5,1.5) (0,2)};
		\draw[smooth, thick] (0,1.3) circle (0.15);
		\draw[smooth, thick] (0.1,0.6) circle (0.07);
		\draw[smooth, thick] plot coordinates{(0.6,0) (0.6,0.5) (1.5,1.5) (1.5,2)};
		\draw[smooth, thick] plot coordinates{(1.2,0) (1.2,0.5) (1.3,0.8) (1.7,0.7) (1.5,0.4) (1.5,0)};
		\draw[smooth, thick] plot coordinates{(0.6,2) (0.6,1.7) (0.8,1.5) (1.2,1.5) (1.3,1.7) (1.2,2)};
		\draw[smooth, thick] (0.8,1.2) circle (0.13);
		\draw[gray] (-0.3,0) rectangle (1.8,2);
		\filldraw (-0.15,1.3) circle (2pt);
		\filldraw (0.3,1.) circle (2pt);
	\end{scope}
	\draw (0.6,1.2) node{$= \quad n_1 n^2$};
	\begin{scope}[scale=1.2, xshift=1.5cm]
		\draw[smooth, thick] plot coordinates{(0,0)(0,2)};
		\draw[smooth, thick] plot coordinates{(0.6,0) (1.5,2)};
		\draw[smooth, thick] plot coordinates{(1.2,0) (1.3,0.3) (1.4,0.3) (1.5,0)};
		\draw[smooth, thick] plot coordinates{(0.6,2) (0.8,1.7) (1,1.7) (1.2,2)};
		\draw[gray] (-0.3,0) rectangle (1.8,2);
		\filldraw (0,1.) circle (2pt);
	\end{scope}
\end{tikzpicture}
\end{center}
This algebraic structure has first appeared in \cite{MartinSaleur} under the name of \textit{blob algebra}, and it was sometimes renamed into \textit{one-boundary Temperley-Lieb algebra} or $1BTL$ since \cite{deGierNichols,deGier2BTL,DJSdense,JS1,JS2}. This algebra has now two parameters $n$ and $n_1$. The generic representation theory follows the one of the $TL$ algebra. The representations are the sets of half-diagrams connecting $N$ points without crossing. Some of the half-loops can now carry blobs. In the case of representations with strings, there are two inequivalent representations, depending on whether the leftmost string is blobbed or unblobbed. There exists a Markov trace over this algebra, and an equivalent of formula $(\ref{eq:Markov})$ involving all the string representations (blobbed and unblobbed). The corresponding coefficients $\D^{\blob}_L$ and $\D^{\unblob}_L$ turn out to satisfy the fusion rule $(\ref{eq:fusionblob})$, or in full generality \cite{JS1,JS2}
\begin{equation}
\label{eq:fusionDblob}
\D^{\blob}_{L} \D_1 = \D^{\blob}_{L+1} + \D^{\blob}_{L-1} \qquad  \D^{\unblob}_{L} \D_1 = \D^{\unblob}_{L+1} + \D^{\unblob}_{L-1}
\end{equation}
and $\D_0=1$, $\D^{\blob}_1= n_1$, $\D^{\unblob}_1=n-n_1$. This can also be written as $\D^{\blob}_L = \frac{\sin (r_1+L) \gamma}{\sin r_1 \gamma}$ and $\D^{\unblob}_L = \frac{\sin (r_1-L) \gamma}{\sin r_1 \gamma}$. Again, with this result at hand, we can write down partition functions very easily. For example, the partition function of the $\On$ model on an annulus with $AS_{\blob}/ Ord$ boundary conditions is simply given by the characters $(\ref{eq:Kas1ord})$ and
\begin{equation}
\label{eq:Zas1ord}
Z_{AS_{\blob} / Ord} = K_0(q) + \sum_{L \geq 1} \D_L^{\blob} K_L^{\blob}(q) +  \sum_{L \geq 1} \D_L^{\unblob} K_L^{\unblob}(q)
\end{equation}
In the case of two boundaries, there is also an equivalent of formula $(\ref{eq:Markov})$, this time involving all the string representation with the different blob status of the leftmost and rightmost strings. More details about this, as well as the expressions for the quantum dimensions $\D^{\blob \; \blobw}_L$, etc. can be found in \cite{DJSdense,JS2}.

\subsubsection{Results for the boundary entropies}
\label{subsec:gfactor}
\begin{figure}
\center
\begin{tikzpicture}
	\begin{scope}
		\draw[thick] (0,0) ellipse (0.5 and 1);
		\draw[thick] (3,-1) arc (-90:90:0.5 and 1);
		\draw[thick] (0,-1) -- (3,-1);
		\draw[thick] (0,1) -- (3,1);
		\draw (-0.8,-0.7) node{$\alpha$};
		\draw (3.8,-0.7) node{$\beta$};
		\draw[thick, ->] (1,-0.7) arc (-65:65:0.3 and 0.7);
		\draw (2,0) node {$q^{L_0-c/24}$}; 
	\end{scope}
	\begin{scope}[xshift=7cm]
		\draw[thick] (0,0) ellipse (0.5 and 1);
		\draw[thick] (3,-1) arc (-90:90:0.5 and 1);
		\draw[thick] (0,-1) -- (3,-1);
		\draw[thick] (0,1) -- (3,1);
		\draw (-0.8,-0.7) node{$\ket{\alpha}$};
		\draw (3.8,-0.7) node{$\ket{\beta}$};
		\draw[thick, ->] (0.8,-0.2) -- (2.8,-0.2);
		\draw (2,0.2) node {$\q^{L_0+\bar{L}_0-c/12}$}; 
	\end{scope}
\end{tikzpicture}
\caption{Open-string channel (on the left) and closed-string channel (on the right).}
\label{fig:channels}
\end{figure}
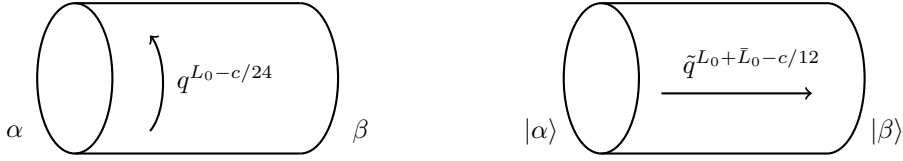

We have just seen that it is now very easy to write down the partition functions of the $\On$ model for the various boundary conditions we are considering. We use these partition functions to compute the boundary entropies \cite{AffleckLudwig} of the corresponding boundary states. Recall that in BCFT, a partition function on the annulus can be viewed either in the open-string channel or in the closed channel \cite{CardyBCFTVerlinde} (figure $\ref{fig:channels}$). So far we have worked only in the open-string channel, where our partition function are Markov traces of the evolution operator $q^{L_0-c/24}$ ($q=e^{-\pi \tau}$). We go to the closed-string channel by performing Poisson resummations $\displaystyle \sum_L \rightarrow \sum_p \int dL e^{i 2 \pi p L}$ on the partition functions. The result is of the form $Z=\left< \alpha \right| \q^{L_0+\bar{L}_0-c/12} \left| \beta \right>$ where $\q=e^{-2\pi/\tau}$ is the modular parameter conjugated to $q$, and $\ket{\alpha}$, $\ket{\beta}$ are the boundary states which implement the boundary conditions $\alpha$ and $\beta$ on both sides of the annulus. In the limit of a very long cylinder ($\q \rightarrow 0$), only the lowest exponent contributes in the closed-string channel. Here the lowest exponent is zero, so $Z \sim \left< \alpha | 0 \right> \q^{-c/12} \left< 0 | \beta \right>$, where $\ket{0}$ is the ground state of $L_0 = \bar{L}_0 -c/12$. The partition function factorizes into contribution from each boundary, and a bulk part which is trivially $\q^{-c/12}$.

The $g$-factors $g_\alpha = \left< \alpha | 0 \right>$ and $g_\beta = \left< \beta | 0 \right>$ count the degeneracy of the ground state in the boundary states $\ket{\alpha}$ and $\ket{\beta}$ \cite{AffleckLudwig}. The free energy $f=-\log Z$ gets a contribution $f_{\rm{l.b}}=- \log g_\alpha$ from the left boundary, and $f_{r.b}=-\log g_\alpha$ from the right one. These quantities are called the $\textit{boundary entropies}$ of the boundary conditions $\alpha$ and $\beta$. The $g$-factors and the boundary entropies are of interest in the study of BCFT. We will use them in section $\ref{part:flows}$ to organize the boundary RG flows between the different boundary conditions of the $\On$ model.

\paragraph{Ordinary boundary condition on both sides:}
the partition function is ($\ref{eq:Zordord}$)
$$
Z_{Ord/Ord}=\frac{q^{-c/24}}{P(q)}  \sum_{L=0}^\infty \frac{\sin (1+L) \gamma}{\sin \gamma} \left( q^{h_{1+L,1}} - q^{h_{1+L,-1}} \right)
$$
and its modular transform is
$$
Z_{Ord/Ord} = \sqrt{\frac{2}{g}} \frac{\q^{-1/12}}{P(\q^2)} \sum_{p \in \Z} \frac{\sin \left( \frac{\gamma }{ g} - \frac{g-1}{g} 2 \pi p \right)}{\sin \gamma} \q^{2 \frac{1}{4g} \left( 2p + g-1 \right)^2}
$$
which gives the boundary entropy $S_{Ord} = - \log g_{Ord}$ where
\begin{equation}
\label{eq:gord}
g_{Ord}= \left( \frac{2}{g} \right)^{1/4} \left( \frac{\sin (\gamma /g)}{\sin \gamma} \right)^{1/2}
\end{equation}

\paragraph{(Isotropic) Special/Ordinary b.c:}
The operator going from ordinary to (isotropic) special b.c is known to be $\Psi_{1,2}$, so the Virasoro character with $L$ strings is
$$
K_{1+L,2}=\frac{q^{-c/24}}{P(q)} \left( q^{h_{1+L,2}} - q^{h_{1+L,-2}} \right)
$$
and the partition function is then
$$
Z_{Sp/Ord}=\frac{q^{-c/24}}{P(q)} \sum_{L=0}^\infty \frac{\sin (1+L) \gamma}{\sin \gamma} \left( q^{h_{1+L,2}}- q^{h_{1+L,-2}} \right)
$$
the modular transform is
$$
Z_{Sp/Ord} = \sqrt{\frac{2}{g}} \frac{\q^{-1/12}}{P(\q^2)} \sum_{p \in \Z} \frac{\sin \left( 2 \frac{\gamma }{ g} + \frac{2-g}{g} 2 \pi p \right)}{\sin \gamma} \q^{2 \frac{1}{4g} \left( 2p + g-1 \right)^2}
$$
which gives the boundary entropy $S_{Sp}=- \log g_{Sp}$ with
\begin{equation}
\label{eq:gSp}
g_{Sp}= \left( \frac{2}{g} \right)^{1/4} \frac{\sin( 2 \gamma /g)}{ \left( \sin \gamma \sin (\gamma/g) \right)^{1/2}}
\end{equation}

\paragraph{Anisotropic special / Ordinary b.c}
with the anisotropic special b.c $AS_{\blob}$ on one side of the annulus and ordinary b.c on the other side, we obtain from $(\ref{eq:Kas1ord})$ and $(\ref{eq:Zas1ord})$
$$
Z_{AS_{\blob}/Ord} = \frac{q^{-c/24}}{P(q)} \sum_{L \in \Z} \frac{\sin (r_1+L) \gamma}{\sin r_1 \gamma} q^{h_{r_1+L,r_1+1}} 
$$
and its modular transform
$$
Z_{AS_{\blob}/Ord} = \sqrt{\frac{2}{g}} \frac{\q^{-1/12}}{P(\q^2)} \sum_{p \in \Z} \frac{\sin \left( (r_1+1) \gamma/g + \frac{1-r_1(g-1)}{g} 2 \pi p \right)}{\sin r_1 \gamma} \q^{2 \frac{1}{4g} \left( 2p + \frac{\gamma}{\pi} \right)^2}
$$
so we have the boundary entropy $S_{AS_{\blob}} = - \log g_{AS_{\blob}}$, with
\begin{equation}
\label{eq:gas1}
g_{AS_{\blob}}=\left( \frac{2}{g} \right)^{1/4} \frac{\sin ((r_1+1) \gamma/g)}{\sin r_1 \gamma} \left( \frac{\sin \gamma}{\sin (\gamma/g)} \right)^{1/2}.
\end{equation}
The boundary entropy $S_{AS_{\unblob}}=-\log g_{AS_{\unblob}}$ can be deduced from this one by the duality transformation $r_1 \rightarrow \pi/\gamma-r_1$. We get
\begin{equation}
g_{AS_{\unblob}} = \left( \frac{2}{g} \right)^{1/4} \frac{\sin (r_1 \gamma/g)}{\sin r_1 \gamma} \left( \frac{\sin \gamma}{\sin (\gamma/g)} \right)^{1/2}.
\label{eq:gas2}
\end{equation}

\label{sec:annulusZ}




\section{Boundary RG flows}
\label{part:flows}
The BCFT machinery developed in section $\ref{partBCFT}$ is now used to explain some features of the phase diagram shown in figure $\ref{fig:phase}$. We also discuss the link with known results of integrable field theory (IFT) and the thermodynamic Bethe Ansatz (TBA).

\subsection{Affleck \& Ludwig's ``$g$-theorem'' and consistency of the phase diagram}

\begin{figure}[t]
\center
\includegraphics[width=0.55\textwidth]{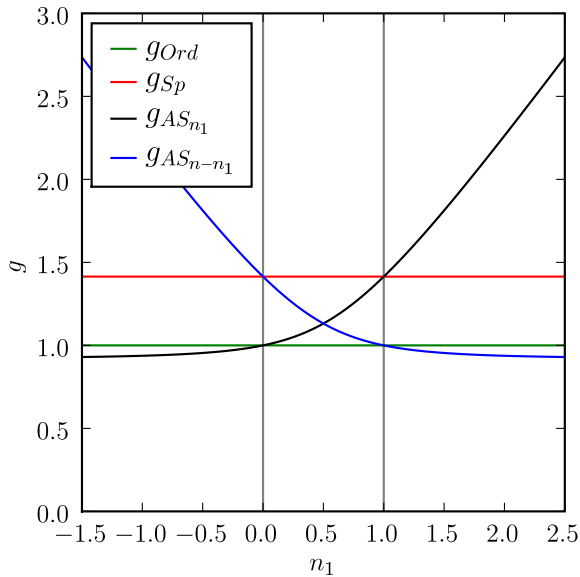}
\caption{$g$-factors of the ordinary, special and anisotropic special boundary conditions for the $\On$ model. The results here are shown for $n=1$. In the ``physical'' region $0 < n_1 <n$, the $g$-factors suggest that the most stable b.c is the ordinary one, and the most unstable is the special one. This is in agreement with the phase diagram shown in figure $\ref{fig:phase}$.}
\label{fig:gfactor}
\end{figure}

In \cite{AffleckLudwig}, it was argued that the $g$-factor (see section $\ref{subsec:gfactor}$) is always decreasing under renormalization from a less stable to a more stable boundary condition. The $g$-factor plays a similar role in BCFT to the one of the central charge in the bulk CFT \cite{cTheorem}. Again, this is true only when the theory is unitary, which is not the case in general for the $\On$ model. However, it turns out that this still makes sense here and that we can organize the boundary RG flows according to the boundary entropies of the different boundary conditions we have considered so far for the $\On$ model. The $g$-factors $(\ref{eq:gord})$, $(\ref{eq:gSp})$, $(\ref{eq:gas1})$, $(\ref{eq:gas2})$ are plotted in figure $\ref{fig:gfactor}$ for $n=1$. We see that the RG flows between the different b.c predicted by Affleck \& Ludwig's argument are in agreement with the phase diagram $\ref{fig:phase}$ in the region $0 < n_1 <n$. The most unstable b.c is the (isotropic) special one. The two anisotropic special b.c are more stable than the special one, but less than the ordinary b.c.
\paragraph{}
Note that outside the region $0<n_1<n$, the $g$-factors are organized differently, so the phase diagram $\ref{fig:phase}$ is probably wrong in this case. Therefore, in this article we restrict our discussion to the ``physical'' region $0<n_1<n$.

\subsection{Boundary perturbing operators : some guesswork}

We want to understand why the ordinary b.c is more stable than the special one when one perturbs the theory in the anisotropic direction. This can be done within the framework of perturbed BCFT. The idea is that, at an RG fixed point, the theory is described (formally) by an action $\mathcal{S}_{BCFT}$. One can perturb this theory by a boundary operator $\Phi_{b}$ with scaling dimension $h_{b}$
\begin{equation}
\mathcal{S}_{BCFT} + \lambda \int_{\rm{boundary}} dl \; \Phi_{b}
\end{equation}
the eigenvalue of the coupling $\lambda$ is then $y_{\lambda} = 1 - h_{b}$ under the RG flow. If one identifies the operator $\Phi_{b}$ properly, one can then say if the perturbation is relevant ($y_\lambda>0$) or not. 
\paragraph{}
In the case of an anisotropic perturbation around an isotropic b.c, the perturbing coupling is $\lambda \propto \Delta$ where $\Delta = w_{\blob}-w_{\unblob}$ is called the \textit{anisotropic coupling}. Now, what follows is very close to what we said about the fractal dimension of the contact set (see previous part) for special and ordinary b.c. Indeed, an operator perturbing in the anisotropic direction should create additional loops on the boundary. We already know what operator creates a piece of loop at a boundary point: it is the $2$-arms operator. In the case of ordinary b.c, it is $\Phi_{3,1}$, which is irrelevant when $0<n\leq 2$. But for special b.c, it is $\Phi_{3,3}$ and this is relevant, with eigenvalue $y_{\Delta}=1-2\frac{(g-1)^2}{g}$. We can reformulate this in terms of the fractal dimensions discussed above. For ordinary b.c, loops in the scaling limit almost surely never touch the boundary. Hence, a very small perturbation at the boundary could hardly affect their critical behaviour. At the special point, however, the loops can touch the boundary in a non-trivial way (on a contact set with a non-trivial scaling limit). For such a boundary condition, a small perturbation at the boundary can affect the critical behaviour in a more crucial way.

\paragraph{}
There is another perturbation of interest around the (isotropic) special point: the perturbation in the isotropic direction. The coupling in that direction is the weight of a boundary monomer $y$ (recall it is also the coupling between two boundary $\On$ spins, see relation $(\ref{eq:ZOnboundary})$).  In that case the perturbing operator has been known for a long time \cite{FendleySaleur2} to be $\Phi_{1,3}$, with eigenvalue $y_y=2 \frac{g-1}{g}$.

\subsection{Shape of the phase diagram around the isotropic special point}
In this section we explain why the phase diagram shown in figure $\ref{fig:phase}$ has a cusp-like shape around the special b.c. To do this, we apply finite-size scaling arguments \cite{CardyBook,Barber}. Consider the boundary free energy $f_{b}(y,\Delta)$ as a function of the isotropic coupling $y$ and the anisotropic coupling $\Delta$ around the special b.c $y=y_S$ and $\Delta=0$. The singular part of $f_b$ has the scaling form
\begin{equation}
f_{b,s}(y,\Delta) = \frac{1}{\xi_b} \Psi \left( \frac{y-y_S}{\xi_b^{y_y}}, \frac{\Delta}{\xi_b^{y_\Delta}} \right)
\end{equation}
where $\xi_b$ is the correlation length along the boundary (the bulk correlation length is infinite because $x=x_c$). Restricting to $y<y_S$, one can apply renormalization group transformations until we reach $(y_S-y) \xi_b^{-y_{y}} = 1$ and then
\begin{equation}
\label{eq:scaling}
f_{b,s}(y,\Delta) = \frac{1}{(y_S-y)^{1/y_y}} \tilde{\Psi} \left( \frac{\Delta}{(y_S-y)^{y_\Delta/y_y}}\right)
\end{equation}
There is a similar function for $y>y_S$ but we already know that in that region one goes from one extraordinary phase to another when $\Delta=0$ and that this is a first order transition (see section $\ref{part:model}$). In the case $y<y_S$ we should cross the anisotropic special transition line for some $\Delta^{+}(y)>0$ and $\Delta^{-}(y)<0$. At these two points $f_{b,s}$ must be singular, whereas it is non-singular in the rest of the domain. We have then $\Delta^{\pm}(y) = C^{\pm} (y_S-y)^{1/\phi}$, with $\phi=y_y/y_\Delta$. According to the previous section $\phi = \frac{1-h_{1,3}}{1-h_{3,3}}<1$ for $0<n \leq 2$, so the anisotropic special transition must have the cusp-like shape drawn in figure $\ref{fig:phase}$.

\begin{figure}[h]
\centering
\includegraphics[height=0.4\textheight]{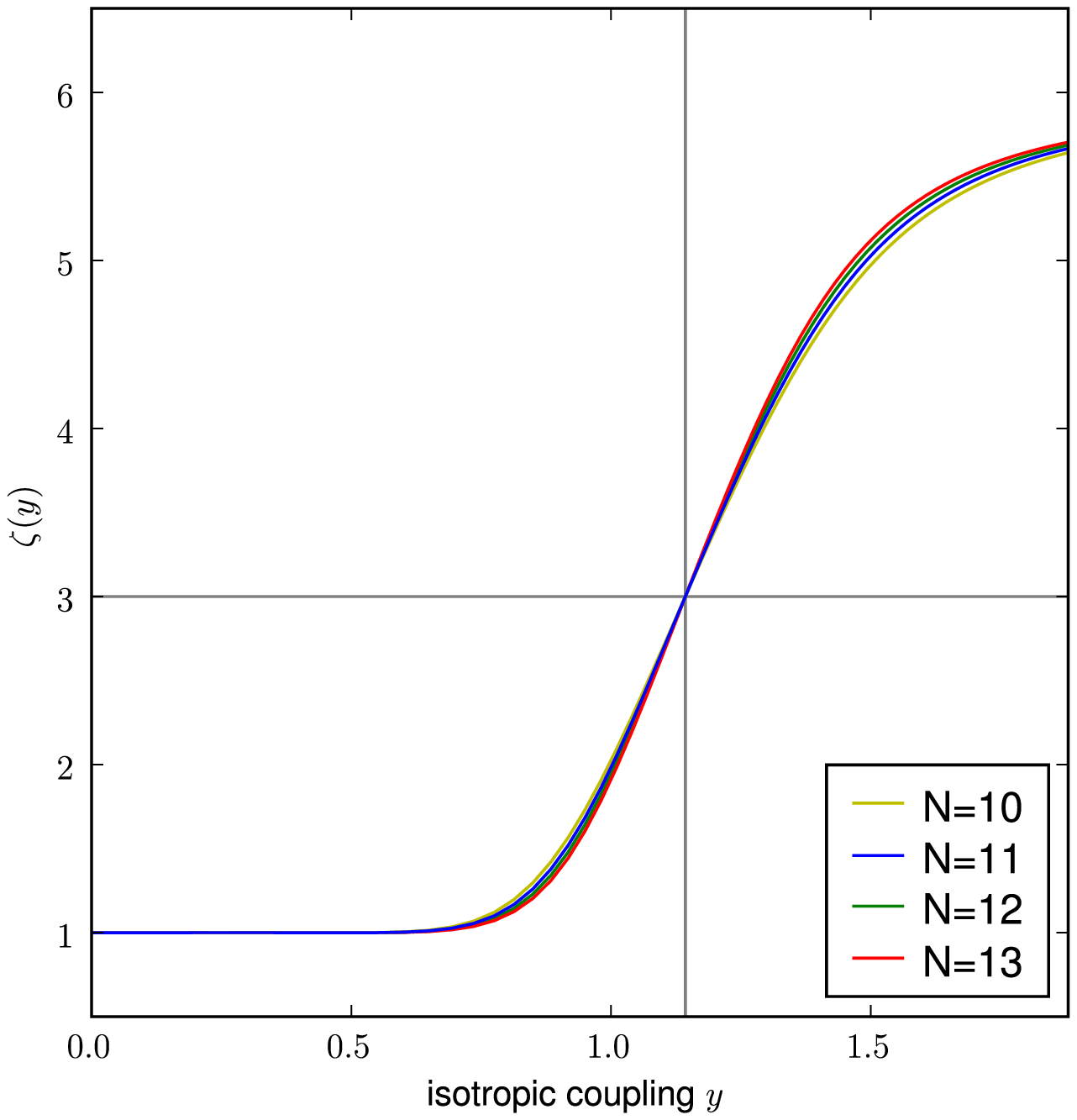} \quad \includegraphics[height=0.4\textheight]{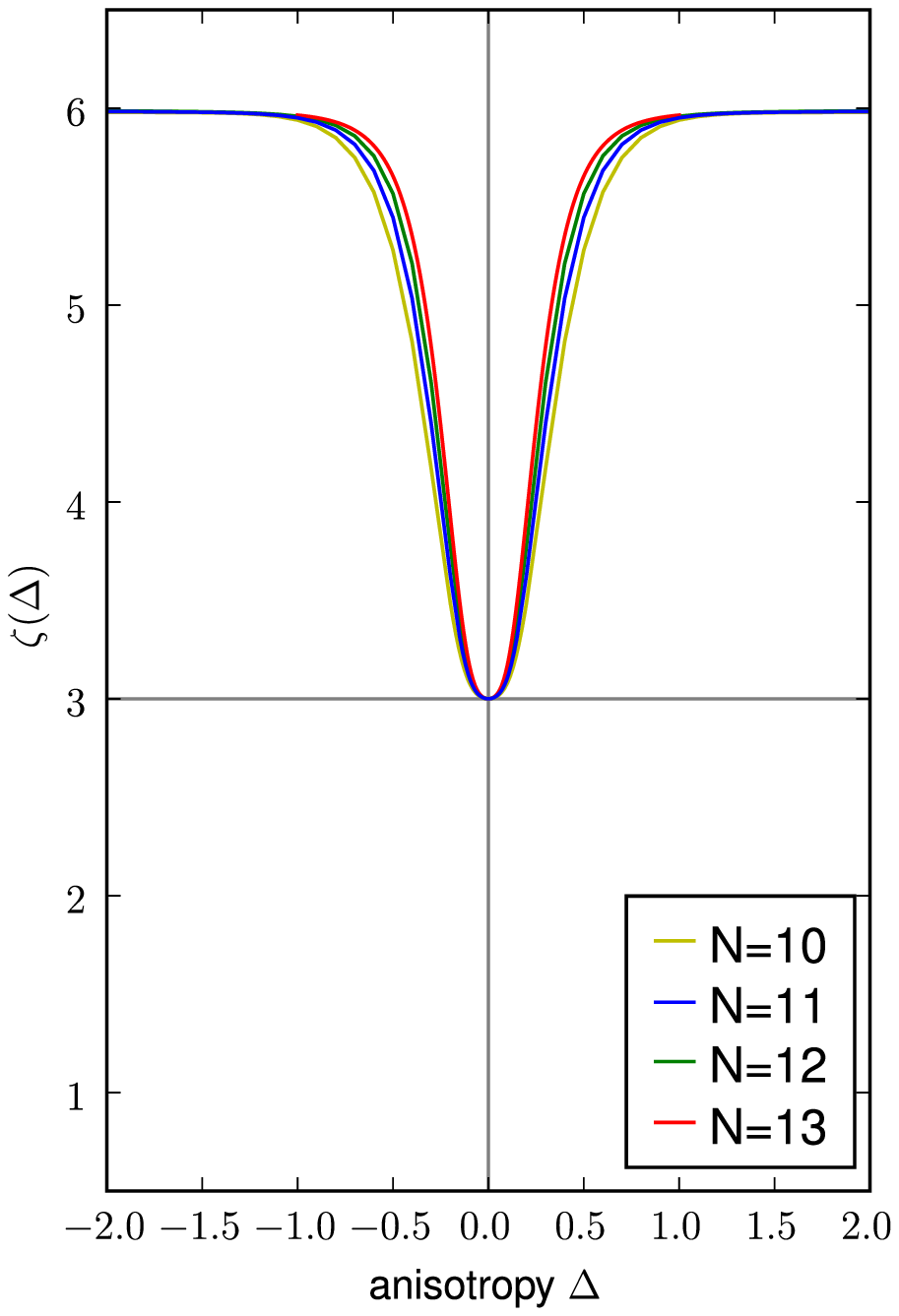}
\caption{The leading exponent $h_0$ in the sector without string is written $h_0=\left( \zeta^2-1\right) \frac{(g-1)^2 }{4 g}$. We plot $\zeta$ vs. $y$ and $\Delta$ around the special point. Here we have $n=\sqrt{2}$, so the central charge is $7/10$ and $g=5/4$. At the special point $h_0=h_{1,2}$ (so here $\zeta = \left| 5\times 1-4 \times 2 \right| =3$). Extraordinary b.c give $h_0=h_{1,1}$ so $\zeta=0$, and extraordinary b.c $h_0=h_{2,1}$ so $\zeta = \left| 5 \times 2 - 4 \times 1 \right|=6$. The results here are in good agreement with $\delta_{y}=0.49\pm0.02$ and $\delta_{\Delta}=1.1\pm0.05$.}
\label{fig:crossing}
\end{figure}

\paragraph{}
We have checked the value of the exponent $\phi$ numerically by transfer-matrix diagonalization. We compute the ground state of the transfer matrix on a strip of width $N$. We compute the free energy per site $f_N$, which should behave as $(\ref{eq:numerics})$. Then we extract the effective central charge $c_{\rm{eff}}= c-24 h_0$ of the theory. Once again we introduce the quantity $\zeta$ related to the ground-state exponent $h_0$ by $h_0=(\zeta^2-1)\frac{(g-1)^2}{4g}$. We expect a scaling behaviour of $\zeta$ of the form (see figure $\ref{fig:crossing}$)
\begin{equation}
\zeta (N,y,\Delta) = \Theta \left(\frac{y-y_S}{N^{\delta_y}}, \frac{\Delta}{N^{\delta_\Delta}} \right).
\end{equation}

It must be possible to relate the exponents $\delta_y$ and $\delta_\Delta$ to the usual exponents of the boundary correlation lengths (parallel and orthogonal), and to $y_y$ and $y_\Delta$. However, because of the mixture of boundary and finite-size effects \cite{Binder,Barber}, this would be complicated. Fortunately we do not have to go through this, because if we simply take the limit $N \rightarrow \infty$, for $y<y_S$ we arrive at the following scaling form
\begin{equation}
\zeta_\infty (y,\Delta) = \tilde{\Theta}\left(\frac{\Delta}{(y_S-y)^{y_\Delta/y_y}} \right).
\end{equation}
where $\tilde{\Theta}$ must now be a piecewise constant function. Then we see that $\phi = \frac{\delta_y}{\delta_\Delta}$, so even if we are not able to relate $y_y$ to $\delta_y$ and $y_\Delta$ to $\delta_\Delta$ independently, we can still compute there ratio $\phi$ numerically. We find
\begin{center}
\begin{tabular}{c|c|c|c|c|c}
c & g & $\delta_{y}$ & $\delta_{\Delta}$ & $\phi=\frac{\delta_{y}}{\delta_{\Delta}}$ & $\phi$ (theor.) \\ 
\hline $0$ & $3/2$ & $0.87\pm0.05$ & $1.00 \pm0.05$ &   $0.87\pm0.1$    & $1$ \\
$1/2$ & $4/3$ & $0.63\pm0.03$ & $1.07 \pm0.05$ &    $0.59 \pm 0.08$   & $3/5=0.6$ \\
$7/10$ & $5/4$ & $0.49\pm0.02$ & $1.10 \pm0.05$ &    $0.45 \pm 0.07$   & $4/9=0.444$ \\
$8/10$ & $6/5$ & $0.41\pm0.02$ & $1.25 \pm0.05$ &    $0.33\pm 0.07$   & $5/14=0.357$
\end{tabular}
\end{center}
The results are in good agreement with the relation $\phi=\frac{1-h_{1,3}}{1-h_{3,3}}$. Note however that the limits $n \rightarrow 0$ and $n \rightarrow 2$ give results which are less accurate than for $n=1$ or $n=\sqrt{2}$.

\subsection{Integrable flows and TBA}
\label{sec:TBA}

So far, we have not discussed the perturbation around a point on the anisotropic special line. Note that the whole line is attracted by one of the two points $AS_{\blob}$ or $AS_{\unblob}$, so it is sufficient to study only the transition at one of these points. In particular, the crossing behaviour along the whole line must be described by the perturbation around $AS_{\blob}$ or $AS_{\unblob}$ in the unstable direction. This is what we study in this section, by means of the Thermodynamic Bethe Ansatz (TBA).

The TBA system for the $O(n)$ model in the bulk and in the rational case $n=2 \cos\left( \frac{\pi}{m} \right)$ was derived in \cite{FendleySaleur1}. In the massless case of interest here, it becomes
\begin{equation}
\label{eq:TBA}
\epsilon_j(\theta) = \delta_{j 1} e^{-\theta} - \int \frac{d \theta'}{ 2\pi} \frac{1}{\cosh (\theta-\theta')} \left( \log (1+e^{-\epsilon_{j-1}(\theta')} + \log (1+e^{-\epsilon_{j+1}(\theta')})  \right)
\end{equation}
where $j=1,\dots,m-2$ and $\epsilon_{0}=\epsilon_{m-1}=\infty$.

\paragraph{}
In \cite{Fendley} Fendley derived the effect of the introduction of a boundary $S$-matrix of arbitrary spin. It turns out to add a term in the free energy of the system
\begin{equation}
\label{eq:impurity}
f_{\rm{boundary}} = -T \int \frac{d \theta'}{2 \pi} \frac{1}{\cosh(\theta - \log(T/T_K))} \log \left( 1+e^{-\epsilon_{2S}(\theta)} \right)
\end{equation}
where, in the language of the Kondo problem, $S$ is the spin of the magnetic impurity and $T_K$ is the Kondo temperature.

\paragraph{}
We analyze this TBA system ($\ref{eq:TBA}$)-($\ref{eq:impurity}$) following \cite{FendleySaleur2}. First, let us focus on the UV limit $T \gg T_K$. The leading contribution to the boundary free energy $(\ref{eq:impurity})$ comes from the region $\theta$ close to infinity. Introducing $x_j = e^{-\epsilon_j(\infty)}$ and taking the limit $\theta \rightarrow \infty$ in $(\ref{eq:TBA})$, one gets 
\begin{equation}
x_j^2 = (1+x_{j-1})(1+x_{j+1})
\end{equation}
for $j=1, \dots,m-2$, $x_0 = x_{m-1}=0$. This yields
\begin{equation}
\label{eq:xj}
1+x_j = \left( \frac{\sin \pi (j+1)/(m+1)}{\sin \pi / (m+1)} \right)^2.
\end{equation}
The leading contribution to the boundary entropy is then
\begin{equation}
\label{eq:fUV}
\frac{f^{(UV)}_{\rm{boundary}}}{T} = - \frac{1}{2} \log (1+x_{2S}) + \mathcal{O}\left( \frac{T_K}{T} \right).
\end{equation}

\paragraph{}
Now, let us turn to the IR limit $T \ll T_K$. In that case the leading contribution to $(\ref{eq:impurity})$ comes from the region $\theta \rightarrow -\infty$. Introducing $y_i = e^{- \epsilon_j(-\infty)}$, the $\theta \rightarrow -\infty$ limit of system $(\ref{eq:TBA})$ gives
\begin{equation}
y_1 = 0 \qquad y_j^2=(1+y_{j-1})(1+y_{j+1})
\end{equation}
with $j=2,\dots,m-2$, $y_{m-1}=0$. This is solved by
\begin{equation}
\label{eq:yj}
1+y_j = \left( \frac{\sin \pi j/m}{\sin \pi / m} \right)^2.
\end{equation}
Now the leading contribution to the boundary entropy is
\begin{equation}
\label{eq:fIR}
\frac{f^{(IR)}_{\rm{boundary}}}{T} = - \frac{1}{2} \log (1+y_{2S}) + \mathcal{O}\left( \frac{T}{T_K} \right).
\end{equation}

\paragraph{}
In this TBA framework, the $g$-factor is related to the boundary free energy by $g=\exp\left( -f_{\rm{boundary}}/T \right)$ times some overall constant which is independent of the scale $T/T_K$. Thus the TBA system $(\ref{eq:TBA})$-$(\ref{eq:impurity})$ leads to
\begin{equation}
\frac{g^{UV}}{g^{IR}} = \frac{(1+x_{2S})^{1/2}}{(1+y_{2S})^{1/2}}=\frac{\sin \pi (2S+1)/(m+1)  \sin \pi/m}{\sin \pi/(m+1) \sin \pi (2S)/m}.
\end{equation}
Comparing this to $(\ref{eq:gord})$ and $(\ref{eq:gas1})$, we see that with $r_1 = 2S$ one has
\begin{equation}
\frac{g^{UV}}{g^{IR}} = \frac{g_{AS_{\blob}}}{g_{Ord}}
\end{equation}
so the TBA system above is describing the flow from the anisotropic special b.c with $r_1=2S$ to the ordinary b.c. The case $r_1=1$ is actually the case of the flow from the (isotropic) special b.c to the ordinary one, and this was studied in \cite{FendleySaleur2}. One can also see here that there is no chance that we get in the end the flow from (isotropic) special to anisotropic special b.c, so this flow is probably not integrable.

\paragraph{}
The TBA system exposed here allows us to compute the dimension of the perturbing operator around the anisotropic special b.c. Again, this can be done following \cite{FendleySaleur2}. Consider the UV limit $T \gg T_K$. Then the TBA system implies the periodicity of the pseudo-energies $\epsilon_j(\theta+(m+1)i \pi) = \epsilon_j(\theta)$, so close to $\theta=\infty$ we can expand
\begin{equation}
\log (1+e^{-\epsilon_j(\theta)}) = \sum_{k=0}^\infty L_j^{(k)} \left( e^{-2\theta/(m+1)} \right)^k
\end{equation}
Plugging this into $(\ref{eq:TBA})$, one can show that the term $k=1$ is zero. Then in the UV limit the boundary free energy can be expanded as
\begin{equation}
\frac{f^{(UV)}_{\rm{boundary}}}{T} = - \frac{1}{2} \log(1+x_{2S}) + \sum_{k=2}^\infty \left( \frac{T_K}{T} \right)^{2k/(m+1)} f_k^{(UV)}
\end{equation}
so the scaling dimension of the energy operator at the anisotropic special transition is always $h_{1,3}=1-\frac{2}{m+1}$. Thus the operator perturbing in the unstable direction around the anisotropic special transition is $\Phi_{1,3}$.




\section{Open boundary conditions}
\label{sec:open}

In this section we focus on "open" boundary conditions for the $\On$ model. At first glance, these conditions are different from the ones we considered so far. However, it turns out that these conditions are related to the anisotropic special b.c in the particular case $n_1=1$. It is then possible to use the foregoing formalism to study these b.c in the BCFT framework. As an application, we derive a crossing formula for the Ising spins clusters on an annulus in section $\ref{sec:crossing}$.

\subsection{Cardy's open boundary conditions, boundary fields in the spin $\On$ model}

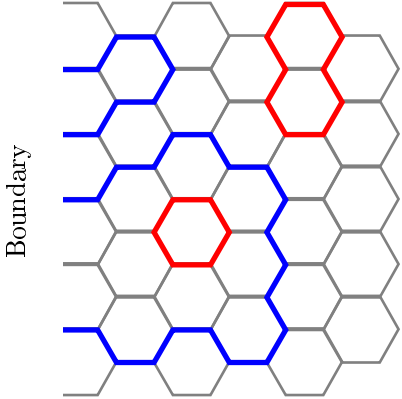
\begin{figure}[h]
\center
\begin{tikzpicture}[scale=0.75]
\begin{scope}
\clip (0.4,-1.8) rectangle (6.5,5.5);
\foreach \x in {0,2,4}
\foreach \y in {-2,0,2,4,6,8}
{
	\honey{\x}{\y}
}
\foreach \x in {1,3,5}
\foreach \y in {-1,1,3,5,7}
{
	\honey{\x}{\y}
}
\draw[line width=2pt, blue, scale=0.6667] (0,0) -- ++(0.5,0.866) -- ++(-0.5,0.866) -- ++(0.5,0.866) -- ++(1,0) -- ++(0.5,0.866) -- ++(1,0) -- ++(0.5,0.866) -- ++(1,0) -- ++(0.5,-0.866) -- ++(1,0) -- ++(0.5,-0.866) -- ++(-0.5,-0.866) -- ++(0.5,-0.866) -- ++(-0.5,-0.866) -- ++(0.5,-0.866) -- ++(-0.5,-0.866) -- ++(-1,0) -- ++(-0.5,0.866) -- ++(-1,0) -- ++(-0.5,-0.866) -- ++(-1,0) -- ++(-0.5,0.866) -- ++(-1,0) -- cycle;
\draw[line width=2pt, blue, scale=0.6667] (0,5.196) -- ++(0.5,0.866) -- ++(1,0) -- ++(0.5,0.866) -- ++(1,0) -- ++(0.5,-0.866) -- ++(-0.5,-0.866) -- ++(-1,0) -- ++(-0.5,-0.866) -- ++(-1,0) -- cycle;
\draw[line width=2pt, red, scale=0.6667] (4.5,2.598) ++(1.5,2.598) -- ++(0.5,0.866) -- ++(-0.5,0.866) -- ++(0.5,0.866) -- ++(1,0) -- ++(0.5,-0.866) -- ++(-0.5,-0.866) -- ++(0.5,-0.866) -- ++(-0.5,-0.866) -- ++(-1,0) -- cycle;
\draw[line width=2pt, red, scale=0.6667] (4.5,2.598) ++(-1.5,-0.866) -- ++(0.5,0.866) -- ++(1,0) -- ++(0.5,-0.866) -- ++(-0.5,-0.866) -- ++(-1,0) -- cycle;
\end{scope}
\draw (-0.4,1.7) node[rotate=90] {Boundary};
\end{tikzpicture}
\caption{Open boundary conditions in the $\On$ model. The closed loops in the bulk still have a fugacity $n$, but the half-loops have a fugacity $\nu_1$.}
\label{fig:openbc}
\end{figure}

In an unpublished note \cite{CardyNote}, Cardy introduced the b.c for the $\On$ model shown in figure $\ref{fig:openbc}$, which we call open b.c. The boundary of the honeycomb lattice is slightly different from the one we were looking at so far. This is important since it is now possible to have half-loops attached to the boundary. A half-loop gets a Boltzmann weight $\nu_1$.

In more physical terms, one can think about these open boundary conditions as a magnetic field $\vec{B}_1$ at the surface, which couples to the boundary spins.
\begin{equation}
Z=\mathrm{Tr} \left\{ \prod_{\mathrm{boundary} } \left( 1+ \vec{S}_i . \vec{B}_1 \right) \prod_{\mathrm{bulk}\;  <ij>} \left( 1 + x_c \vec{S}_i . \vec{S}_j \right) \right\}
\end{equation}
where $\vec{B}_1$ is a boundary field. Proceeding as usual, we expand this expression and take the trace over independent terms to get the familiar loop expansion, with half-loops ending on the left and right boundaries. These loops get weights
\begin{equation}
\nu_1 = \left\| \vec{B}_1 \right\|^2.
\end{equation}

\subsection{Relation with the anisotropic special b.c when $n_1=1$}
\label{sec:openbc}
Open boundary conditions correspond to a surface magnetic field coupling to the boundary spins. We expect such a boundary condition to flow towards fixed boundary conditions in the $\On$ model as soon as $\left\| \vec{B}_1 \right\| > 0$. This leads to the simple result that, in the scaling limit, the b.c condition should not depend on the precise value of $\nu_1$ as soon as $\nu_1 >0$. It is easy to check that fix b.c for the $\On$ model are exactly the open one\footnote{Although it is clear from this argument that the open b.c is simply the fixed b.c in the scaling limit of the $\On$ model, we keep calling it "open" for the consistency of our terminology.} with $\nu_1=1$. Thus the whole critical behaviour for $\nu_1 >0$ is completely caught by the theory at $\nu_1=1$.

\begin{figure}[h]
\center
\begin{tikzpicture}[scale=0.75]
\draw (-6,-2) node {$a.$};
\begin{scope}[xshift=-5cm]
\foreach \x in {0,2,4}
\foreach \y in {-2,0,2,4,6,8}
{
	\honey{\x}{\y}
}
\foreach \x in {1,3,5}
\foreach \y in {-1,1,3,5,7}
{
	\honey{\x}{\y}
}
\draw[line width=2pt, blue, scale=0.6667] (0,0) -- ++(0.5,0.866) -- ++(-0.5,0.866) -- ++(0.5,0.866) -- ++(1,0) -- ++(0.5,0.866) -- ++(1,0) -- ++(0.5,0.866) -- ++(1,0) -- ++(0.5,-0.866) -- ++(1,0) -- ++(0.5,-0.866) -- ++(-0.5,-0.866) -- ++(0.5,-0.866) -- ++(-0.5,-0.866) -- ++(0.5,-0.866) -- ++(-0.5,-0.866) -- ++(-1,0) -- ++(-0.5,0.866) -- ++(-1,0) -- ++(-0.5,-0.866) -- ++(-1,0) -- ++(-0.5,0.866) -- ++(-1,0) -- cycle;
\draw[line width=2pt, blue, scale=0.6667] (0,5.196) -- ++(0.5,0.866) -- ++(1,0) -- ++(0.5,0.866) -- ++(1,0) -- ++(0.5,-0.866) -- ++(-0.5,-0.866) -- ++(-1,0) -- ++(-0.5,-0.866) -- ++(-1,0) -- cycle;
\draw[line width=2pt, red, scale=0.6667] (4.5,2.598) ++(1.5,2.598) -- ++(0.5,0.866) -- ++(-0.5,0.866) -- ++(0.5,0.866) -- ++(1,0) -- ++(0.5,-0.866) -- ++(-0.5,-0.866) -- ++(0.5,-0.866) -- ++(-0.5,-0.866) -- ++(-1,0) -- cycle;
\draw[line width=2pt, red, scale=0.6667] (4.5,2.598) ++(-1.5,-0.866) -- ++(0.5,0.866) -- ++(1,0) -- ++(0.5,-0.866) -- ++(-0.5,-0.866) -- ++(-1,0) -- cycle;
\filldraw[blue, scale=0.6667] (0,0) circle (0.45);
\filldraw[blue, scale=0.6667] (0,1.732) circle (0.45);
\filldraw[blue, scale=0.6667] (0,5.196) circle (0.45);
\draw (-0.7,1.7) node[rotate=90] {Boundary};
\end{scope}
\draw[->, very thick] (2,2) -- (4,2);  
\draw (4,-2) node {$b.$};
\begin{scope}[xshift=5cm]
\clip (0.4,-1.8) rectangle (6.5,5.5);
\foreach \x in {0,2,4}
\foreach \y in {-2,0,2,4,6,8}
{
	\honey{\x}{\y}
}
\foreach \x in {1,3,5}
\foreach \y in {-1,1,3,5,7}
{
	\honey{\x}{\y}
}
\draw[line width=2pt, blue, scale=0.6667] (0,0) -- ++(0.5,0.866) -- ++(-0.5,0.866) -- ++(0.5,0.866) -- ++(1,0) -- ++(0.5,0.866) -- ++(1,0) -- ++(0.5,0.866) -- ++(1,0) -- ++(0.5,-0.866) -- ++(1,0) -- ++(0.5,-0.866) -- ++(-0.5,-0.866) -- ++(0.5,-0.866) -- ++(-0.5,-0.866) -- ++(0.5,-0.866) -- ++(-0.5,-0.866) -- ++(-1,0) -- ++(-0.5,0.866) -- ++(-1,0) -- ++(-0.5,-0.866) -- ++(-1,0) -- ++(-0.5,0.866) -- ++(-1,0) -- cycle;
\draw[line width=2pt, blue, scale=0.6667] (0,5.196) -- ++(0.5,0.866) -- ++(1,0) -- ++(0.5,0.866) -- ++(1,0) -- ++(0.5,-0.866) -- ++(-0.5,-0.866) -- ++(-1,0) -- ++(-0.5,-0.866) -- ++(-1,0) -- cycle;
\draw[line width=2pt, red, scale=0.6667] (4.5,2.598) ++(1.5,2.598) -- ++(0.5,0.866) -- ++(-0.5,0.866) -- ++(0.5,0.866) -- ++(1,0) -- ++(0.5,-0.866) -- ++(-0.5,-0.866) -- ++(0.5,-0.866) -- ++(-0.5,-0.866) -- ++(-1,0) -- cycle;
\draw[line width=2pt, red, scale=0.6667] (4.5,2.598) ++(-1.5,-0.866) -- ++(0.5,0.866) -- ++(1,0) -- ++(0.5,-0.866) -- ++(-0.5,-0.866) -- ++(-1,0) -- cycle;
\filldraw[blue, scale=0.6667] (0,0) circle (0.45);
\filldraw[blue, scale=0.6667] (0,1.732) circle (0.45);
\filldraw[blue, scale=0.6667] (0,5.196) ++(-0.4,-0.4) rectangle ++(0.8,0.8);
\end{scope}
\draw (4.9,1.7) node[rotate=90] {Boundary};
\end{tikzpicture}
\caption{In the special case $n_1=1$, there are no unblobbed loops and the weight of the blobs compensates exactly the weight of the boundary monomers. A loop configuration on the lattice $(a)$ is then equivalent to a configuration on the lattice $(b)$ with open b.c when $\nu_1=1$. Since all the open boundary conditions with $\nu_1 >0$ are expected to renormalize to the same b.c, they must all be described by the theory with $n_1=1$.}
\label{fig:openas}
\end{figure}
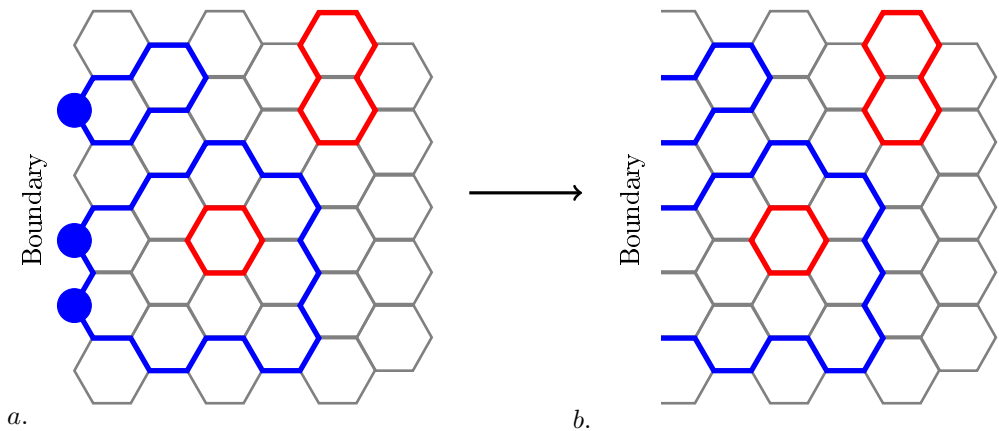

Now consider the anisotropic special b.c $AS_{\blob}$ in the case $n_1=1$. Note that the formulae $(\ref{eq:conjectureAS})$ simplify in that case to $w_{\blob}=1/x_c^2$ and $w_{\unblob}=0$. There are no unblobbed loops because $w_{\unblob}=0$. A loop touching several times the boundary gets a weight $w_{\blob} x_c^2=1$ per pair of boundary monomers (see figure $\ref{fig:openas}$). Then we can just erase the first row of spins of the honeycomb lattice, and we end up with the lattice which gives rise to the open b.c, here with $\nu_1=1$. Because of the above remark, this describes the open b.c for every $\nu_1>0$.

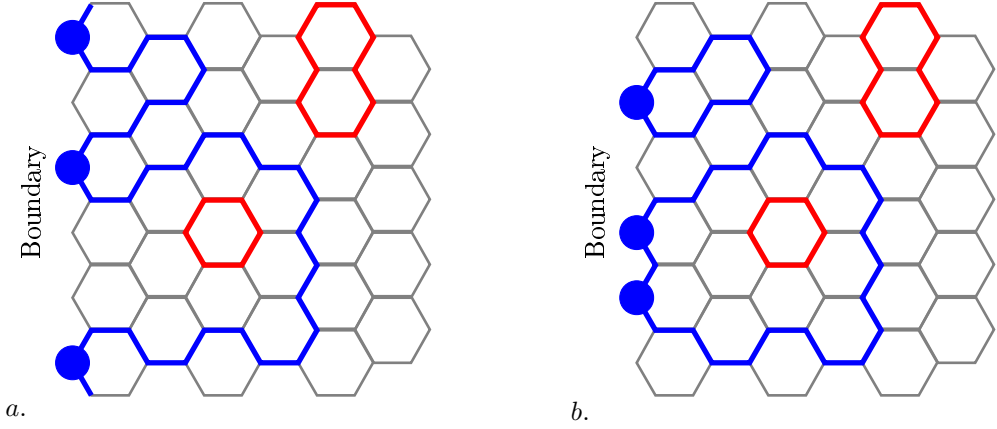
\begin{figure}[h]
\center
\begin{tikzpicture}[scale=0.75]
\draw (-6,-2) node {$a.$};
\begin{scope}[xshift=-5cm]
\foreach \x in {0,2,4}
\foreach \y in {-2,0,2,4,6,8}
{
	\honey{\x}{\y}
}
\foreach \x in {1,3,5}
\foreach \y in {-1,1,3,5,7}
{
	\honey{\x}{\y}
}
\draw[line width=2pt, blue, scale=0.6667] (0,0) ++(0.5,-2.598) -- ++(-0.5,0.866) -- ++(0.5,0.866) -- ++(1,0) -- ++(0.5,-0.866) -- ++(1,0) -- ++(0.5,0.866) -- ++(1,0) -- ++(0.5,-0.866) -- ++(1,0) -- ++(0.5,0.866) -- ++(-0.5,0.866) -- ++(0.5,0.866) -- ++(-0.5,0.866) -- ++(0.5,0.866) -- ++(-0.5,0.866) -- ++(-1,0) -- ++(-0.5,0.866) -- ++(-1,0) -- ++(-0.5,-0.866) -- ++(-1,0) -- ++(-0.5,-0.866) -- ++(-1,0) -- ++(-0.5,0.866) -- ++(0.5,0.866) -- ++(1,0) -- ++(0.5,0.866) -- ++(1,0) -- ++(0.5,0.866) -- ++(-0.5,0.866) -- ++(-1,0) -- ++(-0.5,-0.866) -- ++(-1,0) -- ++(-0.5,0.866) -- ++(0.5,0.866);
\draw[line width=2pt, red, scale=0.6667] (4.5,2.598) ++(1.5,2.598) -- ++(0.5,0.866) -- ++(-0.5,0.866) -- ++(0.5,0.866) -- ++(1,0) -- ++(0.5,-0.866) -- ++(-0.5,-0.866) -- ++(0.5,-0.866) -- ++(-0.5,-0.866) -- ++(-1,0) -- cycle;
\draw[line width=2pt, red, scale=0.6667] (4.5,2.598) ++(-1.5,-0.866) -- ++(0.5,0.866) -- ++(1,0) -- ++(0.5,-0.866) -- ++(-0.5,-0.866) -- ++(-1,0) -- cycle;
\filldraw[blue, scale=0.6667] (0,-1.732) circle (0.45);
\filldraw[blue, scale=0.6667] (0,3.464) circle (0.45);
\filldraw[blue, scale=0.6667] (0,6.928) circle (0.45);
\draw (-0.7,1.7) node[rotate=90] {Boundary};
\end{scope}
\draw (4,-2) node {$b.$};
\begin{scope}[xshift=5cm]
\foreach \x in {0,2,4}
\foreach \y in {-2,0,2,4,6,8}
{
	\honey{\x}{\y}
}
\foreach \x in {1,3,5}
\foreach \y in {-1,1,3,5,7}
{
	\honey{\x}{\y}
}
\draw[line width=2pt, blue, scale=0.6667] (0,0) -- ++(0.5,0.866) -- ++(-0.5,0.866) -- ++(0.5,0.866) -- ++(1,0) -- ++(0.5,0.866) -- ++(1,0) -- ++(0.5,0.866) -- ++(1,0) -- ++(0.5,-0.866) -- ++(1,0) -- ++(0.5,-0.866) -- ++(-0.5,-0.866) -- ++(0.5,-0.866) -- ++(-0.5,-0.866) -- ++(0.5,-0.866) -- ++(-0.5,-0.866) -- ++(-1,0) -- ++(-0.5,0.866) -- ++(-1,0) -- ++(-0.5,-0.866) -- ++(-1,0) -- ++(-0.5,0.866) -- ++(-1,0) -- cycle;
\draw[line width=2pt, blue, scale=0.6667] (0,5.196) -- ++(0.5,0.866) -- ++(1,0) -- ++(0.5,0.866) -- ++(1,0) -- ++(0.5,-0.866) -- ++(-0.5,-0.866) -- ++(-1,0) -- ++(-0.5,-0.866) -- ++(-1,0) -- cycle;
\draw[line width=2pt, red, scale=0.6667] (4.5,2.598) ++(1.5,2.598) -- ++(0.5,0.866) -- ++(-0.5,0.866) -- ++(0.5,0.866) -- ++(1,0) -- ++(0.5,-0.866) -- ++(-0.5,-0.866) -- ++(0.5,-0.866) -- ++(-0.5,-0.866) -- ++(-1,0) -- cycle;
\draw[line width=2pt, red, scale=0.6667] (4.5,2.598) ++(-1.5,-0.866) -- ++(0.5,0.866) -- ++(1,0) -- ++(0.5,-0.866) -- ++(-0.5,-0.866) -- ++(-1,0) -- cycle;
\filldraw[blue, scale=0.6667] (0,0) circle (0.45);
\filldraw[blue, scale=0.6667] (0,1.732) circle (0.45);
\filldraw[blue, scale=0.6667] (0,5.196) circle (0.45);
\draw (-0.7,1.7) node[rotate=90] {Boundary};
\end{scope}
\end{tikzpicture}
\caption{The two configurations for anisotropic special b.c with $n_1=1$ leading to the open b.c configuration shown in figure $\ref{fig:openas}.(b)$. On the annulus (top and bottom of the lattice identified), one contains a blobbed loop $(a)$ winding around the annulus, the other does not $(b)$.}
\label{fig:equivconf}
\end{figure}

It is easy to apply the results of section $\ref{partBCFT}$ to derive new results about the BCFT of the open b.c. The constraint $n_1=1$ fixes $r_1 = \frac{\pi / \gamma -1}{2}$. Then the B.C.C operator going from ordinary b.c to open b.c is $\Phi_{r_1,r_1+1}$, which means that its scaling dimension is
\begin{equation}
h_{\frac{\pi / \gamma -1}{2}, \frac{\pi / \gamma -1}{2}+1} = h_{\frac{\pi / \gamma + 1}{2}, \frac{\pi / \gamma +1}{2}}
\end{equation}
where we used the symmetry of Kac' formula $(\ref{eq:kac})$. This operator, which should be viewed as the B.C.C operator going from Dirichlet to Neumann b.c, was obtained previously by Kostov \textit{et al.} in the context of two-dimensional gravity \cite{KostovSerbanPonsot}. If ones adds strings (say $L$ strings), one has to choose if the leftmost string is blobbed or not. Clearly, if it is blobbed, it gives rise to half-loops when we go to open b.c (see figure $\ref{fig:openas}$), so this case should be the same as the one of $L-1$ strings when the leftmost one is unblobbed. This is in agreement with
\begin{equation}
K_L^{\blob}(q) = K_{L-1}^{\unblob}
\end{equation}
in the case $r_1 = \frac{\pi / \gamma -1}{2}$ (see formula $(\ref{eq:Kas1ord})$). If one wishes to write down a partition function on an annulus with (for example) open/ordinary b.c on each side, one should proceed as follows. First, one should note that to a unique configuration of the loops with open b.c correspond two configurations with $AS_{\blob}$ b.c, as shown in figure $\ref{fig:equivconf}$. One of these configurations contains a blobbed loop which winds around the annulus, the other does not. Then, to avoid double counting, we must count the blobbed loops which wind around the annulus with a fugacity $0$.

This can be done easily, if we recall that the partition function has the Markov trace structure $(\ref{eq:Markov})$ in the case with blobbed loops. The only thing one has to do is to modify the coefficients $\D^{\blob}_L$ and $\D^{\unblob}_L$ in formula $(\ref{eq:Zas1ord})$. One has $\tilde{\D}_0 = 1$, $\tilde{\D}_1^{\blob}=0$, $\tilde{\D}_1^{\unblob}=\tilde{\D}_1=n$ and the relation $(\ref{eq:fusionDblob})$ leads to $\tilde{\D}^{\blob}_L= \frac{\sin(1-L) \gamma}{\sin \gamma}$, $\tilde{\D}_L^{\unblob}= \frac{\sin (1+L)\gamma}{\sin \gamma}$. This relation could also be obtained by specializing the results of \cite{JS1,JS2}. We end up with the partition function
\begin{equation}
\label{eq:Zopenord}
Z_{Open/Ord} = \frac{q^{-c/24}}{P(q)} \sum_{L \in \Z} \frac{\sin (1-L) \gamma}{\sin \gamma} q^{h_{r_1+L,r_1+1}}
\end{equation}
where $r_1 = \frac{\pi/\gamma-1}{2}$. The modular transform of $(\ref{eq:Zopenord})$ is computed once again by performing a Poisson resummation, which yields
\begin{equation}
Z_{Open/Ord} = \sqrt{\frac{2}{g}} \frac{\q^{-c/12}}{P(\q^2)} \sum_{p \in \Z} \frac{\sin \left( \gamma/2 - \pi p \right)}{\sin \gamma} \q^{2 \frac{1}{4 g} \left( 2p+ \frac{\gamma}{\pi} \right)^2}
\end{equation}
The identification of the $g$-factor of the open b.c $g_{Open}$ and the associated boundary entropy $S_{Open}=- \log g_{Open}$ is straightforward
\begin{equation}
g_{Open}= \left( \frac{2}{g} \right)^{1/4} \frac{\sin \left( \gamma/2 \right)}{\left( \sin \gamma  \sin (\gamma/g) \right)^{1/2}}.
\end{equation}

\subsection{Open/open boundary conditions on the strip or the annulus}

The weight of the half-loops $\nu_1$ turns out to be unimportant because in the scaling limit the b.c renormalizes to fixed b.c (\textit{ie} $\nu_1=1$) as soon as $\nu_1>0$. In the case of two boundaries (\textit{eg} on a strip or on the annulus), however there is a non-trivial scaling behaviour of the model. Let us consider the partition function
\begin{equation}
Z=\mathrm{Tr} \left\{ \prod_{\mathrm{boundary \; 1} } \left( 1+ \vec{S}_i . \vec{B}_1 \right) \prod_{\mathrm{boundary \; 2} } \left( 1+ \vec{S}_i . \vec{B}_2 \right) \prod_{\mathrm{bulk}\;  <ij>} \left( 1 + x_c \vec{S}_i . \vec{S}_j \right) \right\}
\end{equation}
where $\vec{B}_1$ and $\vec{B}_2$ are two boundary fields. Expanding this expression to get the loop expansion, one gets a model with half-loops attached on each boundary. A half-loop attached on the boundary $1$ gets a weight $\nu_1$, a half-loop attached on the boundary $2$ gets a weight $\nu_2$, and a half-loop with one end on each boundary has a weight $\nu_{12}$. These parameters are related to the fields $\vec{B}_1$ and $\vec{B}_2$ by 
\begin{equation}
\nu_1 = \left\| \vec{B}_1 \right\|^2 \qquad \nu_2 = \left\| \vec{B}_2 \right\|^2 \qquad \nu_{12} = \vec{B}_1 . \vec{B}_2
\end{equation}
Again, on each boundary the b.c should flow towards fixed b.c as soon as $\left\| \vec{B}_1 \right\| >0$ and $\left\| \vec{B}_2 \right\|>0$. The scaling limit is thus independent of the field strengths. However, it should still depend on the angle between the two fields $\frac{\vec{B}_1 . \vec{B}_2}{ \left\| \vec{B}_1 \right\| \left\| \vec{B}_2 \right\|} = \frac{\nu_{12}}{\sqrt{\nu_1 \nu_2}}$. We can then proceed as in the case with only one boundary, by completing the lattice to get a configuration of the $\On$ model with $AS_{\blob} / AS_{\blobw}$ b.c with $n_1=n_2=1$. The relation between these boundary conditions and the open/open case is exact only if $\nu_1=\nu_2=1$. In that case, a pair of half-loops attached on both boundaries on the open/open lattice gives a doubly blobbed loop with weight $n_{12}=\nu_{12}^2$. When $\nu_1 \neq 1$ or $\nu_2 \neq 1$ there is no exact mapping in the discrete setting between the configurations. However, because the only non-trivial parameter remaining in the scaling limit is the angle between the fields in the open/open case and the loop weight $n_{12}$ in the $AS_{\blob}/AS_{\blobw}$ case, the two models have to be equivalent in the scaling limit if
\begin{equation}
n_{12} = \frac{\nu_{12}^2}{\nu_{1} \nu_2}.
\end{equation}
In particular, this has the direct consequence that in the sector without strings, the Virasoro character\footnote{Note that in the calculation which led to $K_0(q)$ (see section $\ref{sec:Coulomb}$) the number of half-loops with weight $\nu_{12}$ is always even, because they correspond pairwise to a loop with weight $n_{12}$. This calculation has to be modified if one is also interested in the case of odd number of half-loops.} for $open/open$ b.c is ($\ref{eq:K0}$) where the parameters $r_{12}$ and $\nu_{12}$ are related by
\begin{equation}
\label{eq:cardy}
\frac{\nu_{12}}{\sqrt{\nu_1 \nu_2}} \sqrt{n+2} = 2 \cos \left( \frac{r_{12} \gamma}{2} \right)
\end{equation}
which is a consequence of ($\ref{eq:n12}$) with $r_1=r_2=\frac{\pi/\gamma-1}{2}$. The relation $(\ref{eq:cardy})$ was known by Cardy \cite{CardyNote}, who derived it by the following Coulomb gas argument\footnote{We thank J. Cardy for the permission to include this independent argument \cite{CardyNote}, which generalizes his results for percolation \cite{CardyPercolationAnnulus}.}. Recall that in the bulk the oriented loops get weights $e^{\pm i \gamma}$ depending on their orientation. On the open lattice, one counts the boundary monomers with weights $\alpha$, $\beta$, $\delta$, $\mu$ (see figure $\ref{fig:cardyparam}$) and one parameterizes
\begin{equation}
\begin{array}{cc}
\alpha = \rho_1 e^{i \theta}  &  \beta = \rho_1 e^{-i \theta} \\
\delta = \rho_2 e^{i \theta}  &  \mu = \rho_2 e^{-i \theta} \\
\end{array}
\end{equation}
then the unoriented loops get the weights $\nu_1$, $\nu_2$ and $\nu_{12}$ if
\begin{subequations}
\begin{eqnarray}
\nu_1 &=& \alpha \beta \left( e^{i \gamma/2} + e^{-i \gamma/2} \right) \; = \; 2 \rho_1^2 \cos \left(\gamma/2\right) \\
\nu_2 &=& \delta \mu \left( e^{i \gamma/2} + e^{-i \gamma/2} \right) \; = \; 2 \rho_2^2 \cos \left(\gamma/2\right) \\
\nu_{12} &=& \alpha \delta + \beta \mu \; = \; 2 \rho_1 \rho_2  \cos \left(2 \theta \right).
\end{eqnarray}
\end{subequations}
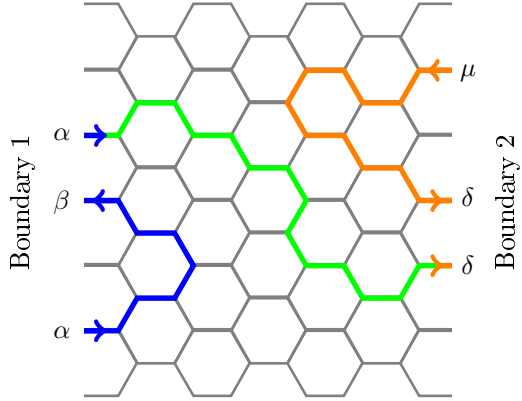
\begin{figure}[h]
\center
\begin{tikzpicture}[scale=0.75]
\begin{scope}
\clip (0.4,-1.8) rectangle (6.9,5.5);
\foreach \x in {0,2,4,6}
\foreach \y in {-2,0,2,4,6,8}
{
	\honey{\x}{\y}
}
\foreach \x in {1,3,5}
\foreach \y in {-1,1,3,5,7}
{
	\honey{\x}{\y}
}
\draw[line width=2pt, blue, scale=0.6667] (0,0) ++(0.5,-2.598) ++(0,0.866) ++(0,0.866) -- ++(1,0) -- ++(0.5,0.866) -- ++(1,0) -- ++(0.5,0.866) -- ++(-0.5,0.866) -- ++(-1,0) -- ++(-0.5,0.866) -- ++(-1,0);
\draw[line width=2pt, blue, scale=0.6667, ->] (0,0) ++(0.5,-2.598) ++(0,0.866) ++(0,0.866) -- ++(0.7,0);
\draw[line width=2pt, blue, scale=0.6667, ->] (0,0) ++(0.5,0) ++(0,0.866) ++(0,0.866) ++(1,0.866) -- ++(-0.7,0);
\draw[line width=2pt, green, scale=0.6667] (0,0) ++(0.5,2.598) ++(0,1.732) -- ++(1,0) -- ++(0.5,0.866) -- ++(1,0) -- ++(0.5,-0.866) -- ++(1,0) -- ++(0.5,-0.866) -- ++(1,0) -- ++(0.5,-0.866) -- ++(-0.5,-0.866) -- ++(0.5,-0.866) -- ++(1,0) -- ++(0.5,-0.866) -- ++(1,0) -- ++(0.5,0.866) -- ++(1,0);
\draw[line width=2pt, blue, scale=0.6667,->] (0,0) ++(0.5,2.598) ++(0,1.732) -- ++(0.7,0);
\draw[line width=2pt, orange, scale=0.6667,>-] (0,0) ++(9.8,0.866) -- ++(1,0);
\draw[line width=2pt, orange, scale=0.6667] (0,0) ++(10.5,0.866) ++(0,1.732) -- ++(-1,0) -- ++(-0.5,0.866) -- ++(-1,0) -- ++(-0.5,0.866) -- ++(-1,0) -- ++(-0.5,0.866) -- ++(0.5,0.866) -- ++(1,0) -- ++(0.5,-0.866) -- ++(1,0) -- ++(0.5,0.866) -- ++(1,0);
\draw[line width=2pt, orange, scale=0.6667, >-] (0,0) ++(9.8,0.866) ++(0,1.732) -- ++(1,0);
\draw[line width=2pt, orange, scale=0.6667, <-] (0,0) ++(9.7,0.866) ++(0,1.732) ++(0,3.464) -- ++(0.4,0);
\end{scope}
\draw (-0.7,1.7) node[rotate=90] {Boundary 1};
\draw (7.9,1.7) node[rotate=90] {Boundary 2};
\draw (0,-0.6) node {$\alpha$};
\draw (0,1.7) node {$\beta$};
\draw (0,2.9) node {$\alpha$};
\draw (7.2,1.8) node {$\delta$};
\draw (7.2,0.6) node {$\delta$};
\draw (7.2,4) node {$\mu$};
\end{tikzpicture}
\caption{Coulomb gas for the open/open b.c \cite{CardyNote} (which generalizes the arguments of \cite{CardyPercolationAnnulus}). As usual, the loops are oriented: in the bulk they get a weight $e^{\pm i \gamma}$ depending on their orientation. The boundary monomers get weight $\alpha$, $\beta$, $\delta$, $\mu$, depending on the orientation.}
\label{fig:cardyparam}
\end{figure}
Noting that $2 \cos \left( \gamma/2 \right)=\sqrt{n+2}$, one has $2 \cos \left( 2 \theta \right) = \frac{\nu_{12}}{\sqrt{\nu_1 \nu_2}} \sqrt{n+2}$, which is the same as relation $(\ref{eq:cardy})$ with $2 \theta = \frac{r_{12} \gamma}{2}$. Now, as in section $\ref{subsubsec:flowgaussian}$, it is argued that $\rho_1$ and $\rho_2$ do not contribute to the universal part of the boundary free energy, since these boundary conditions are expected to flow towards fixed b.c. Then the universal behaviour only depends on the phase factors $e^{\pm i 2 \theta}$ for each loop wrapping around the annulus, and the rest of the argument goes exactly as the calculation of the character $K_0$ (formula ($\ref{eq:K0}$)) performed in section $\ref{subsubsec:flowgaussian}$.

\paragraph{}
Strings could be added using the results for the $AS_{\blob}/AS_{\blobw}$ case. One could also compute full partition functions instead of simple Virasoro characters, using the Markov trace decomposition $(\ref{eq:Markov})$, and treating the blobbed loops which wind around the annulus with care, following what we did in section $\ref{sec:openbc}$. We do not want to go through this here, and we turn now to a simple application of the open/open b.c to the Ising model.

\subsection{Applications to crossing probabilities of Ising clusters on an annulus}
\label{sec:crossing}
\begin{figure}[h]
\center
$$
\begin{array}{l}
K > K_c \; :\\
\includegraphics[width=\textwidth]{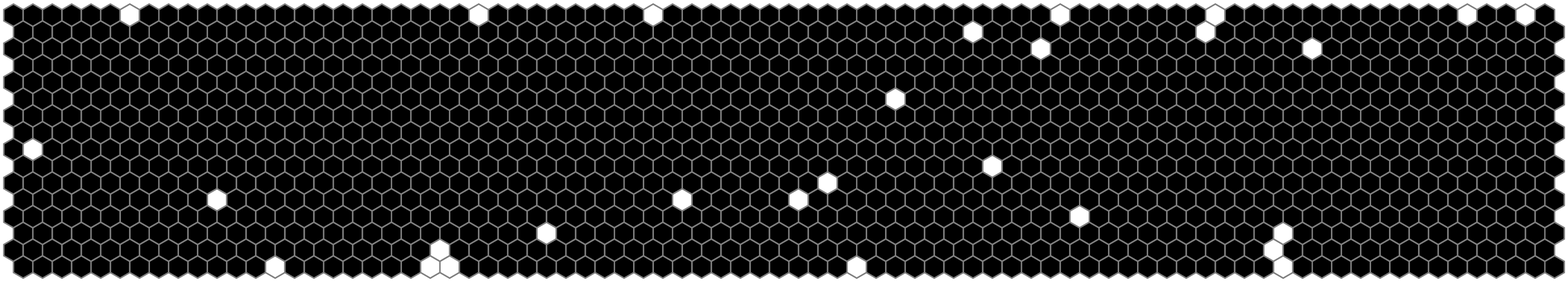} \\
K=K_c  \; :\\
\includegraphics[width=\textwidth]{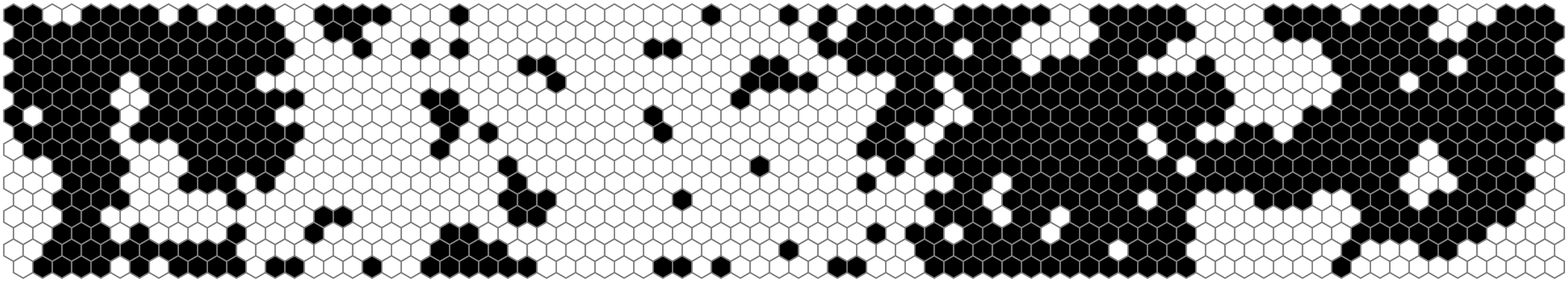} \\
K < K_c \; :\\
\includegraphics[width=\textwidth]{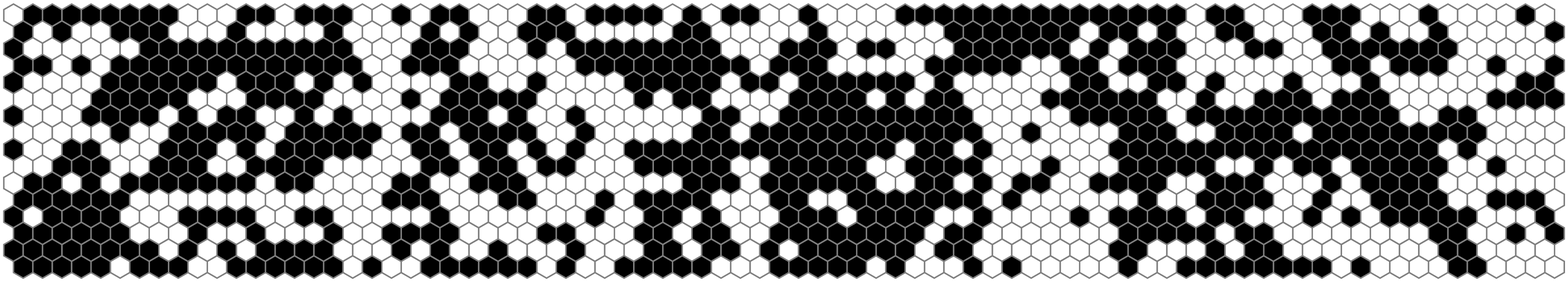} \\
\end{array}
$$
\caption{Ising model on an annulus (periodic boundary conditions on the left and right sides, free on top and bottom) as a function of the coupling $K$. We are interested in the cluster interfaces which cross the annulus from the bottom to the top. In these examples, there is no such interface for $K>K_c$, $2$ interfaces for $K=K_c$ and $6$ for $K<K_c$.}
\label{fig:isingstrip}
\end{figure}

As a particularly simple application of our open/open b.c results, we compute some quantities related to the crossing probabilities of Ising spin clusters on the strip and the annulus. So far in this paper, the $\On$ model was formulated on the honeycomb lattice. When $n=1$ we thus get the Ising model on the honeycomb lattice. This Ising model is dual to the Ising model on the triangular lattice, and this is the main point we use in what follows. The Ising spin cluster on the triangular lattice have boundaries which are exactly the $\mathcal{O}(1)$ loops above with a weight $x$ per monomer. Note that $x$ is related to the coupling energy between two neighbouring spins of the triangular lattice $E=-K \sigma_i \sigma_j$ by $x=e^{2 K}$.

Let us consider the Ising model on an annulus of aspect ratio $\tau = T/L$ where $T$ is the periodic direction. We choose free boundary conditions for the spins. We are interested in the probability $P(\tau)$ that there is at least one cluster boundary crossing the annulus from one boundary to the other (figure $\ref{fig:isingstrip}$). Of course, such a probability behaves non-trivially with the coupling $K$. When $K>K_c$, the model renormalizes towards a trivial theory with all spins frozen in a common state. Thus we expect $P(\tau)=0$ in that case. When $K<K_c$ it renormalizes to a model where all the spins are independent, and they can be in either state with probability $1/2$. This is nothing but critical percolation. The crossing probability for percolation cluster boundaries was first computed by Cardy in \cite{CardyPercolationAnnulus}. It can be written as a ratio of Dedekind eta functions
\begin{equation}
\label{eq:crossingPerco}
P^{\rm{perc}}_c(\tau)=\frac{\eta(i \tau) \eta(i \tau/6)^2}{ \eta(i \tau/2)^2\eta (i \tau/3)}.
\end{equation}
The case we are interested in is of course the case of the critical Ising model $K=K_c$, which should lead to a non-trivial probability $P^{\rm{Ising}}_c(\tau)$. To compute this we need the partition function of the critical Ising model on an annulus with free/free (ordinary/ordinary in the terminology we used above) b.c. We could of course compute this with the above formalism, by plugging $\gamma=\frac{\pi}{3}$, $g=\frac{4}{3}$ into relation $(\ref{eq:Zordord})$ (recall also $(\ref{eq:Kordord})$). This would yield the standard result \cite{CardyBCFTVerlinde,BauerSaleur}
\begin{equation}
Z_{\rm{free/free}}(\tau) = \chi_{1,1}(q) + \chi_{1,3}(q)
\end{equation}
where $\chi_{r,s}$ is the usual Rocha-Caridi character and $q=e^{- \pi \tau}$ as above. We could also have computed this partition function within the open/open b.c framework, because free b.c for the spins on the triangular lattice are equivalent open b.c for the cluster boundaries. We then have $Z_{Open/Open}=Z_{\rm{free/free}}$. The Markov trace structure of $Z_{Open/Open}$ leads to
\begin{equation}
Z_{Open/Open} = K_0(q,n_{12}=1) + \rm{terms \; independent \; of \;} n_{12}.
\end{equation}
In this expression, it is very easy to subtract the configurations which do not contribute to the crossing probability: it is the same expression with $n_{12}=0$ instead of $n_{12}=1$. Since the parameter $n_{12}$ only appears via $r_{12}$ in the character $K_0(q,n_{12})$, one ends up with 
\begin{equation}
\label{eq:crossingP}
P^{\rm{Ising}}_c(\tau)=\frac{q^{-c/24}}{P(q)}\frac{\sum_{n \in \Z} \left(q^{h_{1,1+2 n}} - q^{h_{3,3+2 n}}\right)}{\chi_{1,1}(q)+\chi_{1,3}(q)}
\end{equation}
where we used $(\ref{eq:K0})$ with $r_{12}=1$ ($n_{12}=1$) and $r_{12}=3$ ($n_{12}=0$). The relation $(\ref{eq:crossingP})$ can be reformulated in terms of Dedekind eta function, to get it in a form very close to $(\ref{eq:crossingPerco})$
\begin{equation}
\label{eq:crossingIsing}
P^{\rm{Ising}}_c(\tau) =\frac{\eta(i \tau) \eta(i \tau/12)^2}{ \eta(i \tau/2)^2\eta (i \tau/6)}.
\end{equation}

\paragraph{}
Many other quantities could be computed. Let us mention but one of them before we turn to the numerical check of these results. Instead of computing the probability that there is at least one cluster boundary which crosses the boundary, one could be more precise and ask how many of them there are. One could derive the probability that there are $k \geq 1$ clusters which cross the annulus as a function of the aspect ratio. For critical percolation this was done by Cardy in \cite{CardyPercolationAnnulus}, and we could do it for the critical Ising model. This would turn out to involve more complicated formulae, but we can focus on a simpler quantity. Consider the limit when $\tau \gg 1$, then the annulus looks like a very long strip. We can then ask how many cluster boundaries are crossing the strip per unit length (on average). Again, for percolation this is a result of Cardy \cite{CardyChalker}: there are (on average) $\frac{\sqrt{3}}{2}$ cluster boundaries per unit length which cross the strip if we assume that the width of the strip is $1$. Let us derive the similar result for the critical Ising model here.

\paragraph{}
In the limit $\tau \gg 1$ the partition function $Z_{Open/Open}$ behaves as
\begin{equation}
Z_{Open/Open}(\nu_{12}) \sim q^{h_{r_{12},r_{12}}-c/24}
\end{equation}
where $\nu_{12}$ and $r_{12}$ are related by $(\ref{eq:cardy})$ with $\nu_1=\nu_2=1$. The average number of cluster boundaries per unit length is in this limit
\begin{eqnarray}
\nonumber \frac{\left< N_{\rm{clus. \; bound.}} \right>}{\tau} &=& \frac{1}{\tau} \partial_{\nu_{12}} \log Z_{Open/Open}(\nu_{12})\\
\nonumber & =& - \pi \frac{\partial h_{r_{12},r_{12}}}{\partial \nu_{12}}\\ 
\nonumber &=& \frac{g-1}{2g} \frac{\sqrt{n+2}}{\sin (\gamma/2)}
\end{eqnarray}
which yields $\frac{\sqrt{3}}{4}$ for the critical Ising model.

\paragraph{Numerical check:}

\begin{figure}[h]
\centering
\includegraphics[width=0.65\textwidth]{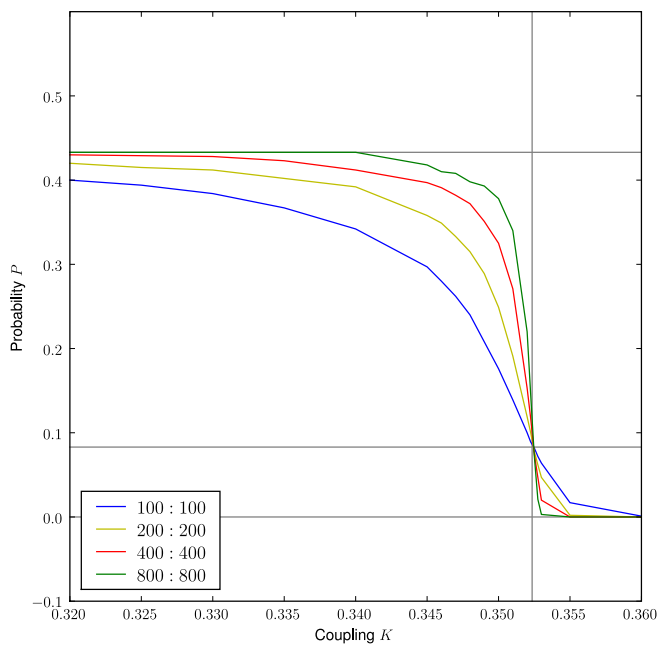}
\caption{Probability that there is at least one contractible spin cluster crossing the annulus for different sizes. The simulations are made on a triangular lattice. The aspect ratio $\tau$ is kept fixed and is equal to $\tau=2/\sqrt{3}$. The horizontal lines correspond to $P=0$, $P=P_{\rm{Crit. \; Ising}}$ and $P=P_{\rm{Crit. \;perco.}}$ from bottom to top. The vertical one is $K=K_c$.}
\label{fig:P_crossing_TL}
\end{figure}

\begin{figure}[h]
\centering
\includegraphics[width=0.85\textwidth]{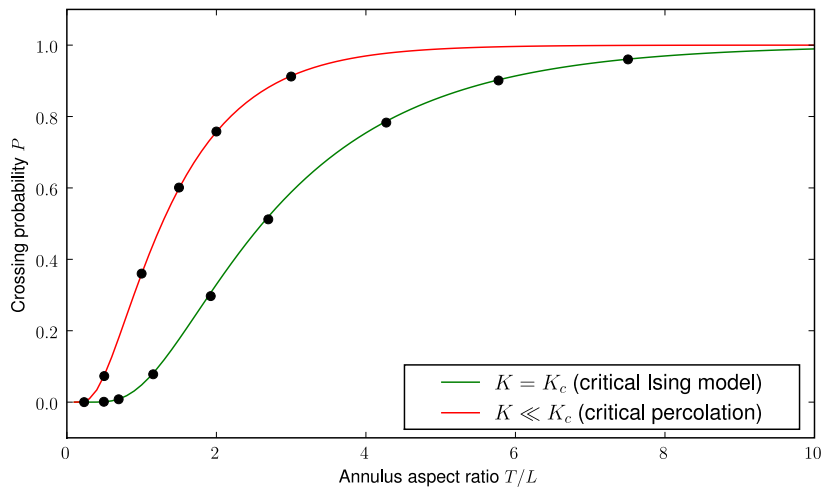}
\caption{Probability that there exists at least one contractible spin cluster crossing the annulus as a function of its aspect ratio. There are two possible non-trivial scaling limits. One is of course when $K=K_c$, then the crossing probability is given by $(\ref{eq:crossingIsing})$. The second one is the low-coupling limit $K \ll K_c$, when the spin clusters look like critical percolation clusters $(\ref{eq:crossingPerco})$.}
\label{fig:P_crossing}
\end{figure}

We checked these results for the Ising model by Monte Carlo simulations. Note that it is very important to be right at the critical point when we compute crossing probabilities numerically, otherwise we are quickly attracted by another critical behaviour (critical percolation if $K<K_c$, frozen spins if $K>K_c$). Because simulations are done in finite size, there is a small shift in the effective critical coupling around $K_c$ \cite{Barber}. It is then a bit difficult to catch the right critical coupling, because of these finite size effects. To define the quantity we actually measure numerically and which is to be compared to our prediction for $P_{\rm{Crit. \; Ising}}$, we proceed as follows. For each aspect ratio $\tau$, we compute the probability $P(\tau,K,N)$ for several $K$ around $K_c$ and different systems of size $N \times (\tau N)$. We plot $P(\tau,K,N)$ as a function of the coupling $K$ for the different sizes $N \times (\tau N)$, and find that the successive curves all intersect in a very small region at some effective critical coupling $K_c(\tau,N)$. This procedure is shown in figure $\ref{fig:P_crossing_TL}$ for $\tau=2/\sqrt{3}$.
The value of the probability at this point is the quantity we compare with our analytic results. For $K<K_c$ we also get a measure for the percolation clusters crossing probability (figure $\ref{fig:P_crossing_TL}$). We plot the results obtained in this way in figure $\ref{fig:P_crossing}$.
For the average number of cluster boundaries crossing a strip per unit length, we simulated a system of aspect ratio $\tau=20$. We find that there are $0.432 \pm 0.002$ cluster boundaries crossing the annulus per unit length (to be compared with the analytical result $\sqrt{3}/4 = 0.4330...$).




\section{Conclusion}

In this article we have introduced a dilute version of the conformal
boundary loop model, the study of which was initiated in \cite{JS1}
and pursued in \cite{JS2,DJSdense}. For clarity, we now recall the
main results obtained in this rather long paper.

\subsection{Summary}

The dilute model is defined by (\ref{eq:ZOnanisotropic}) in terms of
${\cal O}(n)$ type vector spins, and by (\ref{eq:ZOnLoopboundary}) in
terms of an ensemble of self-avoiding loops. In the spin
representation, its most important physical feature is the possibility of
attracting preferentially a subset of $n_1$ indices (or, by duality,
the remaining $n-n_1$ indices) towards the boundary. Correspondingly,
in the loop representation, boundary touching loops come in two
versions (blobbed and unblobbed) with respective weights $n_{\blob}=n_1$
and $n_{\unblob}=n-n_1$.

An essential ingredient is the identification of two sets of
integrable weights in a corresponding lattice model. These read
(\ref{eq:SklyaninBlob1})--(\ref{eq:SklyaninBlob2}) for the square
lattice, and (\ref{eq:conjectureAS}) in the honeycomb limit. It should
be noticed that the integrable weights impose particular weights for
the boundary monomers of blobbed and unblobbed loops.

The ordinary and special transitions (denoted $Ord$ and $Sp$) in the
${\cal O}(n)$ model are well studied in the literature, and correspond
in our setting to the special case where blobbed and unblobbed loops
are indistinguishable. The integrable points found here complete the
picture by defining a pair of anisotropic transitions $AS_{\blob}$ and
$AS_{\unblob}$. The physical interpretation of $AS_{\blob}$ is that
blobbed loops are critically attracted towards the boundary (i.e.,
they stand at a special transition), whereas the unblobbed loops are
repelled from the boundary (i.e., they stand at an ordinary
transition). This interpretation is validated by the results of
section \ref{sec:fractal_dimensions} on the fractal dimensions of the
contact sets of each loop type.  Note that $AS_{\unblob}$ is obtained
from $AS_{\blob}$ by exchanging $n_1$ and $n-n_1$.
The complete phase diagram (in the physical region $0 < n_1 < n$) is
shown in figure \ref{fig:phase}.

Using arguments of boundary conformal field theory (BCFT) and Coulomb
gas we have identified the boundary-condition-changing (B.C.C)
operators corresponding to the above transitions. In particular, we
have $\Phi_{r_1,r_1+1}$ for $(AS_{\blob}|Ord)$, and $\Phi_{r_1,r_1}$
for $(AS_{\unblob}|Ord)$. We have also obtained the corresponding
Virasoro characters (\ref{eq:Kas1ord}) in the sector with $L$
non-contractible loops. There can be combined into conformally
invariant partition functions (\ref{eq:Zas1ord}) which encode the full
operator spectrum of the theory and the corresponding multiplicities.

The boundary entropies of the conformal boundary states identified in
this work are given in (\ref{eq:gord})--(\ref{eq:gas2}). They are
consistent with the $g$-theorem \cite{AffleckLudwig} and the proposed
phase diagram (see figure~\ref{fig:phase}). We have also identified
the operators perturbing the isotropic points along the anisotropic
direction, which are $\Phi_{3,1}$ at $Ord$ and $\Phi_{3,3}$ at
$Sp$. This shows that the anisotropic transition lines of
figure~\ref{fig:phase} form a cusp where they join in $Sp$. Starting
at the anisotropic point $AS_{\blob}$, the operator perturbing in the
unstable direction was identified in section~\ref{sec:TBA} as
$\Phi_{1,3}$.

Finally, we have shown in section~\ref{sec:open} that open boundary
conditions for the ${\cal O}(n)$ model are a special case ($n_1=1$) of
those considered here. They correspond to imposing a magnetic field on
the boundary spins. As a simple geometrical application, we have
derived a new crossing probability (\ref{eq:crossingIsing}) for Ising
domain walls, and found the number of cluster boundaries per unit
length that cross an infinite strip.

\subsection{Outlook}

Following \cite{DJSdense} one may consider a more general model on the
annulus with two distinguished boundaries and loop weights
(\ref{eq:n1})--(\ref{eq:n2}) and (\ref{eq:n12}). In this model, one can
also introduce separate weights for contractible and non-contractible
loops. We have here only alluded briefly to this generalization, which
will be treated more fully elsewhere \cite{DJSprep}.

While this work completes the conformal part of the program set out in
\cite{JS1}, we have omitted here a whole range of algebraic questions.
In particular, one may study the (rational) restrictions of the
conformal boundary loop models when some of the parameters take
particular ``magic'' values. This will also be treated in
\cite{DJSprep}.

Another future direction would be to exploit the integrable solutions
(\ref{eq:SklyaninBlob1})--(\ref{eq:SklyaninBlob2}) of the reflection
equation to set up the corresponding Bethe ansatz
equations. Presumably this would put our results for the critical
exponents and the spectrum generating (partition) functions on a
rigorous basis, and would allow to deduce non-universal quantities
such as the surface free energies.

Also, let us discuss what may happen when boundary loops can be
decorated with $k$ different orthogonal blobs. The weight of an
$i$-blobbed loop (for $i=1,2,\ldots,k$) is $n_i$ and its boundary
monomers have fugacity $w_i$. Obviously, $n = \sum_{i=1}^k n_i$, and
the preceding discussion corresponds to the case $k=2$. When several
$w_i$ coincide, the corresponding blobbed loops are indistinguishable
and their weights $n_i$ may be regrouped.  In particular, when all
$w_i$ coincide we recover the transitions $Ord$ and $Sp$.

Suppose now that the set $\{w_i\}$ takes precisely two different
values, with $k_1$ weights equal to $w$ and $k-k_1$ weights equal to
$\bar{w}$, and $w > \bar{w}$. Regrouping the corresponding loop
weights we are then, in fact, in the situation $k=2$ where the
previous results apply. Correspondingly, we have an anisotropic
special transition, with $k_1$ groups of indices standing at a
special transition, and the remaining $k-k_1$ groups standing at an
ordinary transition. In the full parameter space $\{w_i\}$, this
critical point has $k_1$ unstable directions (corresponding to moving
a $w_i$ away from $w$) and $k-k_1$ stable directions (corresponding to
moving a $w_i$ away from $\bar{w}$). This argument yields a total of
$2^k$ critical points (including $Ord$ and $Sp$). We cannot exclude
the existence of further multi-critical points with finite values of
$\{w_i\}$, but we conjecture that no such point exists.

Finally, we expect that there are applications of this work to other physical problems. One of these is the re-intepretation of  our conformal boundary conditions in terms of boundary degrees of freedom critically coupled to the bulk. While such a reinterpretation involves $U_q[su(2)]$ spins in the case $r_1$ an integer, it is not clear what is the meaning of $r_1$ not integer, and whether this has something to do with $sl(2,\mathbb{R})$. One could also wonder whether our CBCs have any relation with boundary bound states in the S matrix description of the bulk CFTs \cite{Skorik}. 
Finally, we note that the limit $n\to 2$ is deeply related with the Kondo model. What happens to our phase diagram in this case is also an open problem.

\subsection*{Acknowledgements}

We thank J.~Cardy for helpful comments and for correspondence, and J.-E.~Bourgine and I.~Kostov for discussion. This work was supported by the European Community Network ENRAGE (grant
MRTN-CT-2004-005616), by the Agence Nationale de la Recherche (grant
ANR-06-BLAN-0124-03), and by the ESF Network INSTANS.




\end{document}